\documentclass{aa}  
\usepackage{natbib}
\bibpunct{(}{)}{;}{a}{}{,} 
\usepackage[pdfpagelabels]{hyperref}
\usepackage{xcolor} 
\usepackage{comment}
\usepackage{enumitem}
\usepackage{booktabs}
\usepackage{color}
\usepackage{tabularx}
\usepackage{graphicx}
\usepackage{txfonts}
\usepackage{lipsum}
\usepackage{subcaption}         
\usepackage{lscape}             
\usepackage{placeins}           
                                
\begin{document}

   \title{Precise Deep-Learning Reconstruction of Galaxy Cluster Mass Maps and Substructures from Optical Observations}

%
%
%

\author{Atulit Srivastava\inst{1,2}\corrauth{atulit.srivastav@uam.es}
        \and Weiguang Cui\inst{1,2,3}\email{weiguang.cui@uam.es}
        \and Daniel de Andres\inst{4}\email{daniel.deandres@urjc.es}
        \and Guillaume Bonnet\inst{5}
        \and Romeel Dav\'e\inst{3,6}
        \and Stefano Ettori\inst{7}
        \and Massimo Meneghetti\inst{7,8}
        \and Carlo Giocoli\inst{7,8}
        \and Gustavo Yepes\inst{1,2}
        }

   \institute{Departamento de Física Teórica, M-8, Universidad Autónoma de Madrid, Cantoblanco, Madrid 28049, Spain
   \and Centro de Investigación Avanzada en Física Fundamental (CIAFF), Universidad Autónoma de Madrid, Cantoblanco, Madrid 28049, Spain
   \and Institute for Astronomy, University of Edinburgh, Blackford Hill, Edinburgh EH9 3HJ, UK
   \and Nonlinear Dynamics, Chaos and Complex Systems Group, Departamento de Geología, Física y Química Inorgánica, Universidad Rey Juan Carlos, Tulipán s/n, Móstoles, Madrid 28933, Spain
   \and Data Science Department, Euranova, Rue Emile Francqui 4, 1435 Mont-Saint-Guibert, Belgium
   \and Department of Physics and Astronomy, University of the Western Cape, Robert Sobukwe Rd, Cape Town 7535, South Africa
   \and INAF--Osservatorio di Astrofisica e Scienza dello Spazio di Bologna, Via Piero Gobetti 93/3, I-40129 Bologna, Italy
   \and INFN--Sezione di Bologna, Viale Berti Pichat 6/2, I-40127 Bologna, Italy}

   \date{Received September 30, 20XX}


    \abstract
    {Accurately mapping the total mass distribution within galaxy clusters, including dark matter, is fundamental to understanding structure formation and constraining cosmological parameters. Standard mass mapping relies almost exclusively on gravitational lensing, which often yields smooth distributions based on analytical profiles, thereby missing critical internal substructures.}
    {We introduce a complementary, simulation-based inference approach to reconstruct detailed galaxy cluster mass maps and resolve internal substructures directly from optical imaging.}
    {We trained deep learning models on realistic hydro-simulated galaxy clusters to interpret observed cluster images. This framework was then applied to observations of real galaxy clusters obtained by the HST as part of the CLASH program.}
    {The reconstructed mass maps successfully reproduce the overall shapes and radial mass profiles of the CLASH clusters, showing strong agreement with independent gravitational lensing measurements. Using the subhalos derived from these reconstructed maps, we find that their mass functions exhibit a power-law behaviour broadly consistent with predictions from numerical simulations.}
    {Our results demonstrate that combining optical imaging with deep learning provides a scalable, physically meaningful approach to mapping cluster masses. This framework opens a new avenue for utilizing upcoming wide-field imaging surveys to systematically study dark matter substructures, complementing traditional lensing techniques across much larger samples. Furthermore, the model is inherently flexible and can integrate multi-wavelength data, such as X-ray and Sunyaev-Zeldovich observations, to further improve total mass predictions.}
    
    \authorrunning{Srivastava et al.}
    \titlerunning{Deep-Learning Mass and Substructure Mapping of Galaxy Clusters}
    
    \maketitle
\nolinenumbers

\section{Introduction}\label{section:introduction}
Galaxy clusters, which represent the final stage of hierarchical structure formation, are primarily composed of dark matter, contributing $\sim 85\%$ of their total mass, while ordinary baryonic matter (i.e., gas and stars) accounts for only the remaining 15\%\citep{White1993,Fukugita1998}. The dominant dark matter component cannot be directly observed; its presence must be inferred from its gravitational influence on ordinary matter. Therefore, estimating the total mass of clusters (including both dark and baryonic matter) is challenging. However, this is crucial for many astrophysics and cosmology studies. For example, the abundance of galaxy clusters as a function of mass and redshift is highly sensitive to the fundamental cosmological parameters such as the matter density parameter $\Omega_{m}$ and the mass fluctuation amplitude $\sigma_{8}$ on a scale of 8 $h^{-1}$ Mpc. As such, cluster masses can serve as a powerful tool for constraining cosmological parameters \citep{vikhlinin2009chandra,mehrtens2012xmm,ade2016planck}. Hence, it is essential to improve the accuracy of total mass estimates in galaxy clusters (for an extensive review see \citealp{pratt2019galaxy}).

The global masses of galaxy clusters are commonly inferred from X-ray \citep{vikhlinin2006chandra, kravtsov2012formation,ettori2013mass} and Sunyaev–Zel’dovich (SZ) observations \citep{planck2015results,adam2015pressure,eckert2022gravitational} of the intracluster medium, or from the kinematics of member galaxies \citep{aguado2022velocity,biviano2003mass,saro2013toward}. 
These approaches rely on assumptions of hydrostatic or dynamical equilibrium, which are sensitive to non-thermal gas motions, substructure, projection effects, and contamination by interlopers, leading to systematic biases—most notably the hydrostatic mass bias—that can underestimate the true cluster mass \citep{rasia2004dynamical, meneghetti2010weighing, biffi2016nature, old2018galaxy, gianfagna2021exploring}. Cosmological simulations play a key role in robustly quantifying the systematics and biases \citep{piffaretti2008total, bahe2012mock, ferragamo2020biases, munoz2024galaxy}. Nevertheless, these methods are limited to determining total cluster masses within only a few characteristic radii. As such, they cannot recover detailed mass distributions or characterize the subhalo population within galaxy clusters.

Alternatively, strong and weak gravitational lensing \citep{kaiser1994method,bartelmann2001weak,kneib2011cluster,zitrin2016strong, okabe2016locuss} provide a direct probe of the total projected mass distribution of galaxy clusters, which can reveal such detailed information through lensing maps; see \cite{meneghetti2020excess} for the galaxy-galaxy lensing crisis. However, its sensitivity to line-of-sight projection effects such as halo triaxiality and uncorrelated foreground and background structure can limit its ability to precisely recover the mass distribution \citep[e.g.][]{giocoli2025three}.

Recent deep learning (DL)–based mass inference methods trained on cosmologically based hydrodynamical simulations have attracted significant attention. This enables a more direct data-driven modeling approach that is independent of these conventional theoretical assumptions discussed above. DL models trained on simulated X-ray, Sunyaev-Zel’dovich, galaxy kinematics, and weak-lensing data have been shown to recover cluster masses with high fidelity \citep{ntampaka2019deep, ferragamo2023three, kodi2020dynamical, hong2021weak}. By applying the DL method to the observed Planck SZ cluster images, \cite{de2022deep} showed that it can recover cluster masses without a bias. More recently, DL techniques have been shown to successfully reconstruct projected total mass maps of galaxy clusters from multi‑wavelength images generated by hydrodynamic simulations, providing far richer information than scalar mass estimates alone \citep{de2024three, 2025OJAp....8E..92H}. However, that approach has so far been tested only on simulations and has not yet been applied to real observations.

In this work, we build upon \citet{de2024three} by applying a U-Net–based deep learning model to predict projected mass maps as output for galaxy clusters observed with the Hubble Space Telescope (HST), focusing on the Cluster Lensing and Supernova Survey with Hubble (CLASH; \citealt{postman2012cluster}). The CLASH program comprises 25 massive clusters spanning $0.19 \lesssim z \lesssim 0.89$ with high-quality multi-band HST observations, making it well suited for lensing and mass map studies. Furthermore, all CLASH clusters have both reconstructed strong- and weak-lensing mass maps \citep{zitrin2015hubble, umetsu2016clash}, which enable a direct validation of our mass maps inferred via deep learning. 
Note that the original CLASH lensing results are adopted here for consistency considerations; therefore, their latest results from the Frontier Fields programme, as well as the constructed maps from other works, are not included.

To train the U-Net model, we select a sample of 3,273 simulated clusters from The Three Hundred project \citep[The300;][]{cui2018three}, selected to match the CLASH clusters in both mass and redshift. The PyMGal \citep{janulewicz2025pymgal} package is used to mimic the HST ACS/WFC optical-to-NIR filters (F606W, F625W, F775W, F814W, and F850LP), as well as the noise and instrumental characteristics of the real CLASH HST data. Therefore, seven U-Net models in total—five using the individual HST ACS/WFC filters and two using multi-filter input configurations—are trained.  The remainder of this paper is structured as follows. In Section~\ref{The300 simulations}, we introduce the numerical datasets and sample selection criteria from The300 simulation suite. Section~\ref{Mock HST Cluster Observation from The300} outlines the methodology for generating mock HST observations using PyMGal, details our instrumental effects and noise modeling, and describes the creation of the target projected mass maps. In Section~\ref{sec:The Unet algorithms}, we present the architecture of our single-channel and multi-filter U-Net models alongside the data preprocessing and training configurations. Section~\ref{Unet-Test-set} evaluates the performance of the trained U-Net models on the independent simulation test dataset , while Section~\ref{sec:Results} details their application to real HST observations of the CLASH clusters. Here, we validate the deep-learning–inferred mass maps against independent strong- and weak-lensing models \citep{zitrin2015hubble} and X-ray observations \citep{donahue2014clash}. Furthermore, we derive the cluster subhalo masses to verify the well-known power-law mass function predicted by simulations \citep[][and references therein]{gao2004subhalo, de2004substructures, van2005mass, giocoli2008analytical}. Finally, our conclusions are summarized in Section~\ref{sec:Conclusions}.

\section{The Three Hundred Simulations}\label{The300 simulations}
We used a sample of galaxy clusters based on the suite of hydrodynamical simulations from The Three Hundred project \citep{cui2018three}.
The300 utilizes the spherical regions around the 324 most massive cluster regions identified in the dark-matter-only MDPL2 simulation \citep{klypin2016multidark} with a box size of $1 h^{-1}$ Gpc. Each of these selected cluster regions is re-simulated at high resolution using the zoom-in simulation technique within a radius of $15 h^{-1}$ Mpc by adopting the Planck cosmology parameters \citep{planck2016cosmological} of the parent MDPL2 simulation $\Omega_{M}=0.307, \Omega_{B}=0.048, \Omega_{\Lambda}=0.693, h=0.678, \sigma_{8}=0.823, n_{s}=0.96$, including different baryonic physics models within the high-resolution volume. The high-resolution particle masses are m$_{\text{DM}} \approx 1.27 \times 10^{9} h^{-1} M_{\odot}$ for dark matter and m$_{\text{gas}} \approx 2.36 \times 10^{8} h^{-1} M_{\odot}$ for gas. Outside the $15 h^{-1}$ Mpc region, the simulations use lower-resolution particles to capture large-scale tidal fields efficiently.

For this study, we focus on clusters re-simulated with the Gizmo-Simba sub-grid baryonic physics model \citep{dave2019simba,cui2022three}. For each cluster, 128 snapshots are saved with redshifts ranging from $z=16.98$ to $z=0$. The dark matter halos for each cluster region in the The300 project simulation are identified using the Amiga Halo Finder (AHF) \citep{knollmann2009ahf}.
Halos are defined as spherical regions with an average density of 500 times the redshift-dependent critical density, $\rho_\mathrm{crit}(z)$, yielding a mass
\begin{equation}
M_{500c} = \frac{4 \pi}{3} R_{500c}^3 \, 500 \, \rho_\mathrm{crit}(z),
\end{equation}
where $R_{500c}$ is the corresponding halo radius.

\begin{figure}[tbp]
    \centering
    \includegraphics[width=0.8\linewidth]{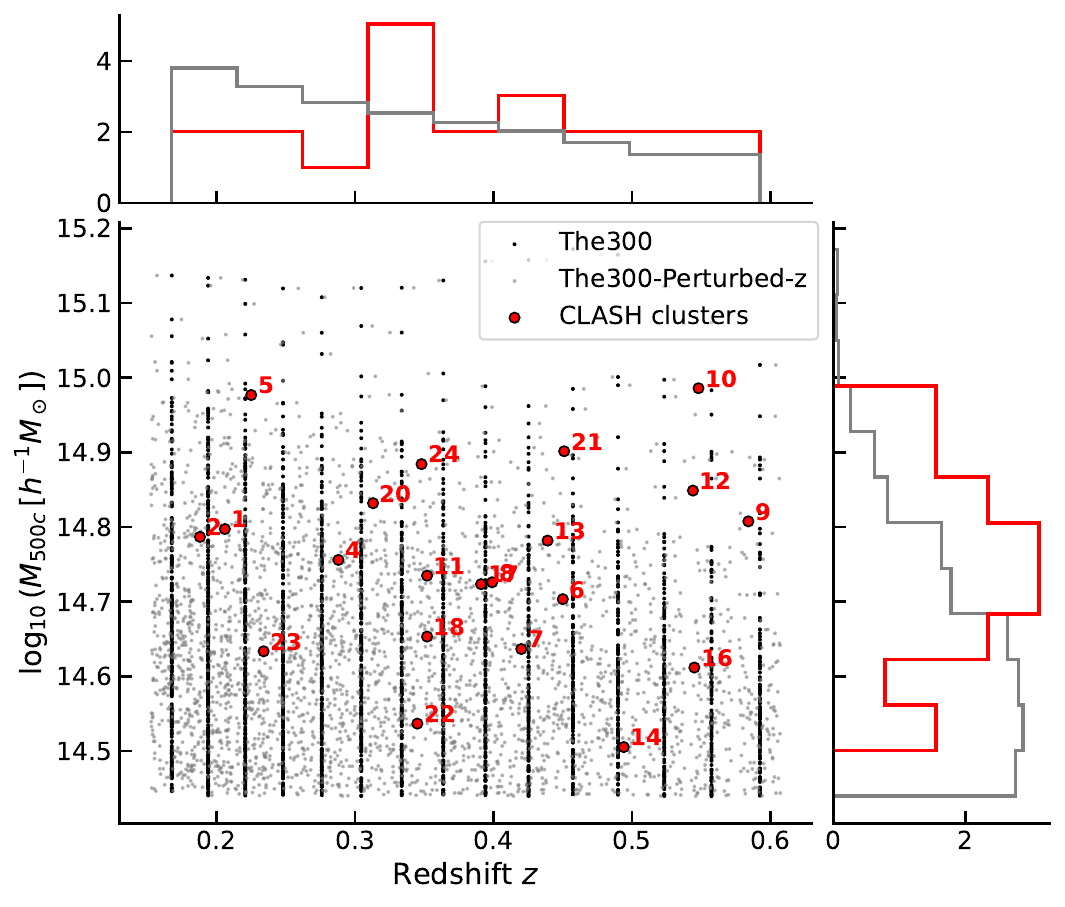}

\caption{Cluster $M_{500c}$ versus redshift $z$ distribution for the selected sample. Black and grey points denote The300 simulated clusters at their exact simulation redshifts and with an applied uniform random perturbation ($\delta z \sim \mathcal{U}(-0.015, 0.015)$), respectively.
Red points and annotations indicate observed \object{CLASH} clusters, labeled by their IDs from Table~\ref{Table:Clash Clusters}.
The marginal histograms display the projected $z$ ($x$-axis) and $M_{500c}$ ($y$-axis) distributions for both The300 (grey) and CLASH (red) samples.}
   \label{fig:cluster_distribution}\label{fig:Dataset}
\end{figure}

The procedure adopted for selecting the simulated halos used to train the deep learning models is as follows:
\begin{itemize}
\item Based on the redshift and lensing-based $M_{500c}$ estimates of the CLASH clusters in Table~\ref{Table:Clash Clusters}, we selected 3,273 halos from The300 simulations that match both the $M_{500c}$ and the redshift of the CLASH sample. These halos span the redshift range $0.152 < z < 0.607$. We also required $f_{\mathrm{Mhires}} > 0.99$, where $f_{\mathrm{Mhires}}$ is the fraction of the halo mass contained in high-resolution particles; this ensures that the halos are uncontaminated by low-resolution particles and lie within the high-resolution zoom-in region.

\item In Fig.~\ref{fig:Dataset}, we present the distribution of the selected The300 halos alongside the CLASH clusters, where the simulated counterparts matched in redshift and $M_{500c}$ are shown as black points, the perturbed halos (with random redshift shifts drawn uniformly within $\pm 0.015$ to break degeneracies) are shown as grey points, and the CLASH clusters are shown as red points annotated following Table~\ref{Table:Clash Clusters}.

\item The redshift histogram indicates that the simulated halos broadly cover the redshift range of the CLASH clusters; however, the $M_{500c}$ histogram reveals that the simulated sample is not uniformly distributed and contains fewer high-mass clusters, whereas a flatter distribution would be preferable to avoid bias.

\item The catalog was split into training, validation, and test sets using a simulation-region-based strategy, in which halos were first grouped by simulation region identifier and the split was performed on the set of unique regions rather than on individual halos. This ensures that all snapshots and evolutionary states of a given halo, including its progenitors and later stages, remain within the same dataset and do not appear in multiple splits.

\item Out of a total of 284 simulation regions, 11\% were allocated to the test set, while 15\% of the remaining regions were used for validation and the rest for training; aggregating the halos within these regions resulted in 2,497 halos for training, 430 halos for validation, and 346 halos for the test set.
\end{itemize}

\section{Mock HST observations and projected mass maps}\label{Mock HST Cluster Observation from The300}
\subsection{Theoretical Mock HST observations using PyMGal}\label{Cleaned HST observation image}
Mock HST observations (i.e. input dataset) for each simulated cluster were generated in units of $\texttt{erg/sec}$ using $\texttt{PyMGal}$ \citep{janulewicz2025pymgal}. $\texttt{PyMGal}$ first uses $\texttt{EzGal}$ \citep{mancone2012ezgal} to assign Spectral Energy Distributions (SEDs) to each star population in the simulation based on a simple Stellar Population Synthesis (SPS) model and the physical properties of the particles, namely their age, metallicity, and mass. The SEDs, together with the filter response files of the respective telescope instruments, are then used to compute magnitudes in the observer’s frame following the $\texttt{EzGal}$ approach. These magnitudes are then converted to luminosities in the observer's frame and finally projected to create synthetic observations. 
While generating our mock HST observations with \texttt{PyMGal}, we adopted the simple stellar population (SSP) models of \cite{bruzual2003stellar} with a Chabrier initial mass function \citep{chabrier2003galactic}, which is agreed to the simulation, and used the HST ACS/WFC F606W, F625W, F775W, F814W, and F850LP filter response curves to produce the mock HST maps.

To augment the dataset, each cluster was rotated around its center along 30 different lines of sight to generate 30 pseudo-independent mock observations per filter. The depth along the line of sight for generating the mock observations is $\pm 5 R_{500c}$.
We generated mock HST observations for each simulated cluster with a fixed physical FOV of $500 \times 500$ kpc$^{2}$.
The number of pixels (N$_{pix}$) depends on the cluster redshift through the angular diameter distance, which is computed assuming Planck cosmology, since the pixel scale is fixed to the HST value of 0.065 arcsec per pixel.

\subsection{Instrumental effects and noise modeling}\label{Noise Modelling}
The synthetic HST observations generated with \texttt{PyMGal} do not include the point spread function (PSF) or other filter-specific instrumental effects. To produce more realistic mock observations that better mimic real HST data, we incorporate these effects in a post-processing step. Specifically, for each filter, the mock maps were first convolved with the corresponding PSF obtained from the CLASH PSF model archive\footnote{\url{[https://archive.stsci.edu/missions/hlsp/clash/psf_models/}}.

To model realistic background noise, we used blank-sky regions from the real CLASH HST data that contain minimal sources and are located far from the cluster centers. For each filter, we extracted multiple patches of approximately $400 \times 400~\text{pix}^2$ from the corresponding HST images and processed them using the \texttt{photutils} package \citep{bradley2016photutils}. Small residual sources in the extracted patches were identified and masked using background estimation combined with $2.5\sigma$ threshold-based segmentation. The masked regions, including the faint wings of detected sources, were then replaced with randomly sampled background pixels drawn from the same image to preserve the intrinsic noise properties.

The resulting source-free noise patches were mean-subtracted and averaged to construct a representative correlated noise template for each filter. These noise templates were converted to physical flux units ($\mathrm{erg s^{-1} cm^{-2}}$) using the \texttt{PHOTFLAM} calibration factor and the corresponding filter bandwidth. To incorporate noise into the mock HST observations, the noise maps were further converted from flux to luminosity units ($\mathrm{erg,s^{-1}}$) using the luminosity distance appropriate for each simulated cluster redshift. The noise was then added to the mock images using \texttt{GalSim} \citep{rowe2015galsim}.

The final noise added to each mock observation corresponds to a random realization of a correlated noise field based on the source-free noise templates derived from the real CLASH data. This \texttt{GalSim}-based procedure ensures that each mock HST observation incorporates a unique noise realization while preserving the underlying two-dimensional correlation properties of the original HST background.

\subsection{Projected Mass Maps from The300 Simulations}\label{Mass-Maps The300}
\begin{figure}[tbp]
    \centering
    \includegraphics[width=\linewidth]{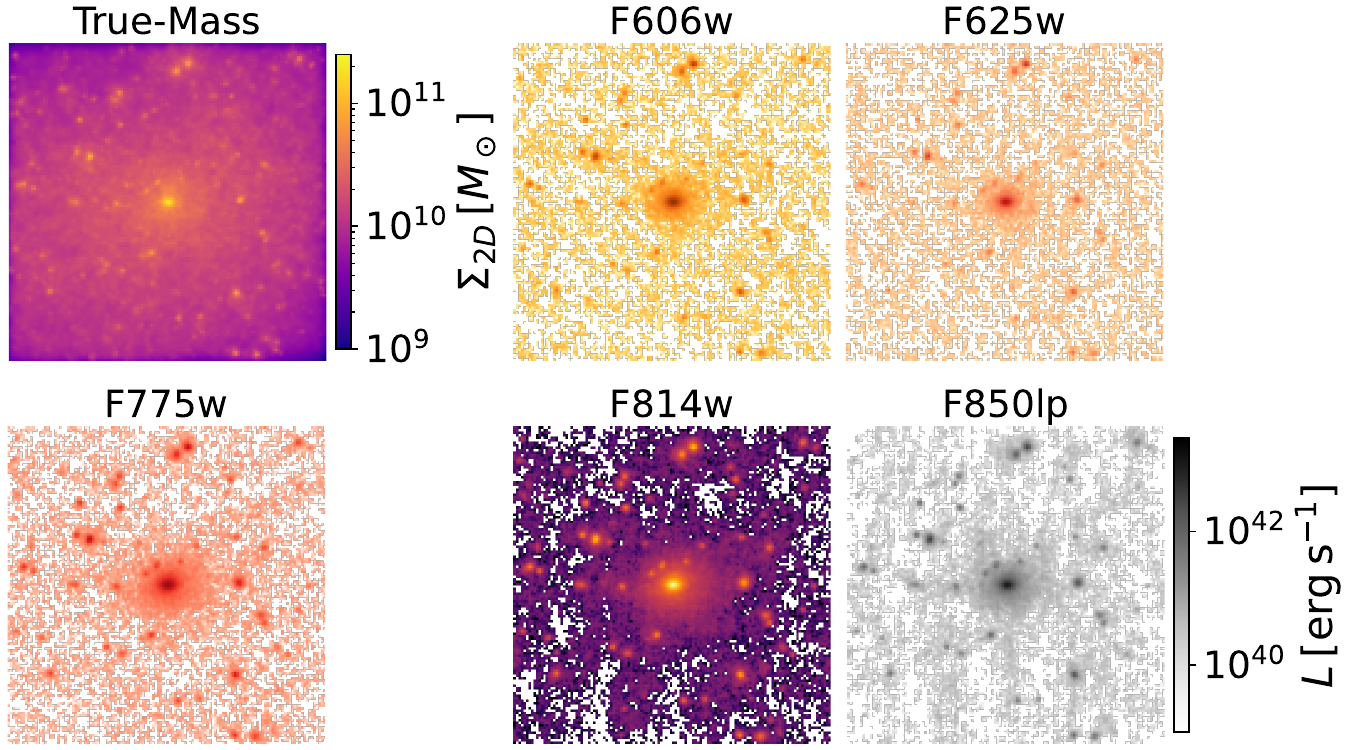}
\caption{Mass and mock HST luminosity maps of an example cluster from The300 simulation. Top left: Projected total mass distribution ($\log_{10} M_\odot$). Remaining panels: Mock HST optical luminosity maps across the F606W, F625W, F775W, F814W, and F850LP filters on a shared logarithmic scale. All panels share the same field of view, line of sight, and corresponding color bar scales.}\label{fig:test-set-300-example}
\end{figure}

The corresponding projected mass maps (i.e. target dataset) for each simulated cluster were also generated by projecting the masses of all particles in the simulation—i.e., gas, star, dark matter, and black hole particles—along the line of sight, using the same pixel resolution scale and FOV as the mock HST observation maps. The depth along the line of sight for generating the mass maps is also $\pm 5 R_{500c}$.
Finally, both the mock HST observations and the mass maps for the simulated clusters were down-sampled to a common pixel size of $128 \times 128$, ensuring that they are conserved during the down-sampling procedure. This ensures that the dimensionality of the maps is consistent and helps stabilize the model during training. 
This also makes the dataset manageable for training on GPUs. In Fig.~\ref{fig:test-set-300-example}, we show the projected total mass map and the mock HST luminosity maps in different filters, expressed in $\mathrm{erg\,s^{-1}}$, for one of the The300 simulated clusters from the test set.

\section{U-Net Architecture and Training}\label{sec:The Unet algorithms}
The deep-learning architecture used to reconstruct the projected total mass maps is based on the U-Net framework, originally developed for biomedical image segmentation \citep{ronneberger2015u} and subsequently applied to a wide range of image-to-image regression tasks. The U-Net model was previously tested on The300 simulations in \cite{de2024three} to reconstruct projected mass maps using idealized X-ray, Sunyaev–Zel’dovich, and stellar mass density maps. However, their study did not include observational noise or systematic effects in the generated maps and therefore did not test the model on realistic observational data.  In this work, we train the U-Net on mock HST observations that incorporate realistic HST instrumental effects and noise for each filter, enabling a direct application of the model to CLASH clusters observed with HST. Below, we describe our four-level single-channel U-Net architecture, which predicts projected mass maps using a single HST/ACS/WFC filter observation (F606W, F625W, F775W, F814W, or F850LP), as illustrated in Fig.~\ref{fig:Single-Channel UNET}.

\begin{figure}[tbp]
\centering
    \includegraphics[width=\linewidth]{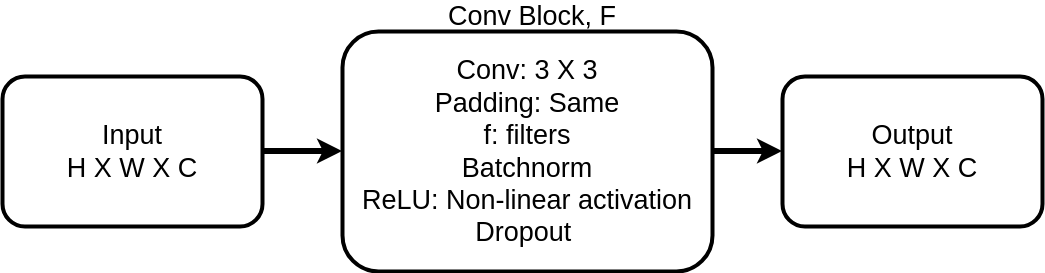}
    
    \vspace{1cm}    
    
    \includegraphics[width=\linewidth]{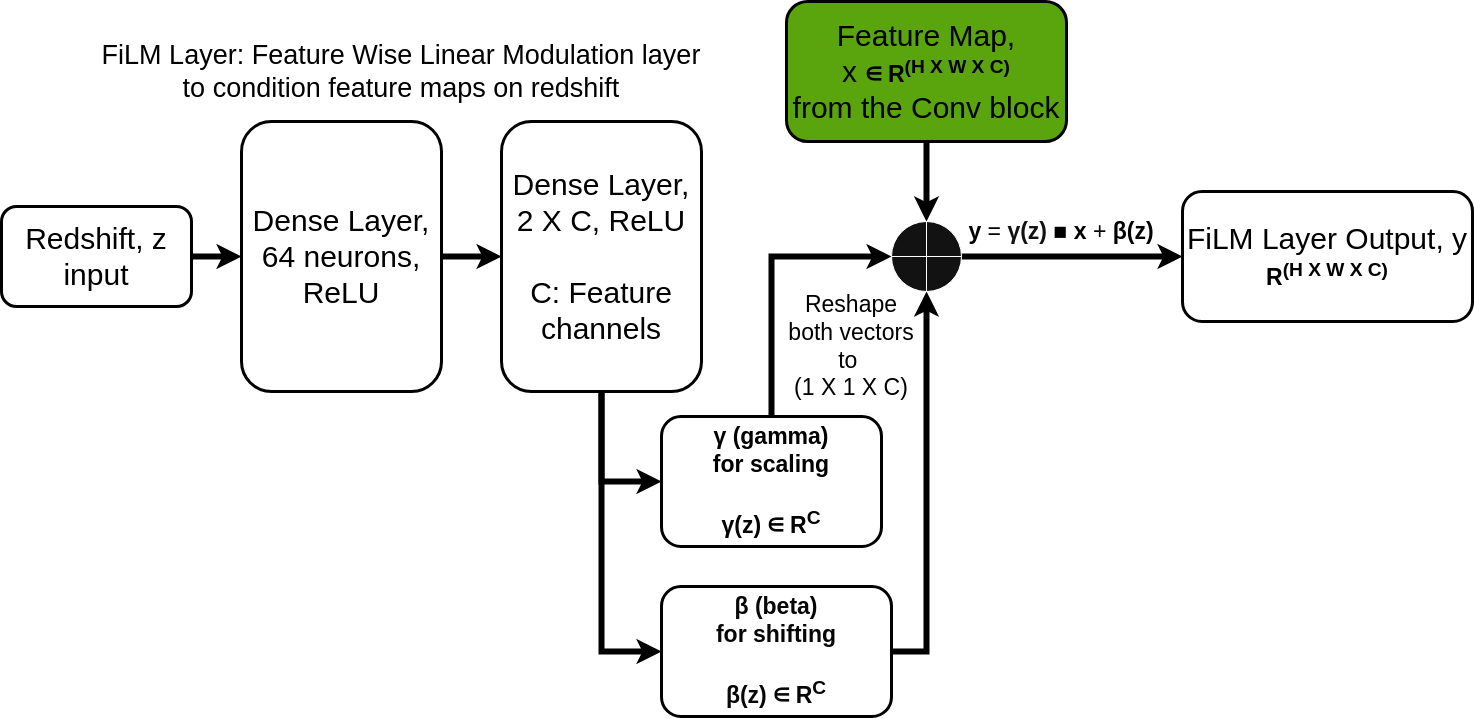}
    
\caption{Schematic of the convolutional block (top) and Feature-wise Linear Modulation (FiLM) layer (bottom) implemented in the U-Net architecture. Each convolutional block contains two sets of convolutional layers ($F=f$ filters), batch normalization, and ReLU activation functions, used across both down-sampling and up-sampling paths. The FiLM layer modulates the bottleneck and decoder feature maps to condition the network on cluster redshift $z$ using learned scale $\gamma(z)$ and shift $\beta(z)$ parameters.}
\label{fig:conv_block_film_layer}
\end{figure}

\begin{figure}[tbp]
    \centering
    \includegraphics[width=.75\linewidth]{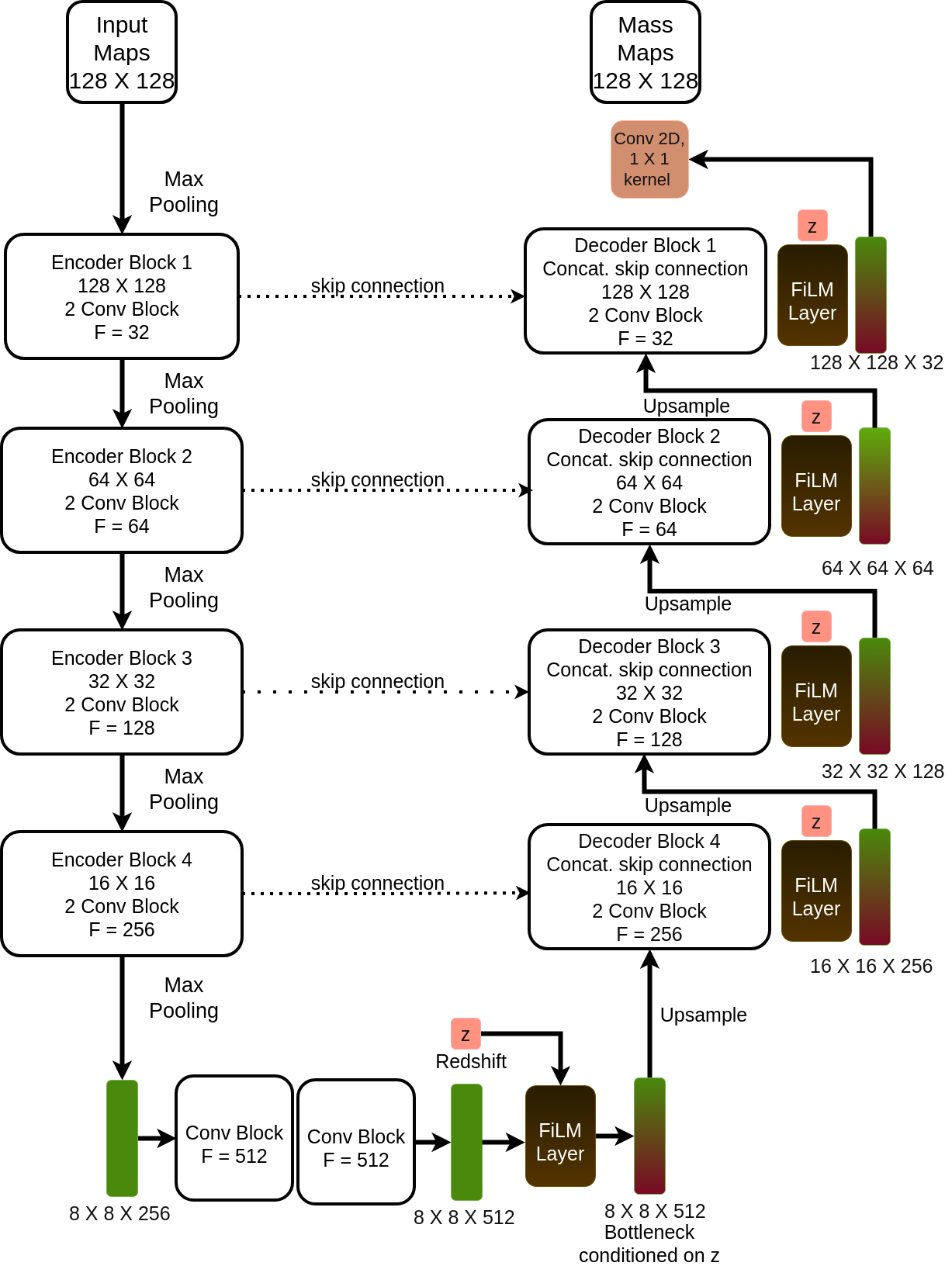}
    \caption{
Single-channel U-Net architecture for processing individual $128 \times 128 \times 1$ HST filter images (F606W, F625W, F775W, F814W, or F850LP). The network compresses the input through a four-stage encoder (starting with $F=32$ filters) to a bottleneck, followed by a symmetric decoder path with skip connections to reconstruct the final projected mass map. Each stage utilizes the convolutional block shown in Fig.~\ref{fig:conv_block_film_layer}. Redshift conditioning is integrated at the bottleneck and throughout the decoder using FiLM layers to modulate the latent feature maps, capturing redshift-dependent features.}
    \label{fig:Single-Channel UNET}
\end{figure}

\begin{enumerate}
\item The encoder (down-sampling) path consists of successive convolutional blocks that progressively reduce the spatial resolution of the input image while increasing the depth of the feature representation. Each block comprises two
3$\times$3 convolutional layers, each followed by rectified linear unit (ReLU) activation \citep{nair2010rectified} and batch normalization \citep{ioffe2015batch}. The number of filters doubles at each level of the encoder, starting from 32 in the first block. The structure of an individual convolutional block is illustrated in the upper panel of Fig.~\ref{fig:conv_block_film_layer} for reference. After each block, a two-dimensional max-pooling operation is applied to reduce the spatial dimensions of the feature maps by half, enabling the network to capture progressively larger contextual features.
\item The output of the final encoder block is passed to the bottleneck, which consists of two additional convolutional blocks, with the number of filters determined by the depth of the U-Net architecture. At this stage, we introduce redshift conditioning using Feature-wise Linear Modulation (FiLM; \citealt{dumoulin2018feature-wise}). This conditioning is motivated by the fact that the relationship between the observed optical properties of galaxy clusters and their underlying projected mass distributions can depend on redshift. Therefore, a network trained across a range of redshifts should be able to adapt its feature representations according to the redshift of the input cluster, rather than learning a single mapping that is independent of redshift.

FiLM provides a mechanism for achieving this by using the cluster redshift as an additional scalar input to modulate the bottleneck feature maps. Specifically, the redshift is passed through a small fully connected network consisting of two dense layers, which generates two sets of channel-wise modulation parameters: a scaling parameter $\gamma(z)$ and a shift parameter $\beta(z)$. Each set contains $C$ parameters, where $C$ is the number of channels in the bottleneck representation. These parameters are broadcast across the spatial dimensions of the feature maps and applied through an affine transformation:
\[
\mathrm{FiLM}(x,z) = \gamma(z)\,x + \beta(z),
\]
where $x$ denotes the bottleneck feature maps and $z$ is the cluster redshift. The scaling parameters control the relative contribution of individual feature channels, while the shift parameters introduce an additional redshift-dependent adjustment to their activations. In this way, the network can adapt its internal representations to the redshift of each cluster while retaining the shared convolutional architecture and learned spatial features. A schematic illustration of the FiLM conditioning mechanism is provided in Fig. \ref{fig:conv_block_film_layer}.

\item The decoder (up-sampling) path mirrors the encoder structure and progressively reconstructs the spatial resolution of the feature maps. At each decoding stage, the feature maps are up-sampled and concatenated with the corresponding encoder feature maps via skip connections, preserving spatial information lost during down-sampling. The concatenated features are then processed by two convolutional blocks. The number of filters is halved at each decoding level to maintain symmetry with the encoder. Redshift conditioning is again applied at each decoder stage using a FiLM layer, which modulates the feature maps according to the input cluster redshift. Through these sequential operations, the decoder reconstructs the final projected total mass map from the bottleneck representation.
\end{enumerate}

The U-Net framework allows multiple HST filters to be incorporated simultaneously. We explored two different extensions:

\begin{enumerate}
\item Multi-channel:  
In this configuration, the network architecture remains identical to the single-channel U-Net, but the input consists of a five-channel image formed by stacking the F606W, F625W, F775W, F814W, and F850LP observations. The resulting input tensor has dimensions $128 \times 128 \times 5$, analogous to an RGB image with additional channels. This configuration follows an early-fusion strategy, where information from all five filters is processed jointly from the initial convolutional layers. This allows the network to learn shared feature representations by combining information across the different photometric bands during the feature extraction process.
\begin{figure}[tbp]
    \centering
    \includegraphics[width=.8\linewidth]{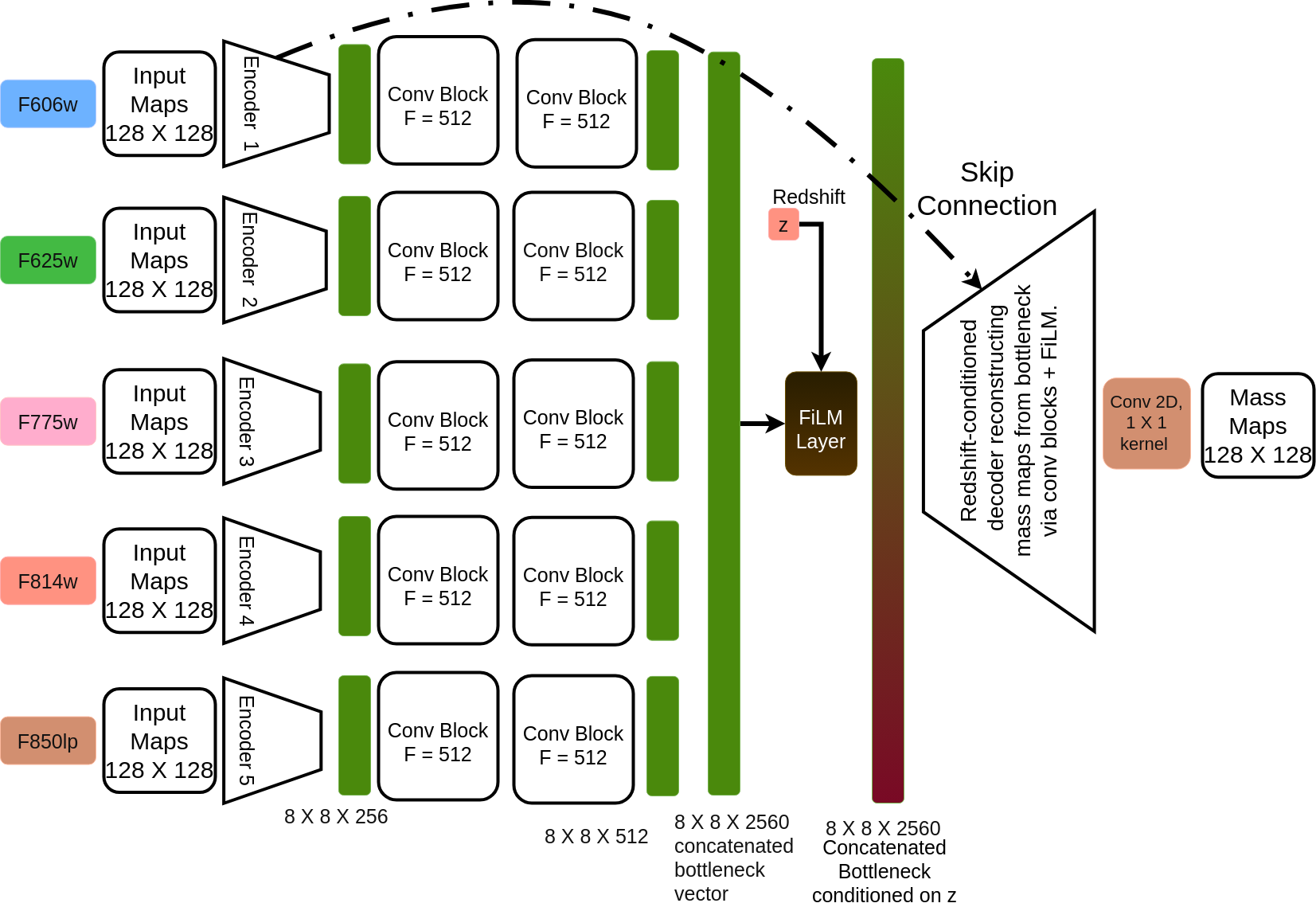}
    \caption{Multi-encoder U-Net architecture for simultaneous processing of five HST filter images (F606W, F625W, F775W, F814W, and F850LP) of the same cluster. Each $128\times128\times1$ input passes through an independent encoder branch ($3\times3$ kernels, base filter size $F=32$). The resulting feature maps are concatenated at the bottleneck into a joint latent representation and conditioned on cluster redshift via a FiLM layer. A shared decoder then reconstructs the final projected mass map, utilizing FiLM modulation at each stage and multi-branch skip connections to preserve spatial information.}
\label{fig:Mutliview-UNET}
\end{figure}
\item Multi-encoder:  
In this configuration, each HST filter is processed through an independent encoder branch, allowing the network to initially extract features from each filter separately. This follows a late-fusion strategy, in which the feature maps extracted by the individual encoders are concatenated at the bottleneck to form a joint multi-filter latent representation. This representation is subsequently conditioned on the cluster redshift via a feature-wise linear modulation (FiLM) layer. A shared decoder then reconstructs the projected mass map from the combined representation, incorporating skip connections from all encoder branches to retain spatial information at different resolutions. Unlike the multi-channel configuration, where information across filters is combined from the initial convolutional layers, the multi-encoder architecture delays the fusion of the independently extracted feature representations until the bottleneck and subsequent decoding stages. Similar parallel-processing and feature-fusion strategies have been explored in multimodal CNN architectures (e.g. \citealp{ait2024multimodality}). This architecture is shown in Fig.~\ref{fig:Mutliview-UNET}.
\end{enumerate}

Before training the models, we applied preprocessing steps to ensure that the U-Net training is stable. Since luminosity maps for each HST filter can contain negative values (arising 
from background subtraction), we apply a signed logarithmic transform,
\begin{equation}
    \tilde{x} =
    \begin{cases}
        \log_{10}(x)       & x > 0, \\
        -\log_{10}\!\left(|x|\right)  & x < 0, \\
        0                  & x = 0,
    \end{cases}
    \label{eq:signed_log}
\end{equation}
which preserves the sign of the original pixel values while symmetrically compressing both 
positive and negative ranges.
By using the signed transform, all information—including negative pixels resulting from background subtraction—is preserved, with negative luminosities retained as meaningful signal without the loss of any signal.

For the mass maps, which are non-negative, we apply 
a standard logarithmic transform,
\begin{equation}
    \tilde{m} =
    \begin{cases}
        \log_{10}(m) & m > 0, \\
        0            & m \leq 0,
    \end{cases}
    \label{eq:log_mass}
\end{equation}
where pixels with zero or nonphysical values are masked to zero. Following these transformations, both maps are standardized using the mean and standard deviation computed exclusively from the training set, and the same statistics are applied to the validation set to prevent any leakage of validation information into the normalization. The normalization statistics for each HST filter and the mass maps are saved and reused at inference time to ensure consistency between training and prediction. For a discussion of the impact of different preprocessing choices on model performance, we refer the reader to Appendix
Section~\ref{sec:Preprocessing}.

The model is optimized using the Adam optimizer \citep{kingma2014adam} starting with a learning 
rate of $10^{-4}$ and a batch size of 64, for a maximum of 150 epochs. To stabilize and 
improve convergence, three callbacks were employed during training. First, a 
\texttt{ReduceLROnPlateau} which halves the learning rate whenever the validation loss fails 
to improve for 20 consecutive epochs, down to a minimum of $10^{-7}$. 
Second, \texttt{EarlyStopping} that monitors the validation loss and terminates training if no 
improvement is observed for 30 epochs, restoring the best weights at the end. Finally, \texttt{ModelCheckpoint} that saves the model weights to disk whenever a new best validation loss is achieved, ensuring that the optimal model state is retained regardless of when training terminates. The model is trained by minimising a per-map mean absolute error (MAE) loss.  All U-Net models were trained on 
NVIDIA A100 GPUs.
\section{U-Net Model Performance on the The300 Test Set}\label{Unet-Test-set}

\begin{figure}[tbp]
\centering
    \includegraphics[width=\linewidth]{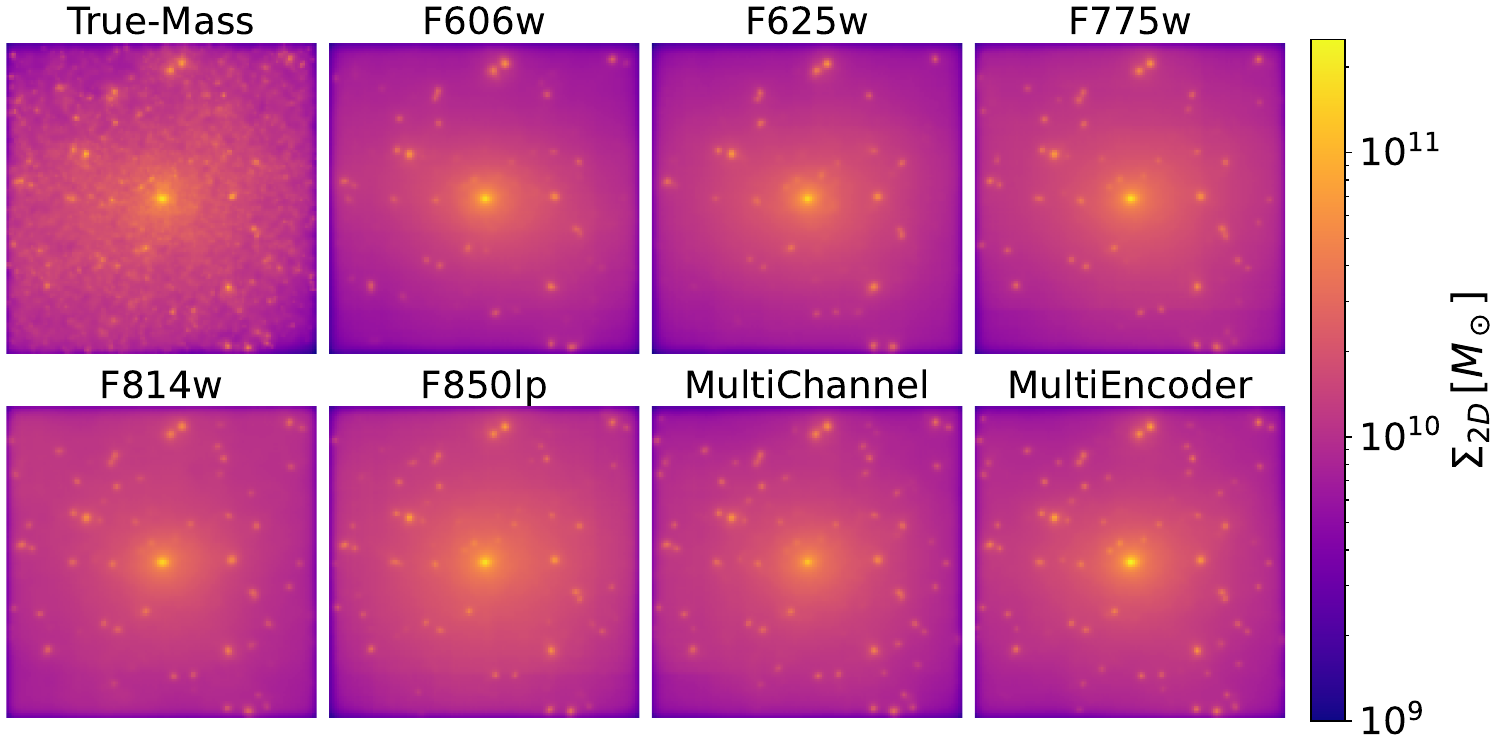}
    \includegraphics[width=\linewidth]{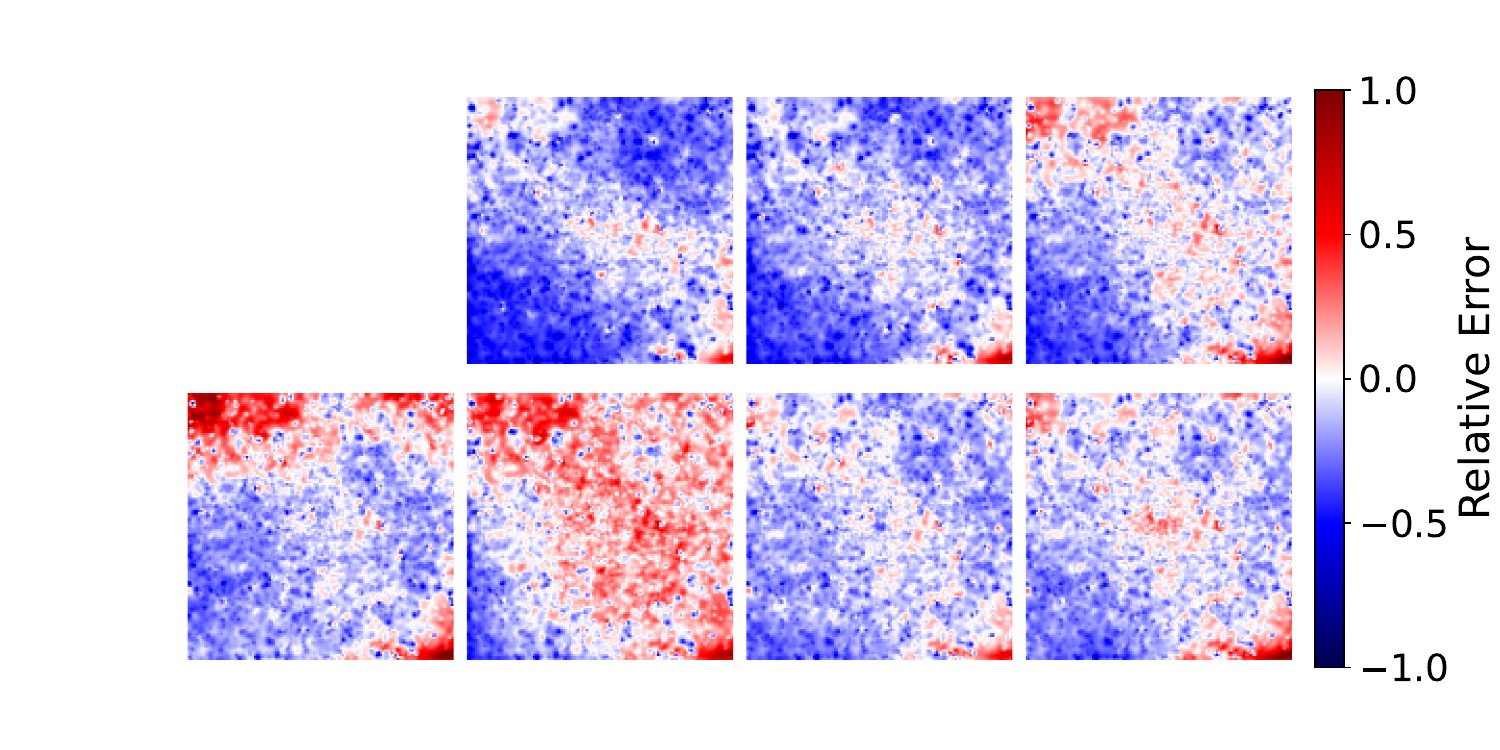}
\caption{Performance of different U-Net models in predicting projected total mass maps for a representative cluster from the simulation test set. Top panel shows the reconstructed maps, and lower panel shows the corresponding relative error residual maps, defined as $(\Sigma_{\text{2D, pred}} - \Sigma_{\text{2D, true}})/\Sigma_{\text{2D, true}}$.}
\label{fig:Unet-Test-Data}
\end{figure}
We show the predictions by the different U-Net models for an example simulated cluster from the test set. The mock HST luminosity maps shown in Fig.~\ref{fig:test-set-300-example} are used as inputs to reconstruct the corresponding projected total mass maps, shown in Fig.~\ref{fig:Unet-Test-Data}. All U-Net models successfully recover the overall morphology of the true projected mass distribution, including small-scale features, by using only the optical information associated with the stellar population component in the simulation. The residual maps corresponding to each U-Net model prediction, quantified in terms of relative errors, are shown in lower panel of Fig.~\ref{fig:Unet-Test-Data}.
\begin{figure}[tbp]
\centering
    \includegraphics[width=\linewidth]{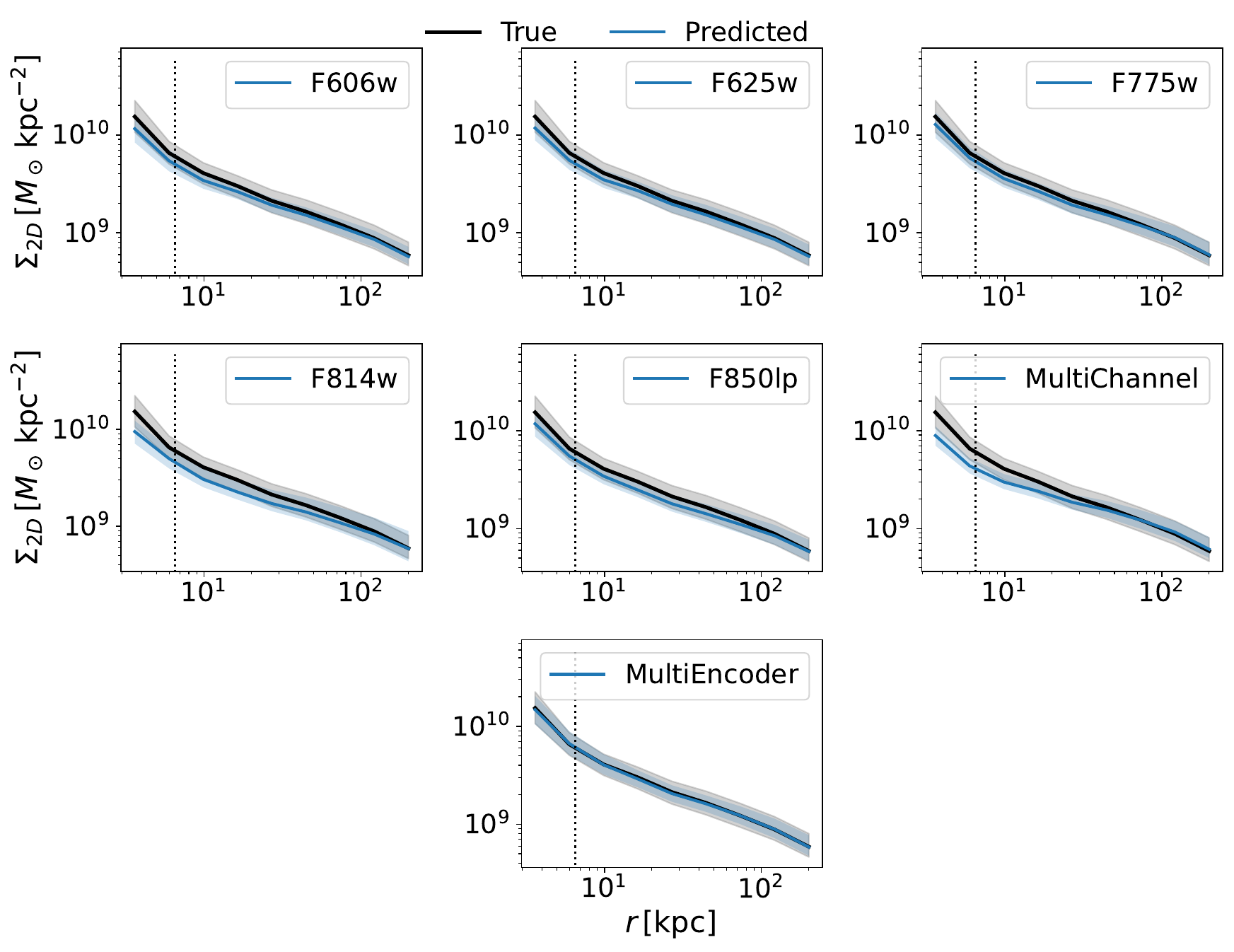}

    \includegraphics[width=\linewidth]{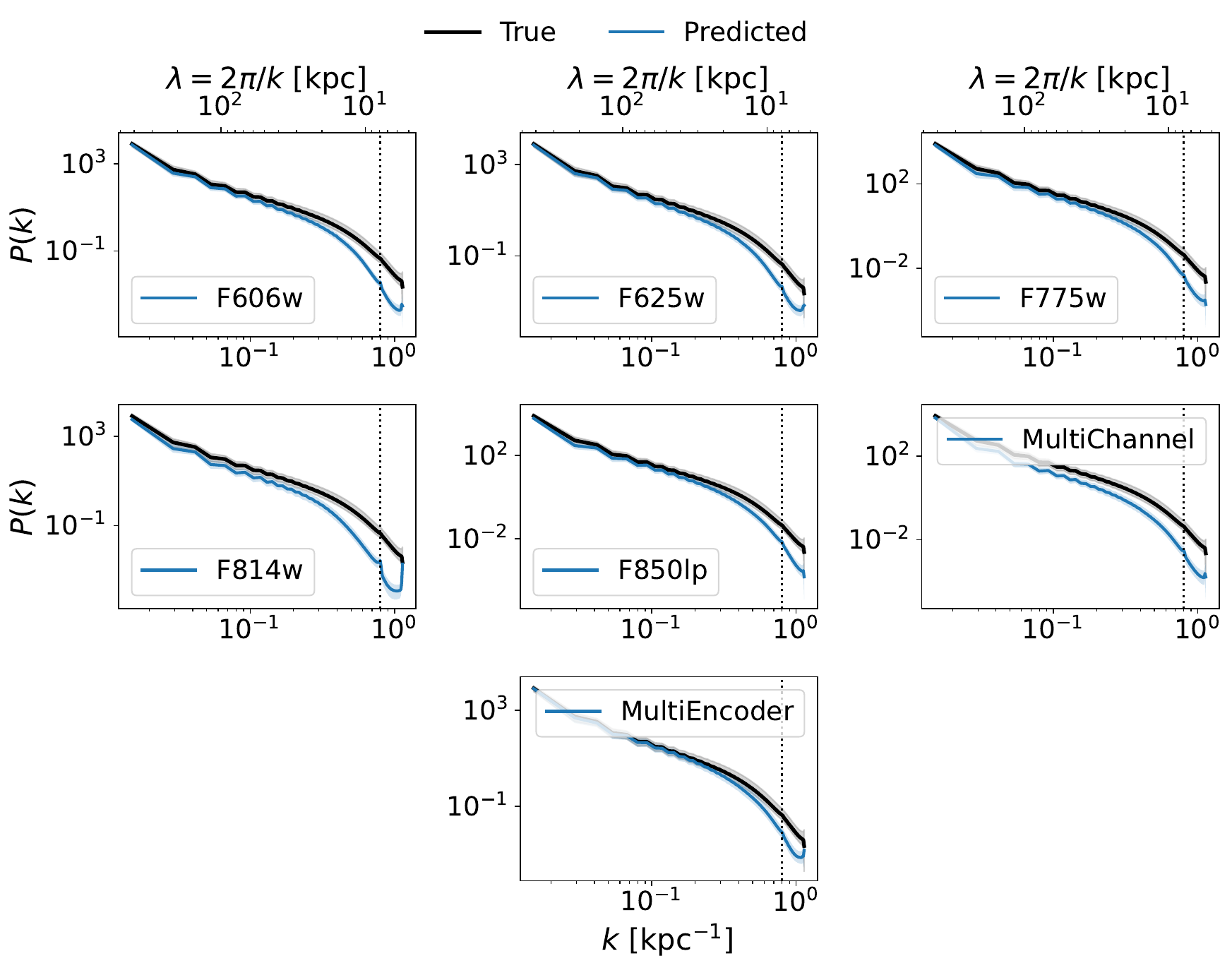}
\caption{Comparison of ground truth and predicted mass maps from different U-Net models over the test dataset. The upper panels show surface mass density radial profiles, whereas, the lower panels show power spectra estimated from the mass maps. Solid lines represent the test set medians, and shaded regions show the $16^{\rm th}\text{--}84^{\rm th}$ percentile range; each subplot compares an individual U-Net architecture against the ground truth. Vertical black dotted lines indicate the spatial resolution of The300 simulations ($\approx 6.5$~kpc, top) and the Nyquist limit of the $128 \times 128$ pixel mass maps ($k \approx 0.8~\text{kpc}^{-1}$ or $\lambda \approx 7.85$~kpc, bottom).}
\label{fig:Unet-radial-and-ps}
\end{figure}
To evaluate the performance of the U-Net-predicted mass maps on the The300 test set, we compare their surface mass density radial profiles and power spectra with those of the ground truth in Fig.~\ref{fig:Unet-radial-and-ps}. In the top panels, the black curves show the median radial profiles of the true projected mass distributions, while the blue curves represent the corresponding median profiles predicted by the different U-Net models. The shaded regions indicate the 16th--84th percentile range, illustrating the cluster-to-cluster variation in the profiles. Overall, the predicted profiles reproduce the shape of the true distributions well over most of the radial range. The best agreement is achieved by the multi-encoder model, where both the median profiles and the scatter closely match the ground truth. Models trained on individual filters show small systematic offsets, suggesting that single-band information is less effective in fully recovering the projected mass distribution. All models, however, show discrepancies in the innermost regions, highlighting the difficulty of accurately reconstructing the mass distribution near the cluster center.

A similar trend is seen in the power spectra shown in the bottom panels of Fig.~\ref{fig:Unet-radial-and-ps}. The predicted mass maps recover the large-scale features of the true mass maps well, as indicated by the good agreement between the ground-truth and reconstructed power spectra at low spatial frequencies. On smaller scales, however, the U-Net reconstructed mass maps' power spectra begin to deviate from the ground truth, indicating that the models are less effective at recovering the fine-scale structure of the mass distribution. Among all models, the multi-encoder configurations again show the closest agreement, demonstrating that combining information from multiple photometric bands improves the reconstruction of both the radial mass profiles and the spatial structure of the projected mass maps.

\begin{figure}[tbp]
\centering
    \includegraphics[width=0.7\linewidth]{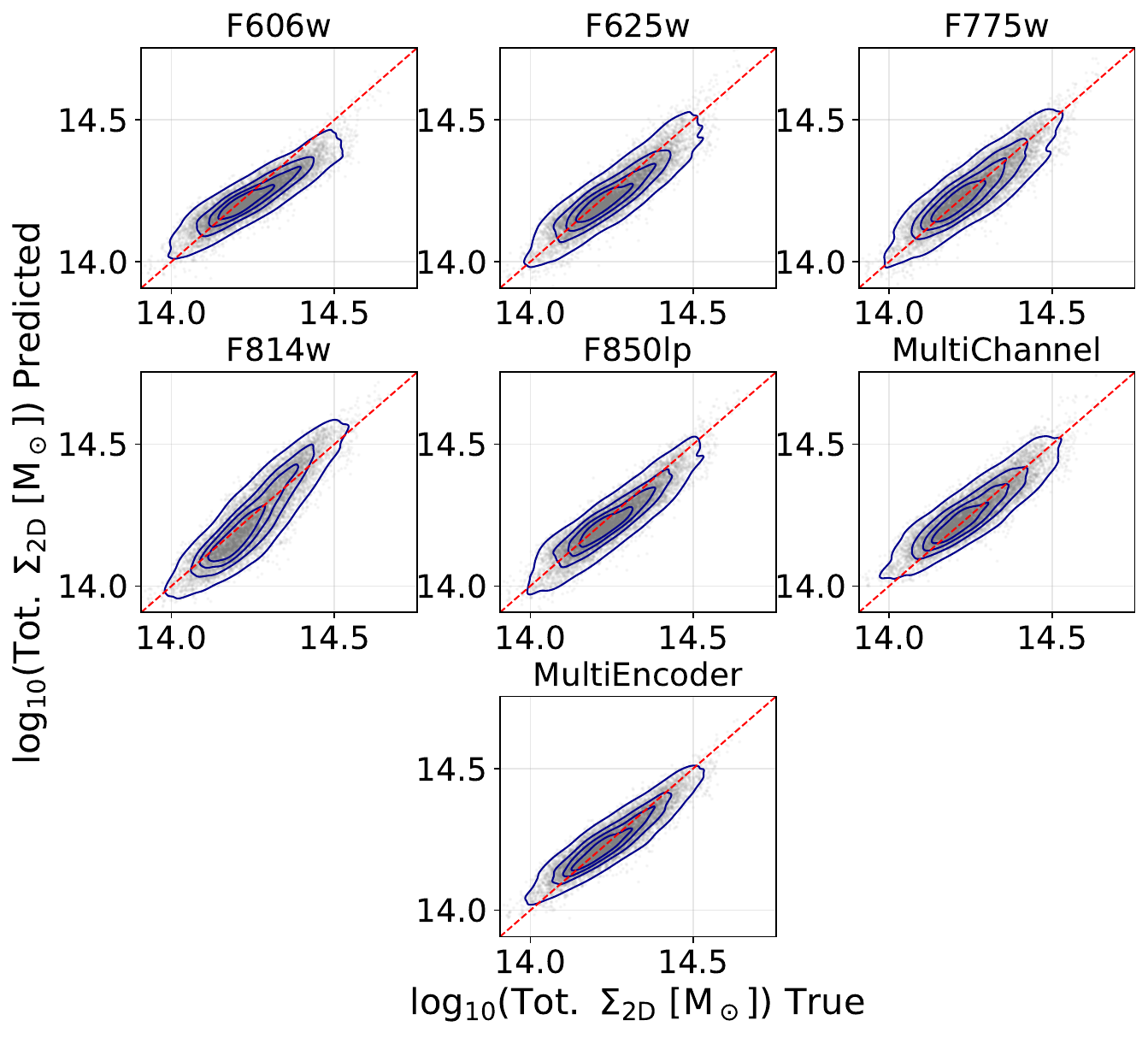}
    
    \includegraphics[width=0.7\linewidth]{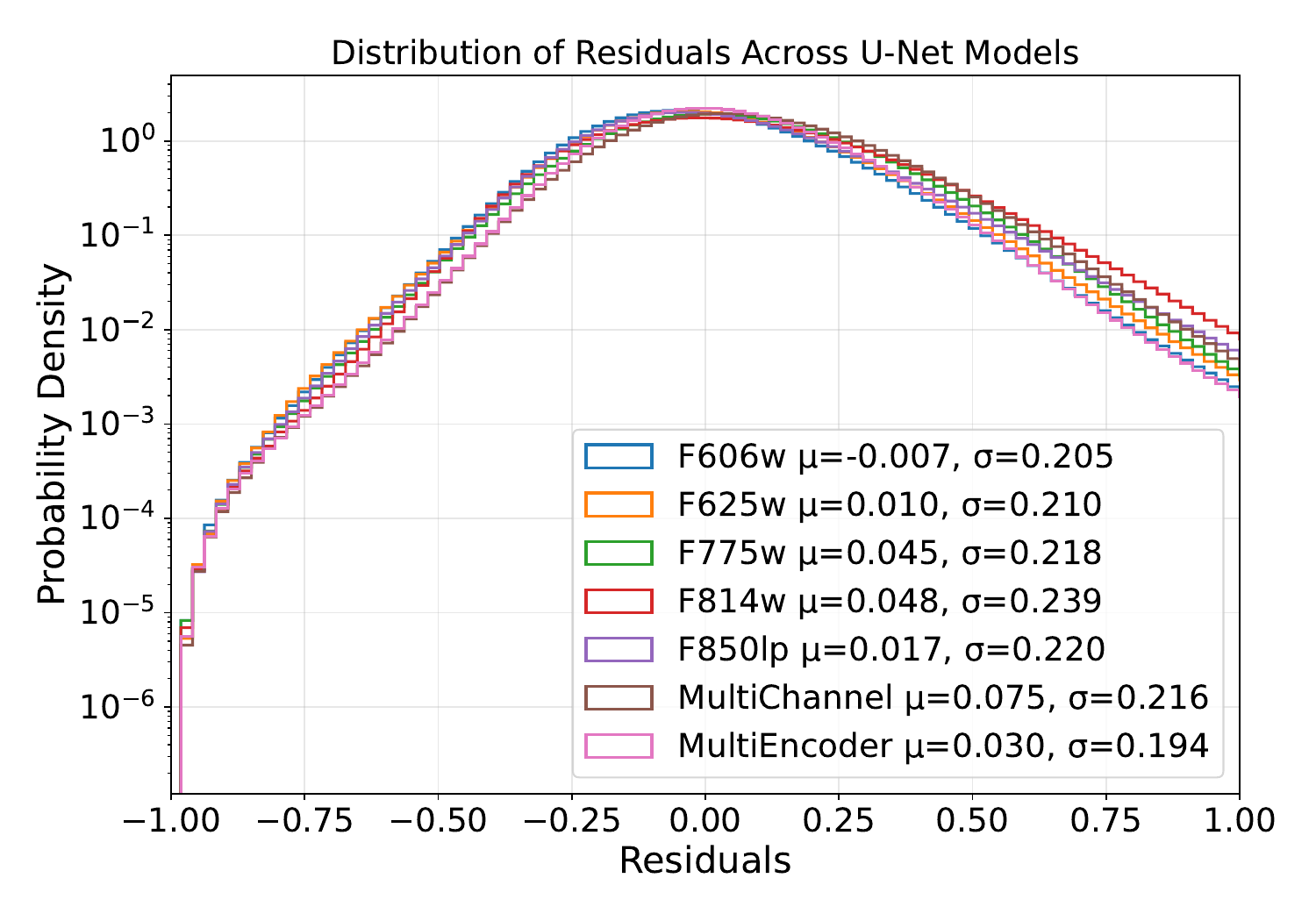}
\caption{Overall performance of the U-Net models in reconstructing total matter distributions. The top panel compares the predicted total projected mass from different U-Net models against the ground truth across the entire simulation test set; red dotted lines denote the ideal 1:1 relation, and blue contours show the kernel density estimation. The bottom panel displays the pixel-wise relative error distribution between predictions and true values for all test set maps, with the legend indicating the mean error and $\pm 1\sigma$ standard deviation for each architecture.}
\label{fig:Unet-Overall-Data}
\end{figure}

In Fig.~\ref{fig:Unet-Overall-Data} (top), we compare the total projected mass predicted by each U-Net model with the true total projected mass from the simulations for the full test set. The predictions broadly follow the one-to-one relation, with moderate scatter and a mild underestimation towards the high-mass end ($\log_{10}(\mathrm{Tot},\Sigma_{\mathrm{2D}}) \gtrsim 14.4$). This behaviour is consistent with the regression-to-the-mean effect, whereby predictions for clusters at the extremes of the target distribution tend to be closer to the values characteristic of the training sample. As illustrated by the cluster mass distribution in Fig.~\ref{fig:cluster_distribution}, the most massive clusters are less well represented in the training set. Consequently, the network may have limited ability to learn the distinctive mass distributions of these clusters, potentially leading to an underestimation of their total projected masses. This effect is particularly noticeable for the models based on the F606w, F625w, and F850lp filters, which show a tendency to under-predict the total projected mass.

We also examine pixel-wise relative residuals between the predicted and true projected mass maps (Fig.~\ref{fig:Unet-Overall-Data} (bottom)). The residuals are defined as $(\Sigma_{\mathrm{2D, pred}} - \Sigma_{\mathrm{2D, true}})/\Sigma_{\mathrm{2D, true}}$ and are computed across all pixels and clusters in the test set. The resulting distributions show a $\pm1\sigma$ scatter of approximately 19–24\%. Models trained on individual filters exhibit minimal mean residuals, with the exception of the F850lp-based model, which shows the largest bias ($\sim$10\%). 

For details regarding the performance of the U-Net models trained on The300 simulations and tested on the TNG cluster suite, we refer the reader to Appendix Section~\ref{TNG-Clusters}. Based on the evaluation of all U-Net architectures across both The300 test set and the TNG cluster (in Appendix Section~\ref{TNG-Clusters}) suite, the multi-encoder model demonstrated the most robust performance. Consequently, we selected this model for the subsequent reconstruction of mass maps from the HST observations of the CLASH sample. 

\section{Results}\label{sec:Results}
We applied our trained U-Net models to 21 CLASH clusters with HST observations. However, we present the results for five of these clusters—Abell~209, MACS~J0416$-$2403, MACS~J1149+2223, MACS~J1206$-$0847, and RXJ~2248.7$-$4431-CLASH-VLT individually, of which the spectroscopic information is available. 
This is essential for accurately masking non-cluster members for correctly identifying and studying subhalos. Furthermore, point sources in the HST images of the clusters are also excluded before applying the U-Net models to predict the total mass maps.
For the remaining 16 CLASH clusters, a similar preprocessing procedure was applied, but with only the CLASH HST photometric-redshift (BPZ) catalog. For assessing the robustness of the U-Net predictions, we compared the inferred projected mass maps with independent lensing-based mass estimates from the CLASH survey. In particular, we adopted the light-traces-mass (LTM) models derived from joint strong and weak lensing analysis \citep{zitrin2015hubble}, which provide convergence maps for all clusters in our sample.
Full details of the preprocessing applied to the CLASH HST images prior to their use as input to the U-Net models, the conversion of LTM $\kappa$ maps to projected mass maps, are provided in Appendix Section \ref{subsection:Preprocessing HST and kappa}.

\begin{figure}[tbp]
    \centering
    \includegraphics[width=.8\linewidth]{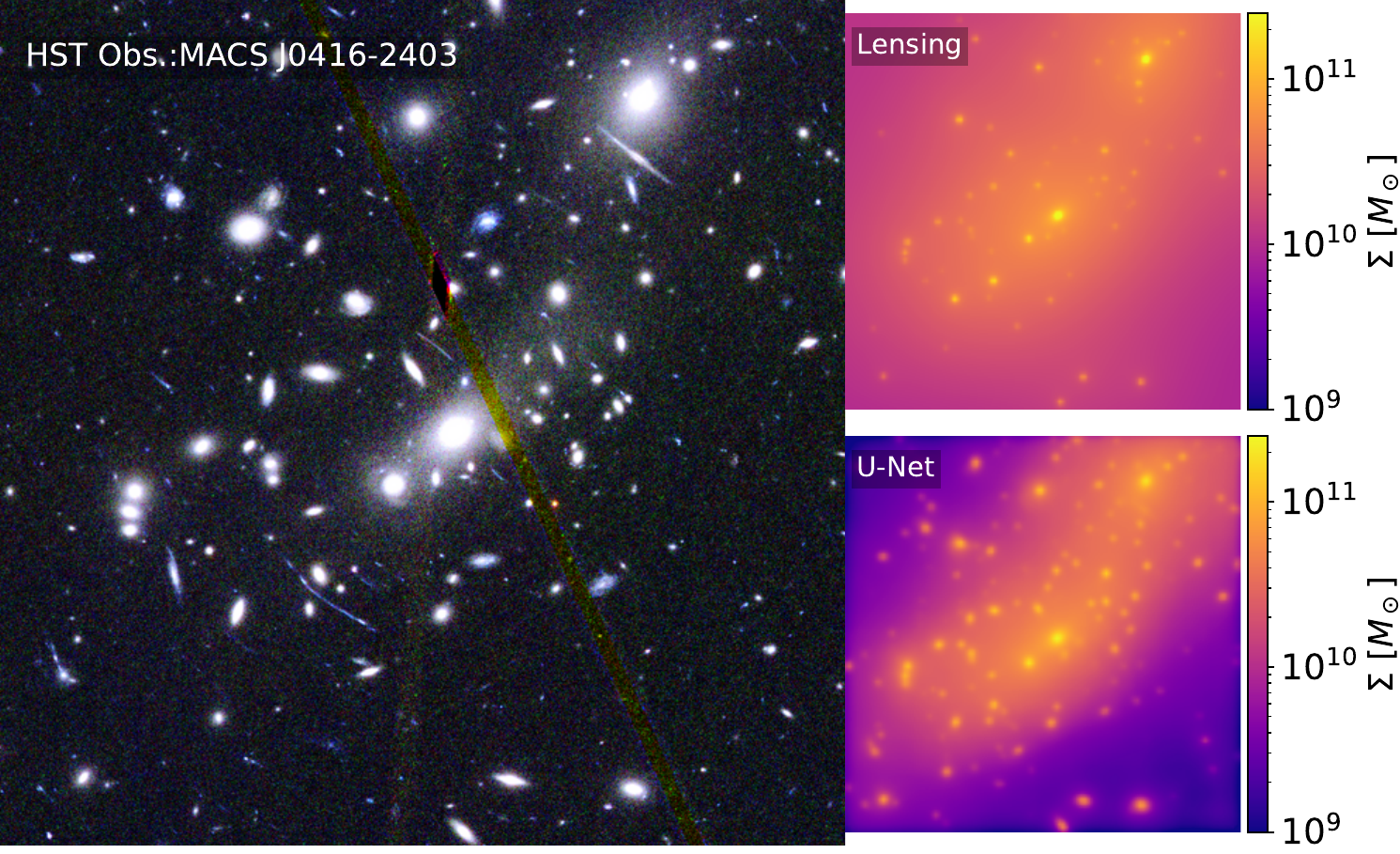}
    \includegraphics[width=.8\linewidth]{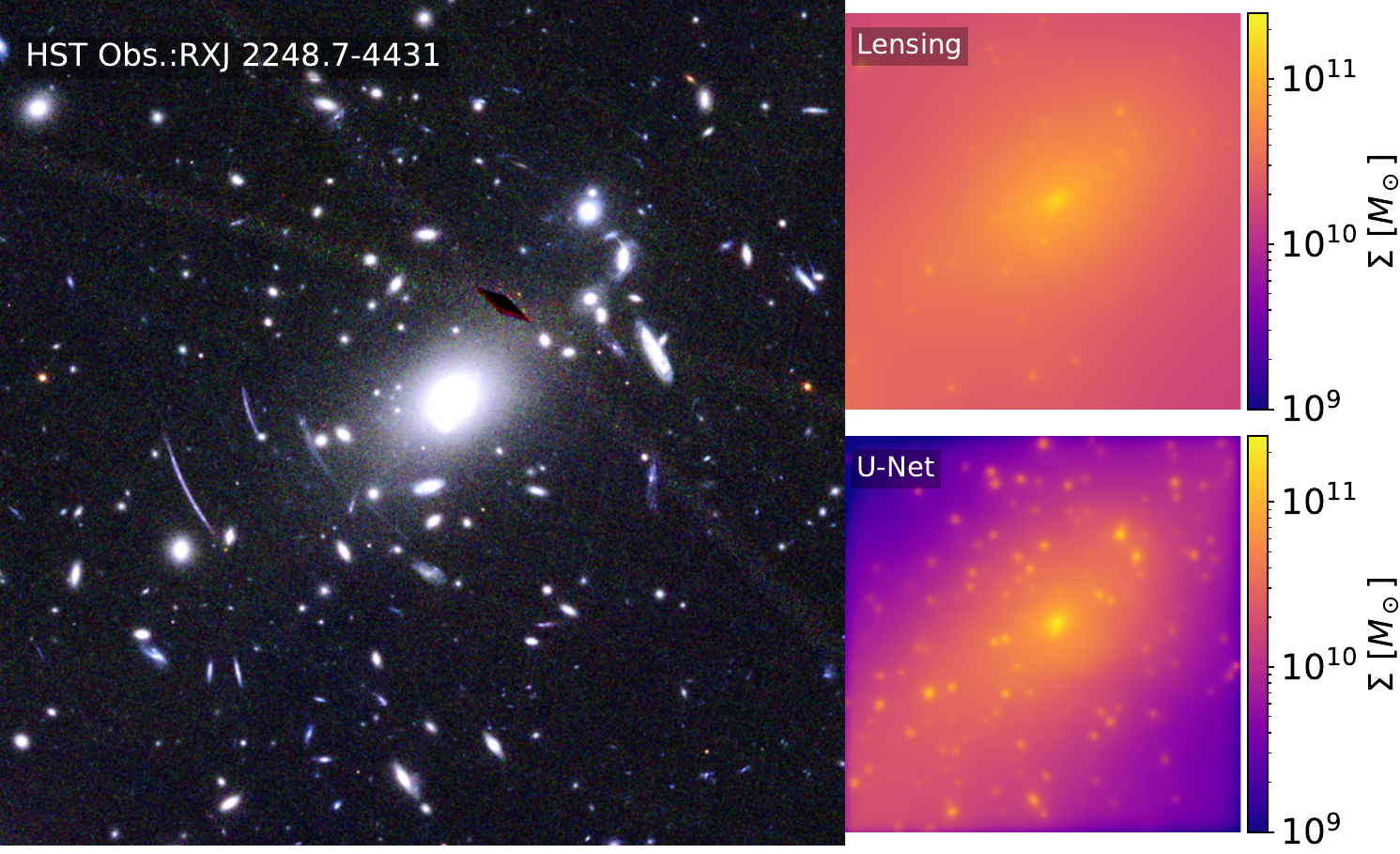}     
\caption{Two example CLASH clusters HST images, the lensing and the multi-encoder U-Net model predicted total mass maps. 
The CLASH clusters MACS~J0416$-$2403 (top, unrelaxed) and RXJ~2248.7$-$4431 (bottom, relaxed). For each cluster, the left column shows the HST observation, while the right column displays the projected total mass maps derived from lensing analysis (upper row) and by the multi-encoder U-Net model (lower row).
}
    \label{fig:UNET-CLASH-Main-clusters}
\end{figure}
Fig.~\ref{fig:UNET-CLASH-Main-clusters} shows two representative CLASH clusters: MACS~J0416$-$2403, a merging (dynamically unrelaxed) system, and RXJ~2248.7$-$4431, a relaxed system.
For each cluster, the left panel shows the HST image, created by combining the F850LP, F775W, and F606W filters as the R, G, and B channels, respectively. The right panels show the lensing-derived mass map (top) and the mass map reconstructed by the multi-encoder U-Net (bottom). For the performance of the remaining U-Net architectures on the CLASH cluster sample, we refer the reader to the project repository\footnote{\url{https://github.com/atulit19/U-Net-Mass-Maps-HST-CLASH}}, where the complete analysis is hosted.

It is clear that the U-Net-based mass maps qualitatively preserve the general shape and structure of the clusters, matching both the HST observations and the lens-derived mass maps. This is not surprising as the LTM models rely on the assumption of light tracing mass, while the U-Net models are solely based on the observed luminosity maps. It seems that the U-Net model predictions are also clumpier and exhibit higher contrast than the lensing results, with more substructures clearly shown. We suspect the reason is that the LTM lensing model adopts an analytical density profile to fill the overall map, while the U-Net relies on the distributions and connections of the galaxy luminosities. 
\begin{figure}[tbp]
    \centering
    \includegraphics[width=.7\linewidth]{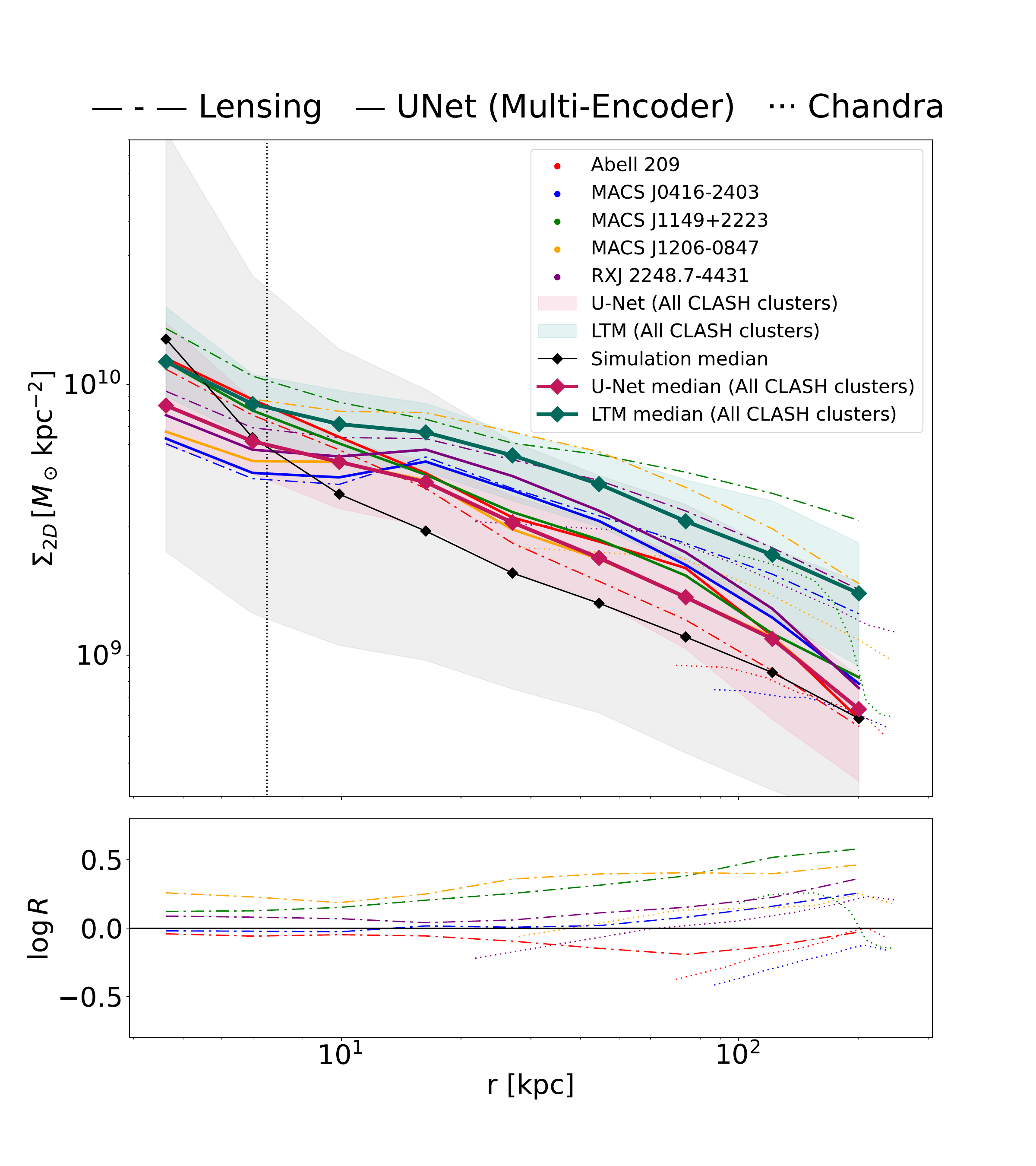}
    \includegraphics[width=.7\linewidth]{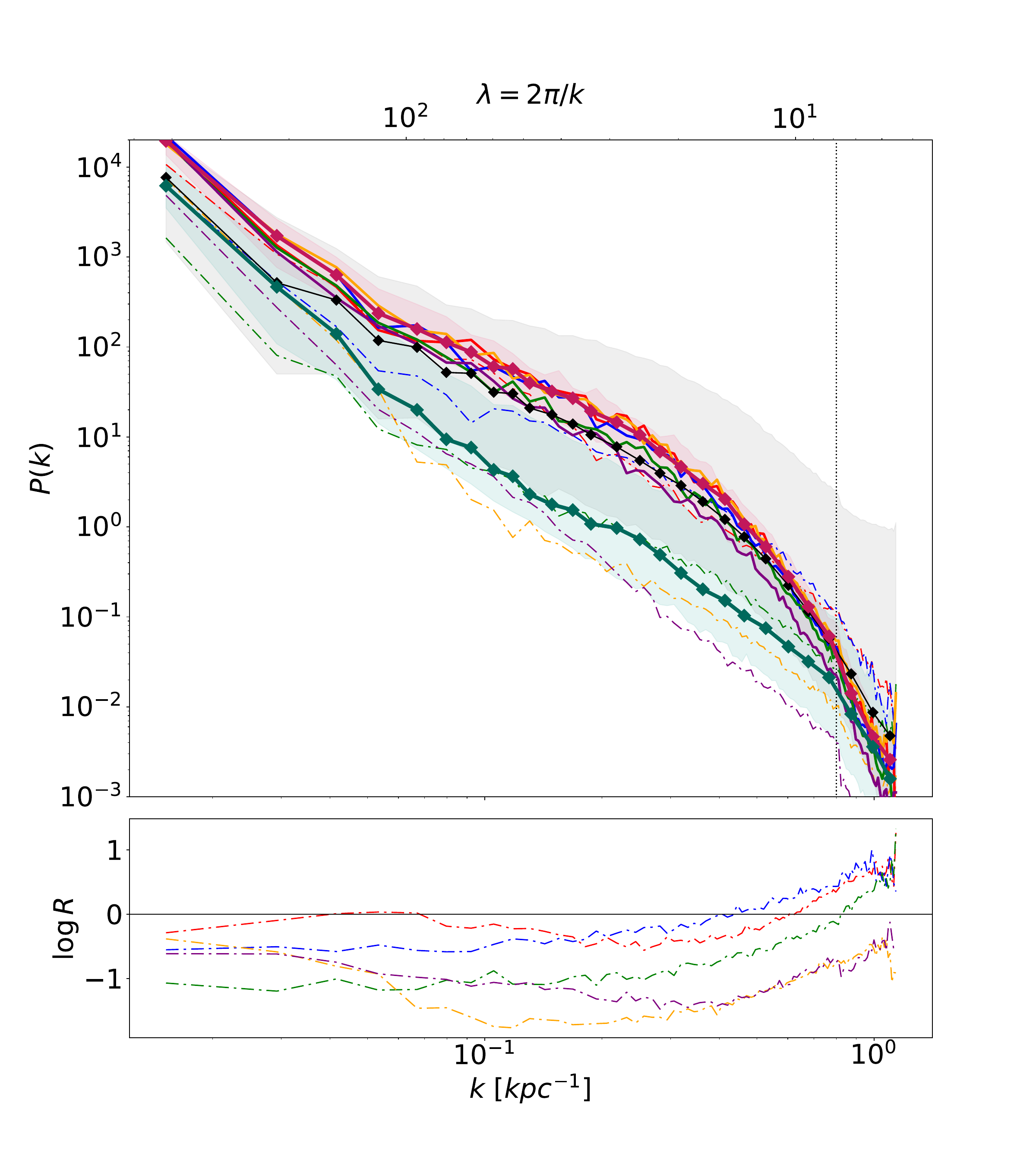}
\caption{Radial profiles of surface mass density (top) and mass power spectra (bottom) for five CLASH clusters with spectroscopic redshifts. Solid lines show multi-encoder U-Net predictions, long-dash-dotted lines denote lensing-based LTM model estimates, and dotted lines represent hydrostatic-equilibrium profiles derived from \textit{Chandra} X-ray data. Colored shaded areas indicate the $16^{\rm th}\text{--}84^{\rm th}$ percentile range across all 21 CLASH clusters, while the grey-shaded region spans the full range of The300 simulations. Median trends are denoted by solid lines with diamond markers with increased line widths; the U-Net and LTM medians are computed over the full CLASH clusters, whereas the simulation median is calculated across the entire simulated cluster sample. Lower panels display the log-ratios calculated with respect to the U-Net predictions: $\log_{10}(\frac{\text{LTM/X-ray}}{\text{U-Net}})$. Power spectra are plotted against wavenumber $k$ (spatial scale $\lambda = 2\pi / k$). Vertical black dotted lines indicate the spatial resolution of The300 simulations ($\approx 6.5$~kpc, top panel) and the Nyquist limit of the $128 \times 128$ pixel mass maps ($k \approx 0.8~\text{kpc}^{-1}$ or $\lambda \approx 7.85$~kpc, bottom panel).}
\label{fig:UNET-CLASH-Main-profiles-ps}
\end{figure}

\begin{figure}[tbp]
    \centering
    \includegraphics[width=.65\linewidth]{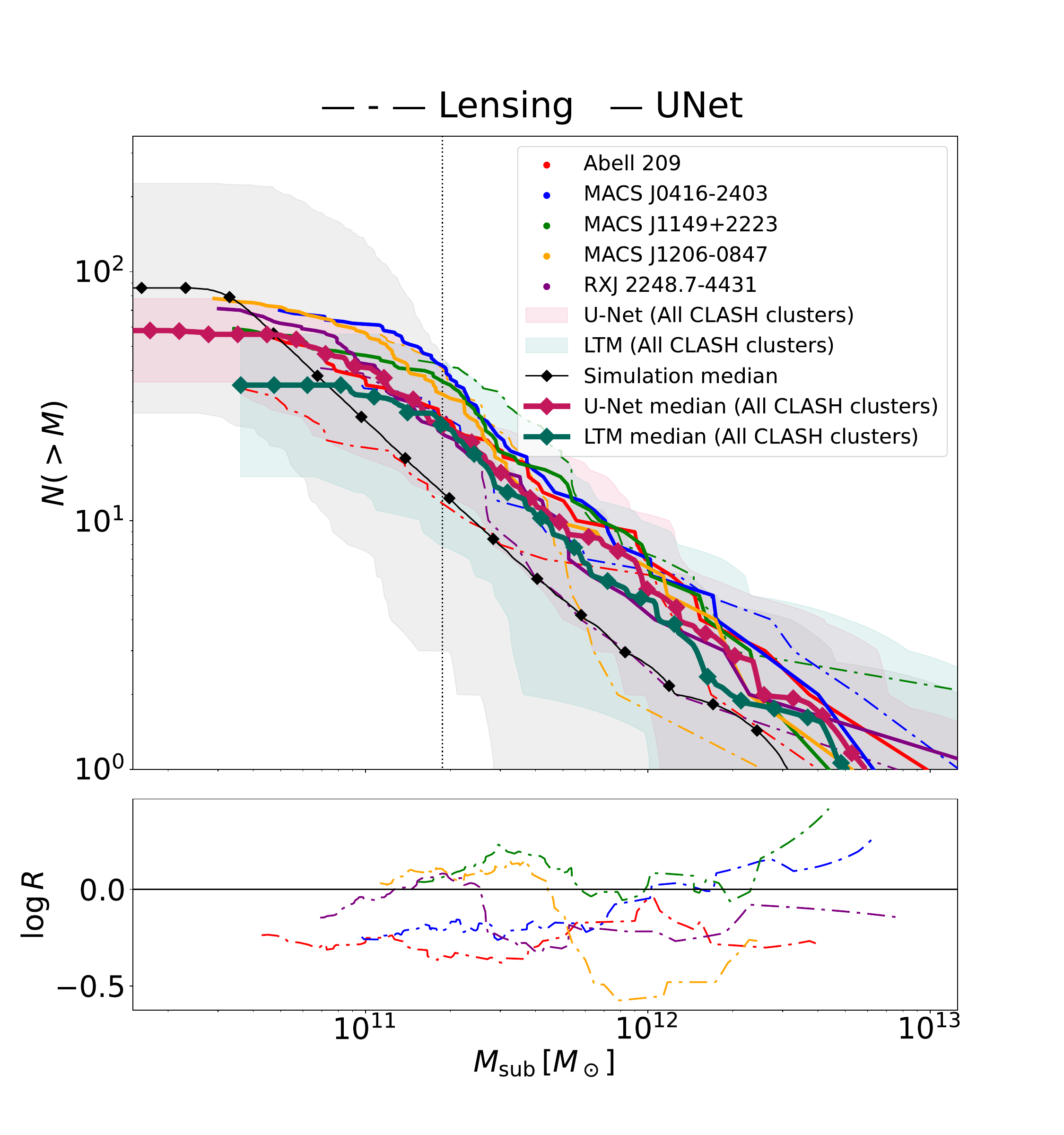}
    \includegraphics[width=.65\linewidth]{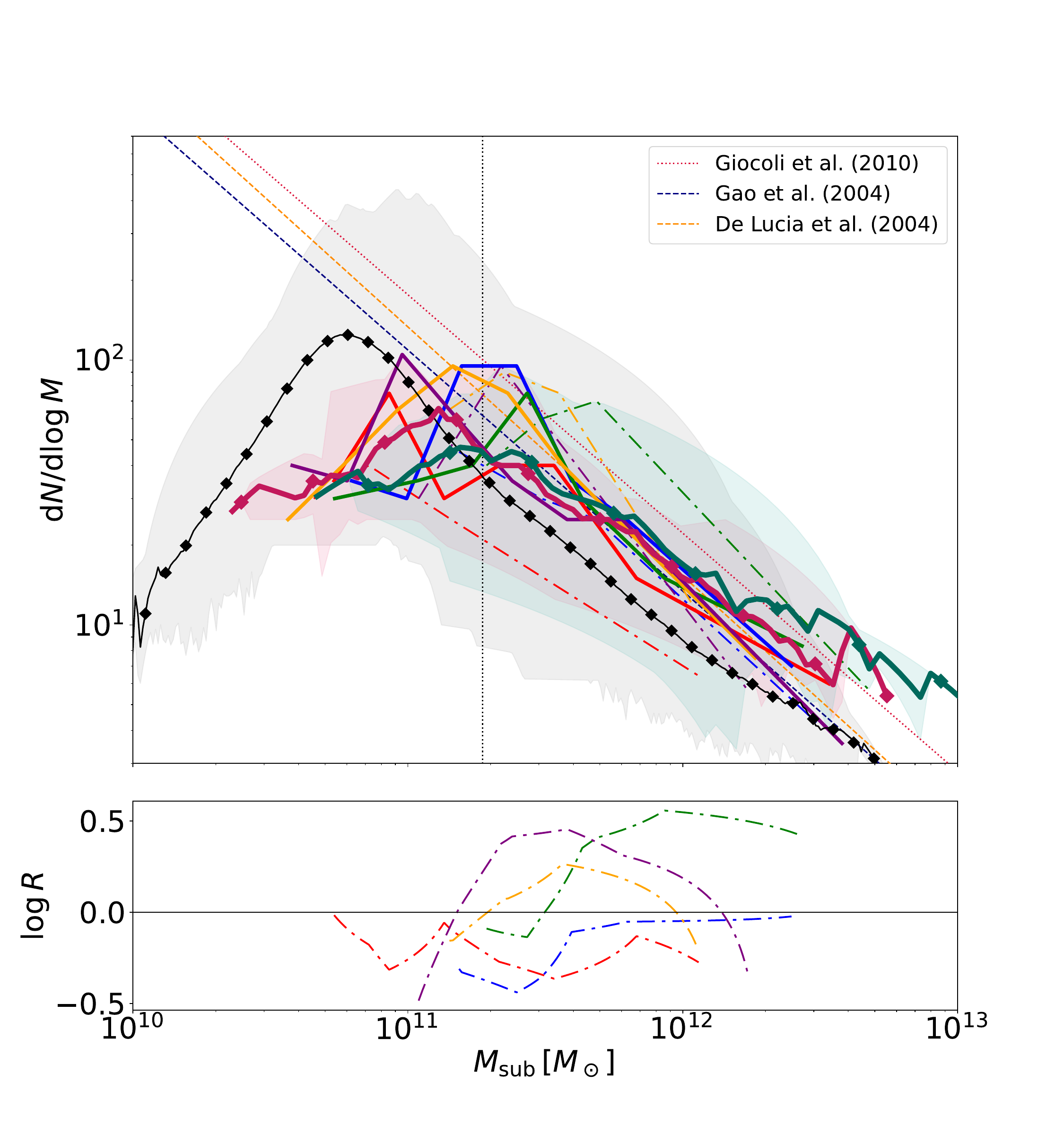}    
\caption{Cumulative (top) and differential (bottom) subhalo mass functions (CSMF and DSMF) derived from the multi-encoder U-Net predictions and lensing-based LTM maps. Solid curves trace individual CLASH clusters with spectroscopic catalogs. The grey-shaded region spans the full scatter of The300 simulations, while the soft pink (U-Net) and light teal (LTM) shaded areas mark the $16^{\rm th}\text{--}84^{\rm th}$ percentile ranges across all 21 CLASH clusters. Median trends are denoted by solid lines with diamond markers and increased line width; the U-Net and LTM medians are computed over the full CLASH clusters, whereas the simulation median is calculated across the entire simulated dataset. Analytical power-law predictions calibrated from simulations are overplotted for comparison \citep{gao2004subhalo,de2004substructures,giocoli2010substructure}. 
Lower panels display the logarithmic differences $\log R = \log(\text{LTM}/\text{U-Net})$. The vertical black dotted line indicates the subhalo mass resolution cutoff at $1.87 \times 10^{11}~M_{\odot}$ determined from the simulation maps.}
\label{fig:UNET-CLASH-Main-MassFunc}
\end{figure}
To quantitatively assess the performance of the multi-encoder U-Net model in reconstructing projected mass maps and information contained at different spatial frequencies, we analyze radial surface mass density profiles and power spectra. Fig.~\ref {fig:UNET-CLASH-Main-profiles-ps} compares U-Net predictions with lensing-based results using radial surface mass density profiles (top panel) and power spectra as a function of spatial scale $\lambda$ (bottom panel).
Furthermore, the radial surface mass density profiles of these clusters, estimated from hydrostatic equilibrium mass profiles derived from Chandra X-ray observations \citep{donahue2014clash}, are included for comparison. The Chandra hydrostatic mass profiles are derived by \cite{donahue2014clash} using the Joint Analysis of Cluster Observations (JACO) code \citep{mahdavi2007joint}, which fits a parametric NFW dark matter profile and a triple $\beta-$model gas density simultaneously to the X-ray spectra under the assumption of hydrostatic equilibrium. Annular bin boundaries are selected to contain at least 1500 source counts each, with a minimum width of several times the Chandra PSF. The best-fit parameters are constrained via a Markov Chain Monte Carlo (MCMC) procedure, from which the cumulative enclosed mass inside shells is estimated. We convert these 3D cumulative mass profiles into projected surface mass density maps by computing the mean 3D density within each spherical shell and applying an analytic Abel projection, assuming spherical symmetry and constant density per shell \footnote{The Chandra surface mass density profiles do not probe all the way to the cluster center. The innermost radius of each profile is set by the midpoint of the first JACO shell, ranging from 6.2 kpc (RXJ~1532.9+3021) to 100.0 kpc (MACS~J1149+2223) across our 21 CLASH clusters, with the number of radial shells ranging from 3 to 22.}
It is well known that the hydrostatic equilibrium mass is biased relative to the true mass by around 20 \%
\citep[e.g.][]{gianfagna2021exploring}, so we simply include that fixed value in the profile to correct this effect.
In the lower panels of both plots in Fig.~\ref{fig:UNET-CLASH-Main-profiles-ps}, we show the logarithmic ratio difference for the surface mass density and power spectra, which more clearly quantifies the differences between the multi-encoder U-Net inference and the lensing-derived results. The shaded gray region in both upper plots indicates the envelope of the projected surface mass density profiles and power spectra derived from the The300 simulated cluster sample used in this study. To estimate the surface mass density radial profiles, the projected mass maps were binned into logarithmically spaced annular regions, and the average surface mass density in each annulus was computed by summing the projected mass of all pixels in the bin and dividing by the annular area. The power spectrum of the total mass maps was computed directly using the PYLIANS library \citep{Pylians} from these 2D mass maps. 

By direct visual inference, we can see that the radial surface mass density profiles from lensing and U-Net–reconstructed mass maps (Fig.~\ref{fig:UNET-CLASH-Main-profiles-ps}) show overall consistency in both slope and normalization across the cluster sample. In the central core regions, approximately within the BCG area, the multi-encoder U-Net predictions are in very good agreement with the lensing-based profiles. Both methods consistently recover a slight central dip, indicating that the network can capture and generalize features that deviate from the median trends observed in simulations. This agreement indicates that the two models can reliably reconstruct the inner mass distribution, where baryon effects are expected to be significant.

At intermediate radii, the differences between clusters become more evident. For MACS~J0416$-$2403 and RXJ~2248.7$-$4431, the multi-encoder U-Net reconstructed profiles reproduce the lensing profiles, whereas for MACS~J1149+2223 and MACS~J1206$-$0847, the multi-encoder U-Net tends to give lower surface mass densities relative to the lensing-based estimates.

In the outer regions, the discrepancies become more prominent. The lensing profiles mostly sit above the U-Net profiles in this regime, as visible in Fig.~\ref{fig:UNET-CLASH-Main-profiles-ps}, except for Abell~209, which shows good agreement in the center ($\lesssim 20$ kpc), but the U-Net profile becomes systematically higher than the lensing estimate at larger radii. In the outer region, more generally, the U-Net predictions tend to lie toward the upper range of The300 cluster samples. We suspect that this is due to a large fraction of non-member galaxies being included: Even with spectroscopic information \citep{wu2022optical, herbonnet2022brightest, zhang2023modelling,nde2025impact}, it is hard to determine the true member galaxies due to the redshift distortion effect. Another possibility is that their true masses at that radius are actually higher by chance; as such, the predictions from U-Net are always in close proximity to the most massive samples in The300 simulation. A third possibility is a mismatch between observations and simulations, potentially arising from differences in baryonic physics or the assumed cosmological model.

Comparisons with X-ray hydrostatic profiles reveal that for MACS-J1149+2223, MACS~J1206$-$0847, and RXJ~2248.7$-$4431, the Chandra profiles lie below the lensing estimates but slightly above the U-Net predictions at outer region, while for Abell~209 and MACS~J0416$-$2403, they remain systematically lower than both LTM and U-Net. Although all the profiles have been corrected to the cosmological parameters of The300, it may still play a role in altering the halo density profiles \citep[see][for example]{ragagnin2021cosmology}. Though the overall distribution of the mass profiles is higher for LTM than U-Net, multiple other photo-z clusters from the CLASH sample do have their U-Net predictions lying in the central part of the shadow region.

The power spectra of these clusters in Fig.~\ref{fig:UNET-CLASH-Main-profiles-ps} (bottom panel) quantify the total mass density fluctuation differences between the two models across spatial scales.
We find that the spectra derived from the U-Net–reconstructed mass maps are systematically higher than those obtained from the lensing-based mass maps, specifically at larger $\lambda\gtrsim 20$. This indicates that U-Net–based mass maps contain more high-spatial-frequency features and, possibly more substructures than their lensing counterparts.
In contrast, for MACS~J0416$-$2403, the U-Net power spectrum exceeds the lensing spectrum at $k<0.31$ kpc $^{-1}$  ($\lambda > 20$  kpc), but the trend inverts at smaller scales, i.e., at $k>0.31$ kpc $^{-1}$ (or at less than 20 kpc). For Abell~209, the lensing-based mass map power spectrum shows a fluctuating trend compared to the U-Net–reconstructed mass map. To consider that statistical result, the agreement is particularly good at small scales, $k \gtrsim $ Nyquist limit; whereas at intermediate and larger spatial scales, the U-Net prediction shows slightly higher power than the lensing-based reconstruction.
Nevertheless, the power spectra from the multi-encoder U-Net model again lie within, or very close to, the power spectrum band of the The300 cluster sample.

From the power spectrum comparison, we find that the U-Net generally reconstructs mass maps with enhanced substructure features. To further quantify how well the U-Net reconstructions recover the abundance and mass distribution of substructure, in Fig.~\ref{fig:UNET-CLASH-Main-MassFunc}, we compare the cumulative (top) and differential (bottom) subhalo mass functions (CSMF and DSMF, respectively) derived from the U-Net–predicted mass maps with those obtained from lensing-based mass maps. The lower panels in both plots show the relative ratios between lensing-derived and U-Net–based subhalo mass functions. Subhalos in both lensing- and U-Net–predicted mass maps are identified using a peak-finding algorithm [Srivastava et al., in prep] that detects local maxima in a background-subtracted map with the subhalo masses are calculated using simulation-calibrated radii. 
Note that the central galaxy is excluded in the identification.
Detailed documentation regarding the algorithmic and technical implementation of the subhalo identification procedure is available in Appendix Section~\ref{sec:peak_finder}.

The gray shaded region indicates the CSMF and DSMF envelopes for the full The300 cluster sample, but from 2D mass maps. The median thick dotted lines for 3D results of The300 clusters are computed in the same way from the projected total mass maps, but simply using the subhalo masses from the corresponding AHF catalog. This provides us with a ground truth to compare against, on which these theoretical subhalo mass functions are based. Here, we use only one line-of-sight projection, and the subhalo selection using the AHF catalog is performed along the same line of sight. The peak detection algorithm recovers, on average, $\sim$97\% of AHF subhalos with $10^{11}\, M_\odot$, evaluated across simulation maps from The300, using a matching radius of 2 pixels. The relative differences between this algorithm and the 3D AHF results are explored further in Srivastava (in prep.) and are also discussed briefly in Appendix Section~\ref{sec:peak_finder}.

From the CSMF(top panel) in Fig.~\ref{fig:UNET-CLASH-Main-MassFunc}, we see that over the subhalo mass range of $\sim10^{11}$–$10^{12}~M_{\odot}$ for Abell$~$209, MACS~J0416$-$2403, and RXJ 2248.7-4431, the U-Net predicts $\sim\times 1.5-2$ more subhalos, mostly at low masses. This is understandable because these low-mass subhalos can not create a deep enough potential to produce detectable lensing signals \citep{vegetti2010detection, despali2022detecting}. As subhalo mass increases, the deviations tend to be smaller. While for the other two clusters, the two models are more-or-less consistent for MACS~J1149$+$2223, but U-net tends to have more massive subhalos at $\gtrsim 10^{12} M_\odot$ for MACS~J1206$-$0847. It is interesting to see that the statistical results from all CLASH clusters are well aligned between the two models all the way down to the vertical line, which indicates the resolution and detection limits. Both the lensing- and U-Net–derived CSMFs largely lie on the The300 upper envelope, which indicates that these observed clusters tend to be subhalo-rich compared with the median from the simulations.

The DSMFs (bottom panel) are noisier due to the limited number of subhaloes within the small sky coverage, exhibiting greater variation compared to the CSMFs. Nevertheless, the overall conclusions for the DSMFs are very similar to the CSMFs shown in the top-panel plot. 
Though the power spectra are much lower from the lensing results than the U-net predictions, the abundance of subhalos in the intermediate mass range is higher for clusters: MACS J1149+2223, MACS J1206-0847 and RXJ 2248.7-4431. This is because these clusters have systematically higher mass maps, which yield relatively larger subhalo masses to push the DSMFs towards the right-hand side. Therefore, their DSMFs can be higher than those from the U-Net results. 

It is interesting to verify these results with theoretical predictions from numerical simulations~ \citep{giocoli2010substructure,gao2004subhalo,de2004substructures} in the lower panel of Fig. \ref{fig:UNET-CLASH-Main-MassFunc}.
The DSMFs from both methods follow a clear power-law trend, with a mean logarithmic slope of $\langle\alpha\rangle = -0.91$ and a standard deviation of $0.13$ across the five CLASH clusters with spectroscopic catalogs. Similar values are obtained both from the LTM models maps and by using all CLASH clusters. This is in line with predictions from numerical simulations, such as \cite{giocoli2010substructure}, which report a slope of $\sim 0.9$, as well as earlier theoretical results from \cite{gao2004subhalo} ($\alpha = -0.91$) and \cite{de2004substructures} ($\alpha = -0.94$), as compiled by \cite{van2005mass}. Furthermore, these lines also have a similar normalization to the literature results. Note that these subhalos are derived from a limited $500\times500$ kpc$^{2}$ field of view; as such, it is not surprising to see the 3D DSMF line is below the other simulation results. Therefore, the normalization agreement is possibly due to two major reasons: (1) as subhalos are mostly concentrated in the cluster central region, and the large simulation depth along the line of sight (10$\times R_{500c}$) compensates for the small projected area; (2)as presented in the previous paragraph -- these observed clusters tend to be more massive and subhalo-rich compared with the median from the simulations. While the normalization and abundance exhibit some variation across the observed clusters, the subhalo mass functions show broad consistency with predictions from numerical simulations, particularly in their power-law behaviour. These results provide encouraging observational evidence supporting the power-law form of the subhalo mass function in galaxy clusters.

\section{Conclusions}\label{sec:Conclusions}
In this work, we apply—and, for the first time, validate on real data—the deep‑learning technique proposed in \cite{de2024three} as an alternative to gravitational–lensing–based reconstructions of the projected total mass in galaxy clusters. Using U‑Net–based architectures trained on mock HST‑like observations generated from The Three Hundred (The300) simulation suite, we reconstruct mass maps for 21 CLASH clusters spanning a broad range of masses and redshifts. These predicted maps not only reproduce the observed mass distributions and morphologies, but also agree quantitatively with independent gravitational‑lensing and X‑ray analysis of density profiles and power spectra. This consistency indicates that the models learn robust correlations between the observed luminosity fields and the underlying projected total mass of the clusters.

Beyond global mass reconstruction, the predicted maps enable a systematic study of cluster substructure. Although subhalos are a well‑established prediction of structure‑formation simulations, confirming their abundance and internal properties observationally remains challenging. 
From the subhalo mass functions of the observed clusters derived using both methods, we find that the observed distributions broadly follow the power-law behaviour predicted by numerical simulations. Although some differences in normalization and cluster-to-cluster variation remain, the agreement in their overall slopes shows that the reconstructed mass maps can recover the expected statistical properties of cluster subhalos.

Using galaxy–galaxy lensing, \cite{meneghetti2020excess} reported a significant discrepancy between simulated and observed subhalo properties. This is further verified by \cite{Natarajan2026}, which found that the inner density profiles and radial distributions of subhalos are strongly discrepant with CDM by nearly an order of magnitude. Our approach provides an independent, image‑based route to identify subhalos directly from observations and, in particular, increases sensitivity to lower‑mass subhalos relative to standard lensing reconstructions, which will allow us to address these questions statistically for different subhalo masses.

Beyond subhalos characterization, the reconstructed cluster mass maps may also provide an opportunity to investigate the asphericity of galaxy clusters. Galaxy clusters generally exhibit triaxiality rather than spherical mass distributions, and their projected shapes depend on their orientation relative to the line of sight \citep{clowe2004effects, limousin2013three}. This is particularly relevant for gravitational lensing, since the orientation of a cluster along the line of sight can affect its projected mass distribution and introduce biases in lensing-based mass estimates \citep{giocoli2014mass}. In particular, the reconstructed two-dimensional mass maps could also be possibly used to investigate the projected shapes of clusters and potentially constrain their three-dimensional geometry and line-of-sight elongation. While these investigations are beyond the scope of the present work, they represent a promising direction for future applications of our deep learning mass maps reconstruction framework.

We also acknowledge the potential systematic differences introduced by domain shift, as our U-Net is trained exclusively on mock observations of The300 clusters and then applied to real HST data. These differences can arise from several sources: (a) the stellar populations, as mocks are based on a single SSP model with a Chabrier IMF applied to Simba star particles, whereas real galaxies may have different star formation histories; (b) the baryonic physics in the Simba model, as also indicated by the TNG test in Appendix \ref{TNG-Clusters}, particularly in the central regions; (c) instrumental modelling, since the adopted PSF models and noise maps may not fully capture the systematics of the real HST observations; and (d) imperfect non-member masking, particularly for the 16 clusters with only photometric redshifts. Another main source of systematic uncertainty is the line-of-sight depth used to build the training data. Both the mock HST maps and the target mass maps integrate over a depth of $\pm 5R_{500c}$, which is shorter than the line-of-sight distance corresponding to the velocity cut of 3000 km/s (approximately $\pm 40$ Mpc at low redshift). This difference may lead to the exclusion of foreground and background structures that would contribute to the projected mass distribution in real observations. These effects may bias the reconstructed masses, mainly at large radii. Future work should investigate both sources of systematics using different simulation suites, alternative stellar and baryonic models, larger or variable projection depths, and domain-adaptation techniques.

This study is a first step toward applying simulation‑based inference to observational imaging at scale. Beyond speed and scale, our framework offers several practical advantages. It delivers non‑parametric mass reconstructions that capture complex, merger‑driven morphologies without adopting fixed halo profiles, and it can operate with photometry alone, providing useful mass proxies even where shear measurements are unavailable or low‑S/N. The framework can be easily extended to forthcoming surveys (e.g., Euclid, CSST, LSST) and deployed to process large datasets with near‑real‑time inference, offering a powerful complement to traditional lensing pipelines. The approach can also be naturally extended to multi‑wavelength fusion (optical/IR, X‑ray, SZ), as has been done in  \cite{de2024three}, providing complementary constraints with distinct systematics. Consequently, it enables the probing of more detailed information at smaller scales than traditional methods allow.

\begin{acknowledgements}
AS, WC, DdA and GY  thank financial support by the Agencia Estatal de Investigación (AEI, Spain)  under project PID2024-156100NB-C21, funded by MCIN/AEI/10.13039/501100011033. AS and WC are also supported by the Atracci\'{o}n de Talento Contract no. 2020-T1/TIC-19882 granted by the Comunidad de Madrid in Spain. WC would like to further thank Agencia Estatal de Investigaci\'{o}n (AEI) for the Consolidaci\'{o}n Investigadora Grant CNS2024-154838. MM was supported by INAF Grants “The Big-Data era of cluster lensing" and "Probing Dark Matter and Galaxy Formation in Galaxy Clusters through Strong Gravitational Lensing", and ASI Grant n. 2024-10-HH.0 "Attività scientifiche per la missione Euclid – fase E". The work presented here emerged out of the The300 collaboration, partially funded by the 'Ayuda para la Organización de Jornadas Científicas en la UAM en el Marco del Programa Propio de Investigación y con el Apoyo del Consejo Social de la UAM'.
The authors acknowledge The Red Espa\~nola de Supercomputaci\'on for granting computing time for running the hydrodynamic simulations of The300 galaxy cluster project in the Marenostrum supercomputer at the Barcelona Supercomputing Centre.
\end{acknowledgements}

\bibliographystyle{aa} 
\bibliography{reference}

\begin{thebibliography}{86}
\expandafter\ifx\csname natexlab\endcsname\relax\def\natexlab#1{#1}\fi

\bibitem[{Adam {et~al.}(2015)Adam, Comis, Mac{\'\i}as-P{\'e}rez, Adane, Ade,
  Andr{\'e}, Beelen, Belier, Beno{\^\i}t, Bideaud, {et~al.}}]{adam2015pressure}
Adam, R., Comis, B., Mac{\'\i}as-P{\'e}rez, J.-F., {et~al.} 2015, Astronomy \&
  Astrophysics, 576, A12

\bibitem[{Ade {et~al.}(2016)Ade, Aghanim, Arnaud, Ashdown, Aumont, Baccigalupi,
  Banday, Barreiro, Bartlett, Bartolo, {et~al.}}]{ade2016planck}
Ade, P., Aghanim, N., Arnaud, M., {et~al.} 2016, Astronomy \& Astrophysics,
  594, A24

\bibitem[{Aguado-Barahona {et~al.}(2022)Aguado-Barahona,
  Rubi{\~n}o-Mart{\'\i}n, Ferragamo, Barrena, Streblyanska, \&
  Tramonte}]{aguado2022velocity}
Aguado-Barahona, A., Rubi{\~n}o-Mart{\'\i}n, J., Ferragamo, A., {et~al.} 2022,
  Astronomy \& Astrophysics, 659, A126

\bibitem[{Ait~Ouahmed {et~al.}(2024)Ait~Ouahmed, Arnouts, Pasquet, Treyer, \&
  Bertin}]{ait2024multimodality}
Ait~Ouahmed, R., Arnouts, S., Pasquet, J., Treyer, M., \& Bertin, E. 2024,
  Astronomy \& Astrophysics, 683, A26

\bibitem[{Bah{\'e} {et~al.}(2012)Bah{\'e}, McCarthy, \& King}]{bahe2012mock}
Bah{\'e}, Y.~M., McCarthy, I.~G., \& King, L.~J. 2012, Monthly Notices of the
  Royal Astronomical Society, 421, 1073

\bibitem[{Bartelmann \& Schneider(2001)}]{bartelmann2001weak}
Bartelmann, M. \& Schneider, P. 2001, Physics Reports, 340, 291

\bibitem[{Bergamini {et~al.}(2023)Bergamini, Acebron, Grillo, Rosati, Caminha,
  Mercurio, Vanzella, Mason, Treu, Angora, {et~al.}}]{bergamini2023glass}
Bergamini, P., Acebron, A., Grillo, C., {et~al.} 2023, The Astrophysical
  Journal, 952, 84

\bibitem[{Biffi {et~al.}(2016)Biffi, Borgani, Murante, Rasia, Planelles,
  Granato, Ragone-Figueroa, Beck, Gaspari, \& Dolag}]{biffi2016nature}
Biffi, V., Borgani, S., Murante, G., {et~al.} 2016, The Astrophysical Journal,
  827, 112

\bibitem[{Biviano \& Girardi(2003)}]{biviano2003mass}
Biviano, A. \& Girardi, M. 2003, The Astrophysical Journal, 585, 205

\bibitem[{Bradley {et~al.}(2016)Bradley, Sipocz, Robitaille, Tollerud, Deil,
  Vin{\'\i}cius, Barbary, G{\"u}nther, Bostroem, Droettboom,
  {et~al.}}]{bradley2016photutils}
Bradley, L., Sipocz, B., Robitaille, T., {et~al.} 2016, Astrophysics Source
  Code Library, ascl

\bibitem[{Bruzual \& Charlot(2003)}]{bruzual2003stellar}
Bruzual, G. \& Charlot, S. 2003, Monthly Notices of the Royal Astronomical
  Society, 344, 1000

\bibitem[{Burke {et~al.}(2015)Burke, Hilton, \& Collins}]{burke2015coevolution}
Burke, C., Hilton, M., \& Collins, C. 2015, Monthly Notices of the Royal
  Astronomical Society, 449, 2353

\bibitem[{Chabrier(2003)}]{chabrier2003galactic}
Chabrier, G. 2003, Publications of the Astronomical Society of the Pacific,
  115, 763

\bibitem[{Clowe {et~al.}(2004)Clowe, De~Lucia, \& King}]{clowe2004effects}
Clowe, D., De~Lucia, G., \& King, L. 2004, Monthly Notices of the Royal
  Astronomical Society, 350, 1038

\bibitem[{Cui {et~al.}(2022)Cui, Dave, Knebe, Rasia, Gray, Pearce, Power,
  Yepes, Anbajagane, Ceverino, {et~al.}}]{cui2022three}
Cui, W., Dave, R., Knebe, A., {et~al.} 2022, MNRAS, 514, 977

\bibitem[{Cui {et~al.}(2018)Cui, Knebe, Yepes, Pearce, Power, Dave, Arth,
  Borgani, Dolag, Elahi, {et~al.}}]{cui2018three}
Cui, W., Knebe, A., Yepes, G., {et~al.} 2018, Monthly Notices of the Royal
  Astronomical Society, 480, 2898

\bibitem[{Dav{\'e} {et~al.}(2019)Dav{\'e}, Angl{\'e}s-Alc{\'a}zar, Narayanan,
  Li, Rafieferantsoa, \& Appleby}]{dave2019simba}
Dav{\'e}, R., Angl{\'e}s-Alc{\'a}zar, D., Narayanan, D., {et~al.} 2019, MNRAS,
  486, 2827

\bibitem[{de~Andres {et~al.}(2022)de~Andres, Cui, Ruppin, De~Petris, Yepes,
  Gianfagna, Lahouli, Aversano, Dupuis, Jarraya, {et~al.}}]{de2022deep}
de~Andres, D., Cui, W., Ruppin, F., {et~al.} 2022, Nature Astronomy, 6, 1325

\bibitem[{de~Andres {et~al.}(2024)de~Andres, Cui, Yepes, De~Petris, Ferragamo,
  De~Luca, Aversano, \& Rennehan}]{de2024three}
de~Andres, D., Cui, W., Yepes, G., {et~al.} 2024, Monthly Notices of the Royal
  Astronomical Society, 528, 1517

\bibitem[{De~Lucia {et~al.}(2004)De~Lucia, Kauffmann, Springel, White, Lanzoni,
  Stoehr, Tormen, \& Yoshida}]{de2004substructures}
De~Lucia, G., Kauffmann, G., Springel, V., {et~al.} 2004, Monthly Notices of
  the Royal Astronomical Society, 348, 333

\bibitem[{Despali {et~al.}(2022)Despali, Vegetti, White, Powell, Stacey,
  Fassnacht, Rizzo, \& Enzi}]{despali2022detecting}
Despali, G., Vegetti, S., White, S.~D., {et~al.} 2022, Monthly Notices of the
  Royal Astronomical Society, 510, 2480

\bibitem[{Donahue {et~al.}(2014)Donahue, Voit, Mahdavi, Umetsu, Ettori, Merten,
  Postman, Hoffer, Baldi, Coe, {et~al.}}]{donahue2014clash}
Donahue, M., Voit, G.~M., Mahdavi, A., {et~al.} 2014, The Astrophysical
  Journal, 794, 136

\bibitem[{Dumoulin {et~al.}(2018)Dumoulin, Perez, Schucher, Strub, Vries,
  Courville, \& Bengio}]{dumoulin2018feature-wise}
Dumoulin, V., Perez, E., Schucher, N., {et~al.} 2018, Distill,
  https://distill.pub/2018/feature-wise-transformations

\bibitem[{Eckert {et~al.}(2022)Eckert, Ettori, Pointecouteau, van~der Burg, \&
  Loubser}]{eckert2022gravitational}
Eckert, D., Ettori, S., Pointecouteau, E., van~der Burg, R., \& Loubser, S.
  2022, Astronomy \& Astrophysics, 662, A123

\bibitem[{Ettori {et~al.}(2013)Ettori, Donnarumma, Pointecouteau, Reiprich,
  Giodini, Lovisari, \& Schmidt}]{ettori2013mass}
Ettori, S., Donnarumma, A., Pointecouteau, E., {et~al.} 2013, Space Science
  Reviews, 177, 119

\bibitem[{Ferragamo {et~al.}(2023)Ferragamo, de~Andres, Sbriglio, Cui,
  De~Petris, Yepes, Dupuis, Jarraya, Lahouli, De~Luca,
  {et~al.}}]{ferragamo2023three}
Ferragamo, A., de~Andres, D., Sbriglio, A., {et~al.} 2023, Monthly Notices of
  the Royal Astronomical Society, 520, 4000

\bibitem[{Ferragamo {et~al.}(2020)Ferragamo, Rubi{\~n}o-Mart{\'\i}n,
  Betancort-Rijo, Munari, Sartoris, \& Barrena}]{ferragamo2020biases}
Ferragamo, A., Rubi{\~n}o-Mart{\'\i}n, J., Betancort-Rijo, J., {et~al.} 2020,
  Astronomy \& Astrophysics, 641, A41

\bibitem[{{Fukugita} {et~al.}(1998){Fukugita}, {Hogan}, \&
  {Peebles}}]{Fukugita1998}
{Fukugita}, M., {Hogan}, C.~J., \& {Peebles}, P.~J.~E. 1998, The Astrophysical
  Journal, 503, 518

\bibitem[{Gao {et~al.}(2004)Gao, White, Jenkins, Stoehr, \&
  Springel}]{gao2004subhalo}
Gao, L., White, S.~D., Jenkins, A., Stoehr, F., \& Springel, V. 2004, Monthly
  Notices of the Royal Astronomical Society, 355, 819

\bibitem[{Gianfagna {et~al.}(2021)Gianfagna, De~Petris, Yepes, De~Luca,
  Sembolini, Cui, Biffi, K{\'e}ruzor{\'e}, Mac{\'\i}as-P{\'e}rez, Mayet,
  {et~al.}}]{gianfagna2021exploring}
Gianfagna, G., De~Petris, M., Yepes, G., {et~al.} 2021, Monthly Notices of the
  Royal Astronomical Society, 502, 5115

\bibitem[{Giocoli {et~al.}(2025)Giocoli, Despali, Meneghetti, Rasia,
  Moscardini, Borgani, Lesci, Marulli, Cui, \& Yepes}]{giocoli2025three}
Giocoli, C., Despali, G., Meneghetti, M., {et~al.} 2025, Astronomy \&
  Astrophysics, 697, A184

\bibitem[{Giocoli {et~al.}(2014)Giocoli, Meneghetti, Metcalf, Ettori, \&
  Moscardini}]{giocoli2014mass}
Giocoli, C., Meneghetti, M., Metcalf, R.~B., Ettori, S., \& Moscardini, L.
  2014, Monthly Notices of the Royal Astronomical Society, 440, 1899

\bibitem[{Giocoli {et~al.}(2008)Giocoli, Pieri, \&
  Tormen}]{giocoli2008analytical}
Giocoli, C., Pieri, L., \& Tormen, G. 2008, Monthly Notices of the Royal
  Astronomical Society, 387, 689

\bibitem[{Giocoli {et~al.}(2010)Giocoli, Tormen, Sheth, \& Van
  Den~Bosch}]{giocoli2010substructure}
Giocoli, C., Tormen, G., Sheth, R.~K., \& Van Den~Bosch, F.~C. 2010, Monthly
  Notices of the Royal Astronomical Society, 404, 502

\bibitem[{Herbonnet {et~al.}(2022)Herbonnet, Crawford, Avestruz, Rasia,
  Giocoli, Meneghetti, von Der~Linden, Cui, \& Yepes}]{herbonnet2022brightest}
Herbonnet, R., Crawford, A., Avestruz, C., {et~al.} 2022, Monthly Notices of
  the Royal Astronomical Society, 513, 2178

\bibitem[{Hong {et~al.}(2021)Hong, Park, Jee, Bak, \& Cha}]{hong2021weak}
Hong, S.~E., Park, S., Jee, M.~J., Bak, D., \& Cha, S. 2021, The Astrophysical
  Journal, 923, 266

\bibitem[{{Hsu} {et~al.}(2025){Hsu}, {Ho}, {Lin}, {Markey}, {Ntampaka}, {Trac},
  \& {P{\'o}czos}}]{2025OJAp....8E..92H}
{Hsu}, A., {Ho}, M., {Lin}, J., {et~al.} 2025, The Open Journal of
  Astrophysics, 8, 92

\bibitem[{Ioffe \& Szegedy(2015)}]{ioffe2015batch}
Ioffe, S. \& Szegedy, C. 2015, in International conference on machine learning,
  pmlr, 448--456

\bibitem[{Janulewicz \& Cui(2025)}]{janulewicz2025pymgal}
Janulewicz, P. \& Cui, W. 2025, RAS Techniques and Instruments, 4, rzaf027

\bibitem[{Jouvel {et~al.}(2014)Jouvel, Host, Lahav, Seitz, Molino, Coe,
  Postman, Moustakas, Ben{\`\i}tez, Rosati, {et~al.}}]{jouvel2014clash}
Jouvel, S., Host, O., Lahav, O., {et~al.} 2014, Astronomy \& Astrophysics, 562,
  A86

\bibitem[{Kaiser {et~al.}(1994)Kaiser, Squires, \&
  Broadhurst}]{kaiser1994method}
Kaiser, N., Squires, G., \& Broadhurst, T. 1994, arXiv preprint
  astro-ph/9411005

\bibitem[{Kingma \& Ba(2014)}]{kingma2014adam}
Kingma, D.~P. \& Ba, J. 2014, arXiv preprint arXiv:1412.6980

\bibitem[{Klypin {et~al.}(2016)Klypin, Yepes, Gottl{\"o}ber, Prada, \&
  Hess}]{klypin2016multidark}
Klypin, A., Yepes, G., Gottl{\"o}ber, S., Prada, F., \& Hess, S. 2016, Monthly
  Notices of the Royal Astronomical Society, 457, 4340

\bibitem[{Kneib \& Natarajan(2011)}]{kneib2011cluster}
Kneib, J.-P. \& Natarajan, P. 2011, The Astronomy and Astrophysics Review, 19,
  47

\bibitem[{Knollmann \& Knebe(2009)}]{knollmann2009ahf}
Knollmann, S.~R. \& Knebe, A. 2009, The Astrophysical Journal Supplement
  Series, 182, 608

\bibitem[{Kodi~Ramanah {et~al.}(2020)Kodi~Ramanah, Wojtak, Ansari, Gall, \&
  Hjorth}]{kodi2020dynamical}
Kodi~Ramanah, D., Wojtak, R., Ansari, Z., Gall, C., \& Hjorth, J. 2020, Monthly
  Notices of the Royal Astronomical Society, 499, 1985

\bibitem[{Kravtsov \& Borgani(2012)}]{kravtsov2012formation}
Kravtsov, A.~V. \& Borgani, S. 2012, Annual Review of Astronomy and
  Astrophysics, 50, 353

\bibitem[{Limousin {et~al.}(2013)Limousin, Morandi, Sereno, Meneghetti, Ettori,
  Bartelmann, \& Verdugo}]{limousin2013three}
Limousin, M., Morandi, A., Sereno, M., {et~al.} 2013, Space Science Reviews,
  177, 155

\bibitem[{Mahdavi {et~al.}(2007)Mahdavi, Hoekstra, Babul, Sievers, Myers, \&
  Henry}]{mahdavi2007joint}
Mahdavi, A., Hoekstra, H., Babul, A., {et~al.} 2007, The Astrophysical Journal,
  664, 162

\bibitem[{Mancone \& Gonzalez(2012)}]{mancone2012ezgal}
Mancone, C.~L. \& Gonzalez, A.~H. 2012, Publications of the Astronomical
  Society of the Pacific, 124, 606

\bibitem[{Mehrtens {et~al.}(2012)Mehrtens, Romer, Hilton, Lloyd-Davies, Miller,
  Stanford, Hosmer, Hoyle, Collins, Liddle, {et~al.}}]{mehrtens2012xmm}
Mehrtens, N., Romer, A.~K., Hilton, M., {et~al.} 2012, Monthly Notices of the
  Royal Astronomical Society, 423, 1024

\bibitem[{Meneghetti {et~al.}(2020)Meneghetti, Davoli, Bergamini, Rosati,
  Natarajan, Giocoli, Caminha, Metcalf, Rasia, Borgani,
  {et~al.}}]{meneghetti2020excess}
Meneghetti, M., Davoli, G., Bergamini, P., {et~al.} 2020, Science, 369, 1347

\bibitem[{Meneghetti {et~al.}(2010)Meneghetti, Rasia, Merten, Bellagamba,
  Ettori, Mazzotta, Dolag, \& Marri}]{meneghetti2010weighing}
Meneghetti, M., Rasia, E., Merten, J., {et~al.} 2010, Astronomy \&
  Astrophysics, 514, A93

\bibitem[{Merten {et~al.}(2009)Merten, Cacciato, Meneghetti, Mignone, \&
  Bartelmann}]{merten2009combining}
Merten, J., Cacciato, M., Meneghetti, M., Mignone, C., \& Bartelmann, M. 2009,
  Astronomy \& Astrophysics, 500, 681

\bibitem[{Merten {et~al.}(2015)Merten, Meneghetti, Postman, Umetsu, Zitrin,
  Medezinski, Nonino, Koekemoer, Melchior, Gruen, {et~al.}}]{merten2015clash}
Merten, J., Meneghetti, M., Postman, M., {et~al.} 2015, The Astrophysical
  Journal, 806, 4

\bibitem[{Mu{\~n}oz-Echeverr{\'\i}a {et~al.}(2024)Mu{\~n}oz-Echeverr{\'\i}a,
  Mac{\'\i}as-P{\'e}rez, Artis, Cui, de~Andres, De~Luca, De~Petris, Ferragamo,
  Giocoli, Hanser, {et~al.}}]{munoz2024galaxy}
Mu{\~n}oz-Echeverr{\'\i}a, M., Mac{\'\i}as-P{\'e}rez, J., Artis, E., {et~al.}
  2024, Astronomy \& Astrophysics, 682, A124

\bibitem[{Nair \& Hinton(2010)}]{nair2010rectified}
Nair, V. \& Hinton, G.~E. 2010, in Proceedings of the 27th international
  conference on machine learning (ICML-10), 807--814

\bibitem[{{Natarajan} {et~al.}(2026){Natarajan}, {Chiang}, \&
  {Dutra}}]{Natarajan2026}
{Natarajan}, P., {Chiang}, B.~T., \& {Dutra}, I. 2026, The Astrophysical
  Journal, 1001, L12

\bibitem[{Nde {et~al.}(2025)Nde, Wu, Cao, Kamau, Tamosiunas, To, \&
  Zhou}]{nde2025impact}
Nde, T.~N., Wu, H.-Y., Cao, S., {et~al.} 2025, arXiv preprint arXiv:2510.00753

\bibitem[{Nelson {et~al.}(2024)Nelson, Pillepich, Ayromlou, Lee, Lehle, Rohr,
  \& Truong}]{nelson2024introducing}
Nelson, D., Pillepich, A., Ayromlou, M., {et~al.} 2024, Astronomy \&
  Astrophysics, 686, A157

\bibitem[{Ntampaka {et~al.}(2019)Ntampaka, ZuHone, Eisenstein, Nagai,
  Vikhlinin, Hernquist, Marinacci, Nelson, Pakmor, Pillepich,
  {et~al.}}]{ntampaka2019deep}
Ntampaka, M., ZuHone, J., Eisenstein, D., {et~al.} 2019, The Astrophysical
  Journal, 876, 82

\bibitem[{Okabe \& Smith(2016)}]{okabe2016locuss}
Okabe, N. \& Smith, G.~P. 2016, Monthly Notices of the Royal Astronomical
  Society, 461, 3794

\bibitem[{Old {et~al.}(2018)Old, Wojtak, Pearce, Gray, Mamon, Sif{\'o}n,
  Tempel, Biviano, Yee, de~Carvalho, {et~al.}}]{old2018galaxy}
Old, L., Wojtak, R., Pearce, F., {et~al.} 2018, Monthly Notices of the Royal
  Astronomical Society, 475, 853

\bibitem[{Piffaretti \& Valdarnini(2008)}]{piffaretti2008total}
Piffaretti, R. \& Valdarnini, R. 2008, Astronomy \& Astrophysics, 491, 71

\bibitem[{Pillepich {et~al.}(2018)Pillepich, Nelson, Hernquist, Springel,
  Pakmor, Torrey, Weinberger, Genel, Naiman, Marinacci,
  {et~al.}}]{pillepich2018first}
Pillepich, A., Nelson, D., Hernquist, L., {et~al.} 2018, Monthly Notices of the
  Royal Astronomical Society, 475, 648

\bibitem[{{Planck Collaboration} {et~al.}(2016){Planck Collaboration}, Ade,
  Aghanim, {et~al.}}]{planck2015results}
{Planck Collaboration}, Ade, P. A.~R., Aghanim, N., {et~al.} 2016, Astronomy \&
  Astrophysics, 594 [\eprint[arXiv]{1502.01598}]

\bibitem[{{Planck Collaboration}(2016)}]{planck2016cosmological}
{Planck Collaboration}, e.~a. 2016, Astronomy \& Astrophysics, 594, A13

\bibitem[{Postman {et~al.}(2012)Postman, Coe, Benitez, Bradley, Broadhurst,
  Donahue, Ford, Graur, Graves, Jouvel, {et~al.}}]{postman2012cluster}
Postman, M., Coe, D., Benitez, N., {et~al.} 2012, The Astrophysical Journal
  Supplement Series, 199, 25

\bibitem[{Pratt {et~al.}(2019)Pratt, Arnaud, Biviano, Eckert, Ettori, Nagai,
  Okabe, \& Reiprich}]{pratt2019galaxy}
Pratt, G., Arnaud, M., Biviano, A., {et~al.} 2019, Space Science Reviews, 215,
  25

\bibitem[{Ragagnin {et~al.}(2021)Ragagnin, Saro, Singh, \&
  Dolag}]{ragagnin2021cosmology}
Ragagnin, A., Saro, A., Singh, P., \& Dolag, K. 2021, Monthly Notices of the
  Royal Astronomical Society, 500, 5056

\bibitem[{Rasia {et~al.}(2004)Rasia, Tormen, \&
  Moscardini}]{rasia2004dynamical}
Rasia, E., Tormen, G., \& Moscardini, L. 2004, Monthly Notices of the Royal
  Astronomical Society, 351, 237

\bibitem[{Ronneberger {et~al.}(2015)Ronneberger, Fischer, \&
  Brox}]{ronneberger2015u}
Ronneberger, O., Fischer, P., \& Brox, T. 2015, in International Conference on
  Medical image computing and computer-assisted intervention, Springer,
  234--241

\bibitem[{Rowe {et~al.}(2015)Rowe, Jarvis, Mandelbaum, Bernstein, Bosch, Simet,
  Meyers, Kacprzak, Nakajima, Zuntz, {et~al.}}]{rowe2015galsim}
Rowe, B.~T., Jarvis, M., Mandelbaum, R., {et~al.} 2015, Astronomy and
  Computing, 10, 121

\bibitem[{Saro {et~al.}(2013)Saro, Mohr, Bazin, \& Dolag}]{saro2013toward}
Saro, A., Mohr, J.~J., Bazin, G., \& Dolag, K. 2013, The Astrophysical Journal,
  772, 47

\bibitem[{Springel(2010)}]{springel2010pur}
Springel, V. 2010, Monthly Notices of the Royal Astronomical Society, 401, 791

\bibitem[{Umetsu {et~al.}(2016)Umetsu, Zitrin, Gruen, Merten, Donahue, \&
  Postman}]{umetsu2016clash}
Umetsu, K., Zitrin, A., Gruen, D., {et~al.} 2016, The Astrophysical Journal,
  821, 116

\bibitem[{Van Den~Bosch {et~al.}(2005)Van Den~Bosch, Tormen, \&
  Giocoli}]{van2005mass}
Van Den~Bosch, F.~C., Tormen, G., \& Giocoli, C. 2005, Monthly Notices of the
  Royal Astronomical Society, 359, 1029

\bibitem[{Vegetti {et~al.}(2010)Vegetti, Koopmans, Bolton, Treu, \&
  Gavazzi}]{vegetti2010detection}
Vegetti, S., Koopmans, L., Bolton, A., Treu, T., \& Gavazzi, R. 2010, Monthly
  Notices of the Royal Astronomical Society, 408, 1969

\bibitem[{Vikhlinin {et~al.}(2009)Vikhlinin, Kravtsov, Burenin, Ebeling,
  Forman, Hornstrup, Jones, Murray, Nagai, Quintana,
  {et~al.}}]{vikhlinin2009chandra}
Vikhlinin, A., Kravtsov, A., Burenin, R., {et~al.} 2009, The Astrophysical
  Journal, 692, 1060

\bibitem[{Vikhlinin {et~al.}(2006)Vikhlinin, Kravtsov, Forman, Jones,
  Markevitch, Murray, \& Van~Speybroeck}]{vikhlinin2006chandra}
Vikhlinin, A., Kravtsov, A., Forman, W., {et~al.} 2006, The Astrophysical
  Journal, 640, 691

\bibitem[{{Villaescusa-Navarro}(2018)}]{Pylians}
{Villaescusa-Navarro}, F. 2018, {Pylians: Python libraries for the analysis of
  numerical simulations}, Astrophysics Source Code Library, record
  ascl:1811.008

\bibitem[{{White} {et~al.}(1993){White}, {Navarro}, {Evrard}, \&
  {Frenk}}]{White1993}
{White}, S. D.~M., {Navarro}, J.~F., {Evrard}, A.~E., \& {Frenk}, C.~S. 1993,
  Nature, 366, 429

\bibitem[{Wu {et~al.}(2022)Wu, Costanzi, To, Salcedo, Weinberg, Annis, Bocquet,
  da~Silva~Pereira, DeRose, Esteves, {et~al.}}]{wu2022optical}
Wu, H.-Y., Costanzi, M., To, C.-H., {et~al.} 2022, Monthly Notices of the Royal
  Astronomical Society, 515, 4471

\bibitem[{Zhang {et~al.}(2023)Zhang, Wu, Zhang, Frieman, To, DeRose, Costanzi,
  Wechsler, Adhikari, Rykoff, {et~al.}}]{zhang2023modelling}
Zhang, Z., Wu, H.-Y., Zhang, Y., {et~al.} 2023, Monthly Notices of the Royal
  Astronomical Society, 523, 1994

\bibitem[{Zitrin \& Broadhurst(2016)}]{zitrin2016strong}
Zitrin, A. \& Broadhurst, T. 2016, The Astrophysical Journal, 833, 25

\bibitem[{Zitrin {et~al.}(2015)Zitrin, Fabris, Merten, Melchior, Meneghetti,
  Koekemoer, Coe, Maturi, Bartelmann, Postman, {et~al.}}]{zitrin2015hubble}
Zitrin, A., Fabris, A., Merten, J., {et~al.} 2015, The Astrophysical Journal,
  801, 44

\end{thebibliography}

\begin{appendix}
\nolinenumbers
\section{Preprocessing of CLASH HST observations and convergence maps}\label{subsection:Preprocessing HST and kappa}

In Table \ref{Table:Clash Clusters}, we give all the CLASH clusters with their corresponding redshift and mass M${_{500c}}$ \footnote{The mass M$_{{500c}}$ is defined as the mass contained within a sphere of radius R$_{500c}$ corresponding to a spherical over-density of 500 times the critical density of the Universe.} based on \cite{merten2015clash}. These masses were derived with the \texttt{SaWLens} \citep{merten2009combining} code, which consistently combines weak and strong lensing in a non-parametric fashion on adaptively refined grids.
\begin{table}[htbp]
\caption{The CLASH cluster sample with their redshifts and $M_{500c}$ masses estimated from lensing \citep{merten2015clash}. We excluded four clusters (MACS~J2129$-$07, Abell 1423, MACS~J0744$+$39, and CL~J1226$+$3332) due to limited analysis data or redshifts falling outside our restricted range ($0.152 < z < 0.607$). Bold entries indicate clusters with spectroscopic redshift catalogs discussed in Section \nameref{sec:Results}.}
\label{Table:Clash Clusters}
\centering
\setlength{\tabcolsep}{4.5pt} 
\small 
\begin{tabular}{rlccc}
\toprule
ID & Cluster Name & Redshift ($z$) & $M_{500c}$ ($10^{14}\,M_\odot/h$) & Used \\
\midrule
1  & \textbf{Abell 209} & 0.206 & 6.27 & \checkmark \\
2  & Abell 383 & 0.188 & 6.12 & \checkmark \\
3  & Abell 1423 & 0.213 & -- & $\times$ \\
4  & Abell 611 & 0.288 & 5.70 & \checkmark \\
5  & Abell 2261 & 0.225 & 9.48 & \checkmark \\
6  & MACS~J0329$-$02 & 0.450 & 5.05 & \checkmark \\
7  & \textbf{MACS~J0416$-$2403} & 0.420 & 4.33 & \checkmark \\
8  & MACS~J0429$-$02 & 0.399 & 5.32 & \checkmark \\
9  & MACS~J0647+70 & 0.584 & 6.42 & \checkmark \\
10 & MACS~J0717+37 & 0.548 & 9.68 & \checkmark \\
11 & MACS~J1115+01 & 0.352 & 5.43 & \checkmark \\
12 & \textbf{MACS~J1149+2223} & 0.544 & 7.06 & \checkmark \\
13 & \textbf{MACS~J1206$-$0847} & 0.439 & 6.05 & \checkmark \\
14 & MACS~J1311$-$03 & 0.494 & 3.20 & \checkmark \\
15 & MACS~J2129$-$07 & 0.570 & -- & $\times$ \\
16 & MACS~J1423+24 & 0.545 & 4.09 & \checkmark \\
17 & MACS~J1720+35 & 0.391 & 5.29 & \checkmark \\
18 & MACS~J1931$-$26 & 0.352 & 4.50 & \checkmark \\
19 & MACS~J0744+39 & 0.686 & 4.88 & $\times$ \\
20 & MS~2137.3$-$2353 & 0.313 & 6.79 & \checkmark \\
21 & RXJ~1347$-$1145 & 0.451 & 7.97 & \checkmark \\
22 & RXJ~1532.9+3021 & 0.345 & 3.44 & \checkmark \\
23 & RXJ~2129+0005 & 0.234 & 4.30 & \checkmark \\
24 & \textbf{RXJ~2248.7$-$4431} & 0.348 & 7.66 & \checkmark \\
25 & CL~J1226+3332 & 0.890 & 10.80 & $\times$ \\
\bottomrule
\end{tabular}
\end{table}
We analyzed 21 clusters out of the 25 CLASH clusters for mass maps prediction using different U-Net models, see Table \ref{Table:Clash Clusters} for details.
For five of these CLASH clusters—namely Abell~209, MACS~J0416$-$2403, MACS~J1149+2223, MACS~J1206$-$0847, and RXJ~2248.7$-$4431—the corresponding CLASH-VLT spectroscopic catalogs are available and are used to identify cluster membership based on spectroscopic redshifts.
This identification step is required before applying the U-Net models to real CLASH HST observations, in order to replace the pixel values of point sources and non-cluster members with a background estimate.

Membership for CLASH clusters with spectroscopic redshift catalog was assigned using a velocity-based selection criterion,
\begin{equation}
\Delta v = c \left|\frac{z_{\mathrm{spec}} - z_{\mathrm{cl}}}{1 + z_{\mathrm{cl}}}\right| \leq 3000~\mathrm{km,s^{-1}},
\end{equation}
where \(z_{\mathrm{spec}}\) is the spectroscopic redshift from the catalog and \(z_{\mathrm{cl}}\) is the cluster redshift. The identified cluster non-members (i.e. those satisfying \(\Delta v > 3000~\mathrm{km,s^{-1}})\) were cross-matched with the CLASH HST photometric-redshift (BPZ) catalog. 
Point sources were also identified using the same CLASH HST photometric-redshift catalog, for which we required a stellarity index \(\geq 0.1\). The circular aperture for each cluster non-member and point source was determined from the catalog isophotal aperture area.

Bright foreground sources with strong diffraction spikes that were not captured by the catalog were added manually. We generated a circular-region record file containing the pixel coordinates and masking radii corresponding to each identified cluster non-member, foreground contamination and point source. This region file was then used to mask these regions and replace the overlapping pixels with a synthetic background estimate. During masking, each radius was scaled by a factor of 1.5 to remove extended faint light and diffraction features.

We used \texttt{Photutils} \citep{bradley2016photutils} to construct a two-dimensional background model for each HST image. A 2.5\(\sigma\) iterative sigma-clipping procedure with a \(50 \times 50\) pixel box size was employed to estimate the local median background and the corresponding background RMS. After estimating the background statistics, we iterated over each object in the region file and replaced the pixel values within its masking aperture with the local background-model value at that pixel position plus Gaussian noise drawn from the local background RMS.

For the remaining 16 clusters, for which CLASH-VLT spectroscopic catalogs are not available, we rely on just using the CLASH HST photometric-redshift (BPZ) catalog. For these clusters, cluster membership is assigned to objects whose photometric redshifts lie within $3\sigma$ of the cluster redshift, where the $1\sigma$ photometric-redshift uncertainty is given by $0.04(1+z_{\mathrm{object}})$ \citep{jouvel2014clash, burke2015coevolution}. This corresponds to the condition
$
|z_b - z_c| < 3 \times 0.04(1 + z_b),
$
where $z_c$ is the cluster redshift and $z_b$ is the photometric redshift of the object.
The point-source identification criteria for these clusters are the same as for the five clusters with spectroscopic coverage. The masking and background-replacement procedure also remains unchanged.

The U-Net models were trained on mock HST observations of The300 simulated clusters as inputs, with the corresponding projected mass maps as targets. To apply these models to real CLASH observations, the HST data must be converted into units consistent with those used in the simulations.
The CLASH images, originally provided in count rates, were converted into observer-frame luminosities (in $\mathrm{erg,s^{-1}}$) using \texttt{PHOTFLAM}—the inverse-sensitivity calibration factor for each instrument/filter—together with the corresponding filter bandwidth and the luminosity distance computed at the cluster redshift assuming a Planck cosmology.
These luminosity maps were then cropped to a $500 \times 500~\mathrm{kpc}^{2}$ field of view centered on the cluster's optical center and down-sampled to $128 \times 128$ pixels, matching the format of the simulated HST maps used during training.

To compare the projected mass maps inferred using our U-Net models with an independent observational estimate, we used lensing and X-ray-based mass measurements of the CLASH clusters. For lensing-based mass estimates, the CLASH survey provides convergence ($\kappa$) maps derived from joint strong and weak lensing analysis based on \citet{zitrin2015hubble}. They provide two different mass-model parametrization: (i) a light-traces-mass (LTM) model and (ii) a pseudo-isothermal elliptical mass distribution combined with an elliptical NFW halo (PIEMD+eNFW). We adopt the LTM models for comparison, as their underlying assumption—that the total matter distribution traces the observed light—is more closely aligned with ours. For further details between the two models, as well as their latest result, \citep{bergamini2023glass}, we refer interested readers to their papers.

In the LTM framework, each cluster galaxy is assigned a power-law surface density profile scaled by its luminosity; these are summed and smoothed with a Gaussian kernel to represent the diffuse dark-matter component. The galaxy and dark-matter components are then combined with a relative weighting factor to generate a continuous projected mass distribution. To compare the LTM outputs with our U-Net predictions, we converted the $\kappa$ maps into physical projected mass maps using
\begin{equation}
\Sigma~[M_{\odot}] = \kappa \times \Sigma_{\text{crit}} \times A_{\text{pix}},
\end{equation}
where $\Sigma_{\text{crit}}$ is the critical surface density,
\begin{equation}
\Sigma_{\text{crit}} = \frac{c^{2}}{4 \pi G}\frac{D_{S}}{D_{L} D_{LS}},    
\end{equation}
expressed in $M_{\odot}~\text{Mpc}^{-2}$, with angular-diameter distances $D_{L}$, $D_{S}$, and $D_{LS}$ computed for the lens and source redshifts. Finally, the pixel area $A_{\text{pix}}$ is obtained by converting the HST pixel scale of $0.065~\text{arcsec pixel}^{-1}$ into physical units using the cluster redshift and Planck cosmology, yielding $\text{Mpc}^{2}$ per pixel.

\section{Performance validation with TNG-Clusters}\label{TNG-Clusters}
We further evaluate the U-Net models by applying them to an independent set of simulated galaxy clusters drawn from the TNG-Cluster simulation suite \citep{nelson2024introducing}, which differs from The300 in terms of numerical implementation, resolution, and baryonic physics model. This provides a test of the U-Nets' ability to generalize the prediction of projected mass maps across different simulations, and as such, it can be applicable to real observations. TNG-Cluster consists of 352 high-resolution zoom-in simulations of massive galaxy clusters ($M_{200c} \gtrsim 10^{14.5}, M_{\odot}$), selected from a $1~\mathrm{Gpc}^{3}$ parent volume and evolved with the \texttt{AREPO} code \citep{springel2010pur}. The simulations adopt the IllustrisTNG galaxy formation model \citep{pillepich2018first} and a 
Planck-consistent $\Lambda$CDM cosmology \citep{planck2016cosmological},
same as The300,
with mass and spatial resolutions comparable to the highest-resolution TNG300 realization. Specifically, the baryonic and dark-matter particle masses are $m_{\mathrm{bar}} \simeq 1.1 \times 10^{7} M_{\odot}$ and $m_{\mathrm{DM}} \simeq 6.1 \times 10^{7} M_{\odot}$, respectively, with a gravitational softening length of $\sim 1.5$ kpc at $z=0$ for collisionless components. From the TNG-clusters simulation dataset, we select a subset of 42 clusters covering $z \simeq 0.2$–$0.5$ and $M_{200c} \sim (3$–$10)\times10^{14}, M_\odot$, matching the mass and redshift range of the CLASH sample. For each cluster, mock observations in different HST filters and the corresponding projected total mass maps were generated along a single line of sight and a field of view of 500 $\times$ 500 kpc$^2$ using the same pipeline adopted for The300 clusters (see, Sections~\ref{Mock HST Cluster Observation from The300}). As the TNG clusters were not used during training, this sample enables a direct assessment of the U-Net models’ ability to generalize across simulations when reconstructing projected mass maps.

\begin{figure}[tbp]
\centering
    \includegraphics[width=0.7\linewidth]{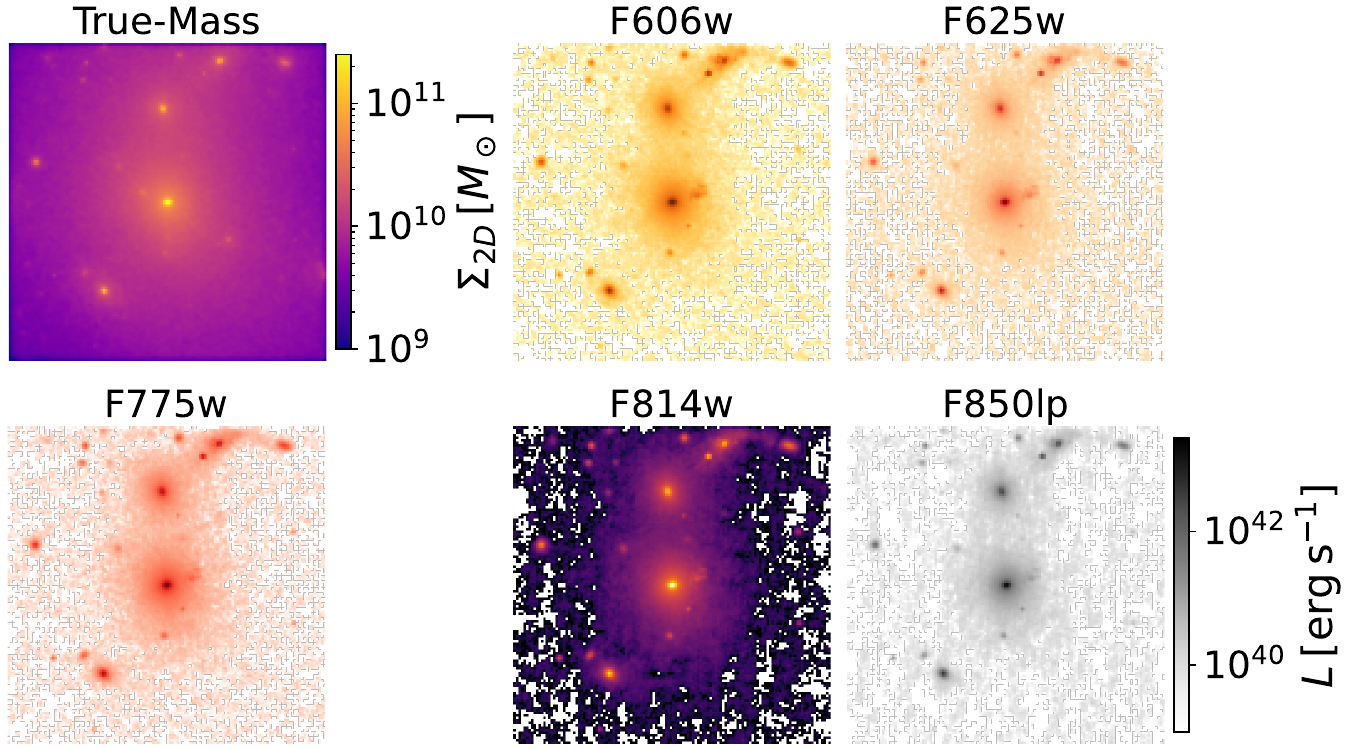}
    \includegraphics[width=0.7\linewidth]{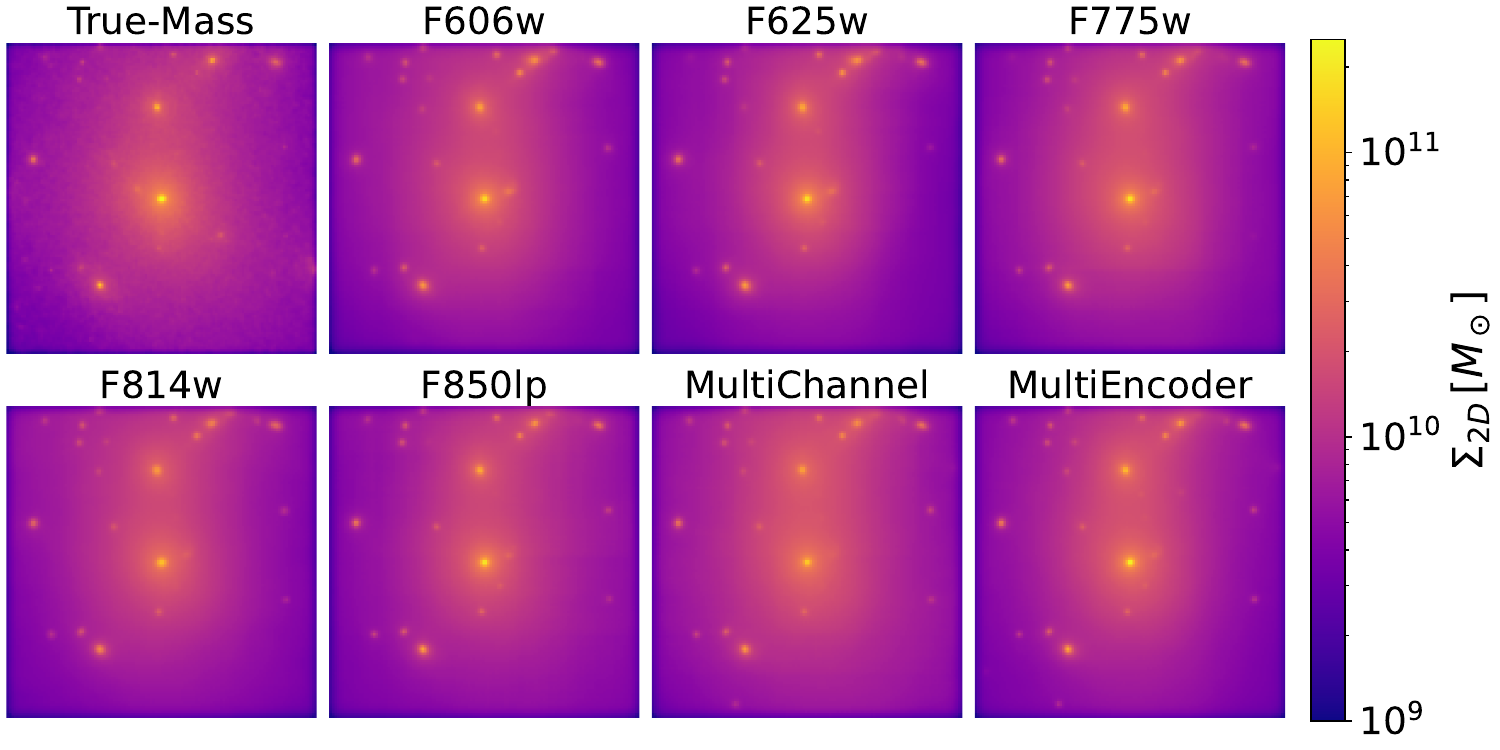}
    \includegraphics[width=0.7\linewidth]{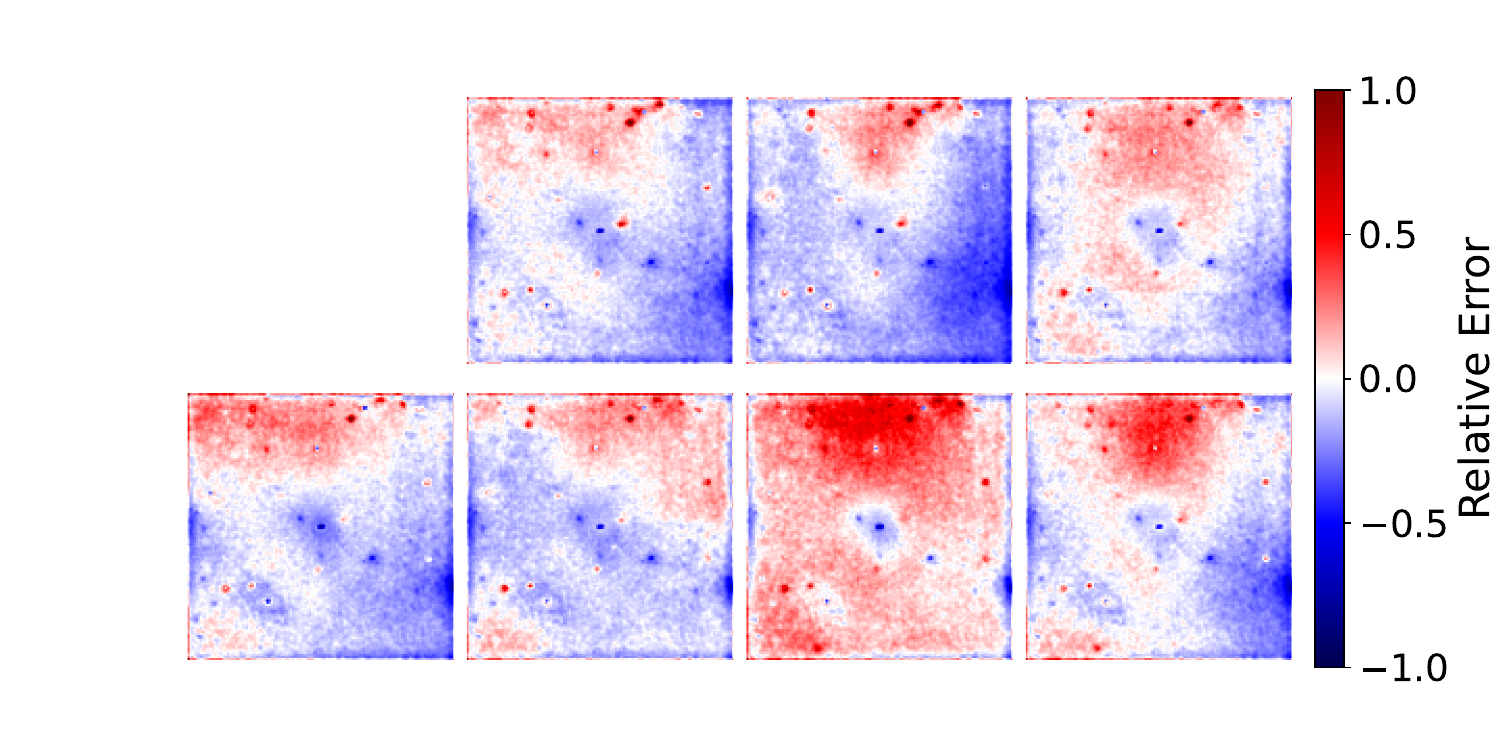}

\caption{Cross-simulation generalization to TNG clusters. Example reconstruction for a TNG cluster omitted from the training phase. The top panel displays the true projected mass distribution alongside the corresponding mock HST luminosity maps across multiple filters. The middle panel shows the projected mass maps predicted by U-Net models trained exclusively on The300 simulations, and the bottom panel displays the corresponding pixel-wise relative error residuals. All subplots share an identical field of view.}
\label{fig:TNG-Data-Example}
\end{figure}

In Fig.~\ref{fig:TNG-Data-Example}, we show a typical example cluster from the TNG simulation suite. Top panel of Fig.~\ref{fig:TNG-Data-Example} presents the true projected mass map together with the corresponding HST-like luminosity maps in different filters, while the middle panel~\ref{fig:TNG-Data-Example} shows the projected mass reconstructions obtained with the U-Net models trained on The300 simulations. The models successfully recover the overall morphology of the cluster and many of the small-scale features of the projected mass distribution. The corresponding residual maps are shown in bottom panel of Fig.~\ref{fig:TNG-Data-Example}. Although the models recover the small-scale features, the residual plots indicate that the amplitudes (or masses) associated with these features are either overestimated or underestimated, implying a bias in the reconstructed projected mass maps.

\begin{figure}[tbp]
\centering

    \includegraphics[width=\linewidth]{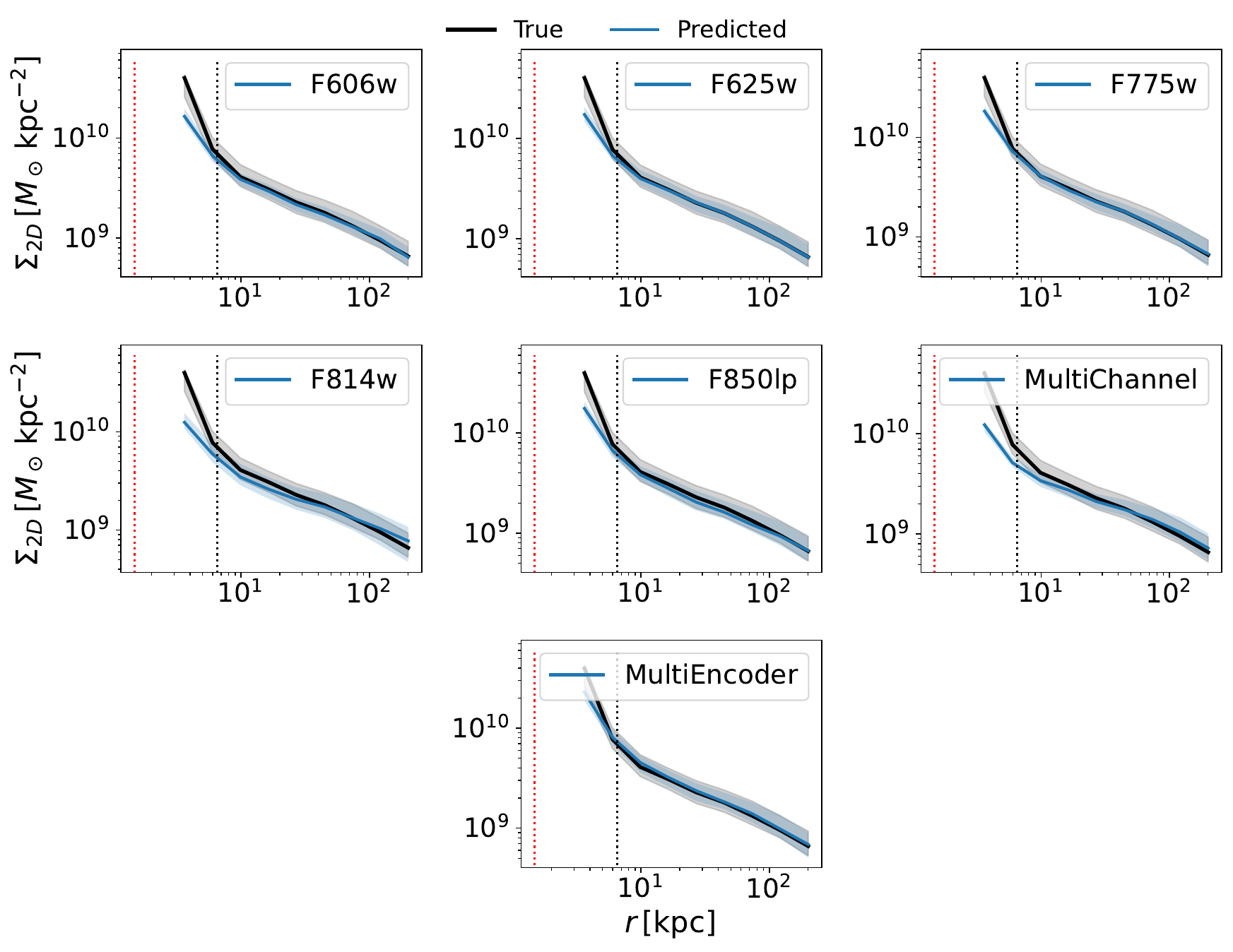}
    \includegraphics[width=\linewidth]{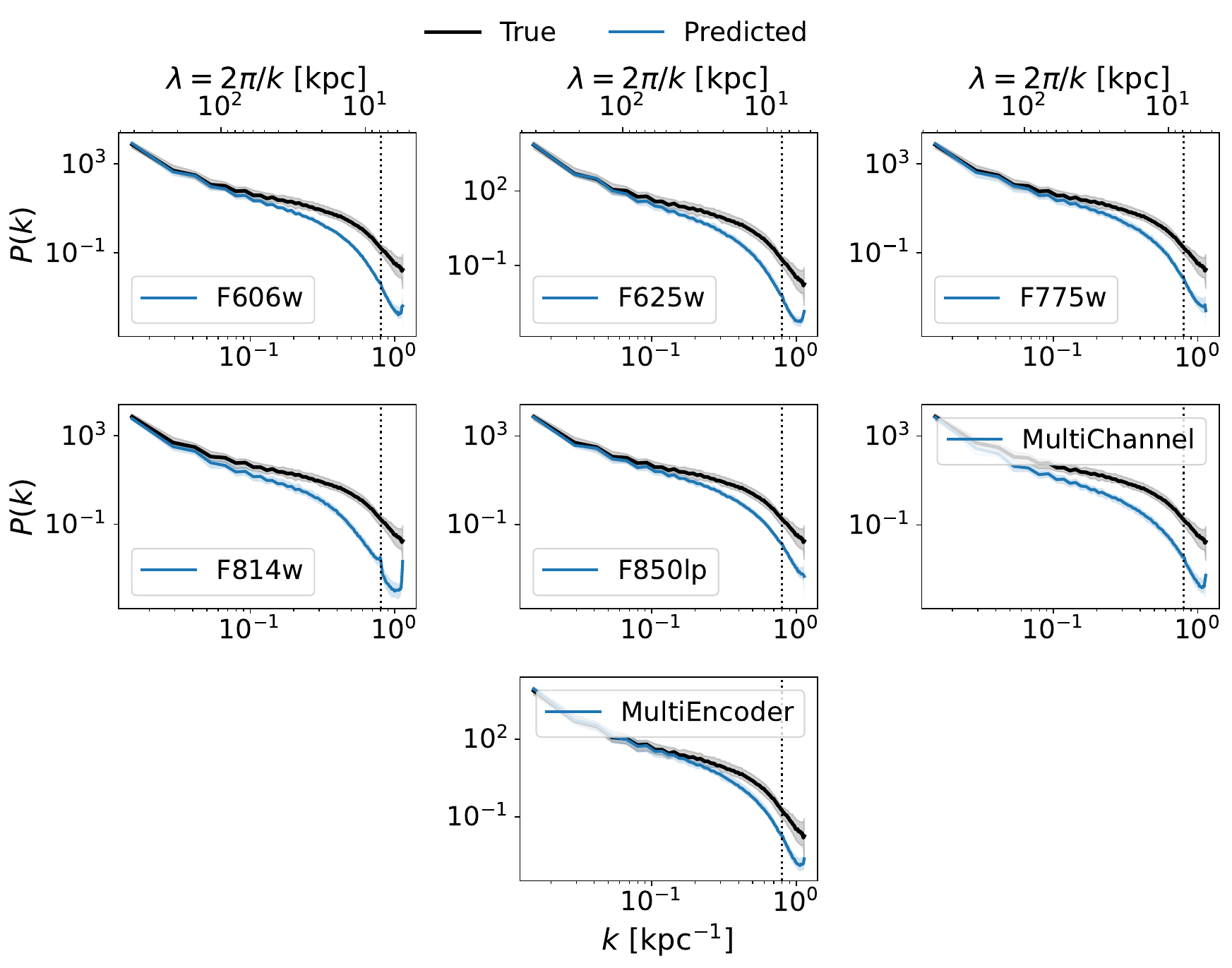}

\caption{Generalization performance of U-Net models trained on The300 simulations when applied to TNG cluster mass maps. The top panel displays the surface mass density radial profiles, and the bottom panel shows the mass map power spectra. In both panels, black lines correspond to the true median, red lines indicate predicted median values, and shaded areas represent the $16^{\rm th}\text{--}84^{\rm th}$ percentile scatter across the TNG cluster sample. Subplots compare individual U-Net architectures. In the radial profile plots, the vertical red and black dotted lines indicate the spatial resolutions of the TNG ($\approx 1.48$~kpc) and The300 ($\approx 6.5$~kpc) simulations, respectively. In the power spectrum plots, the vertical black dotted line at $k \approx 0.8~\text{kpc}^{-1}$ ($\lambda \approx 7.85$~kpc) denotes the Nyquist limit imposed by the $128 \times 128$ pixel grids.}
\label{fig:TNG-radial-and-ps}
\end{figure}

We compare the surface mass density radial profiles and power spectra in Fig.~\ref{fig:TNG-radial-and-ps} to quantify the performance of the projected mass maps across all 42 clusters in the TNG sample. The median predicted surface mass density profiles agree well with the true profiles at intermediate and large radii, but show noticeable discrepancies in the innermost regions. These central differences likely reflect variations in the modeling of central galaxies between the two simulation suites, The300 and TNG. Since the U-Net models are trained on The300, they are naturally better tuned to reproduce the properties of central galaxies as represented in that simulation framework. Outside the central regions, however, the azimuthally averaged projected mass profiles are in significantly better agreement and remain robust across the two simulations. The scatter, quantified by the 16th–84th percentile range, is comparable to that of the ground-truth profiles, indicating stable performance and minimal bias introduced by applying the model across simulations. A similar trend is seen in the power spectra shown in Fig.~\ref{fig:TNG-radial-and-ps}, consistent with the results found for the The300 test sample. For the TNG clusters, the power spectra of the ground-truth and U-Net reconstructed mass maps are consistent on large scales across all U-Net models, as also seen for the The300 test set. On smaller scales, however, the reconstructed power spectra decline more rapidly and diverge from the ground truth. Overall, we can again infer that the multi-encoder U-Net model performs the best compared to the single filter and multi-channel models.
\begin{figure}[tbp]
\centering

    \includegraphics[width=0.7\linewidth]{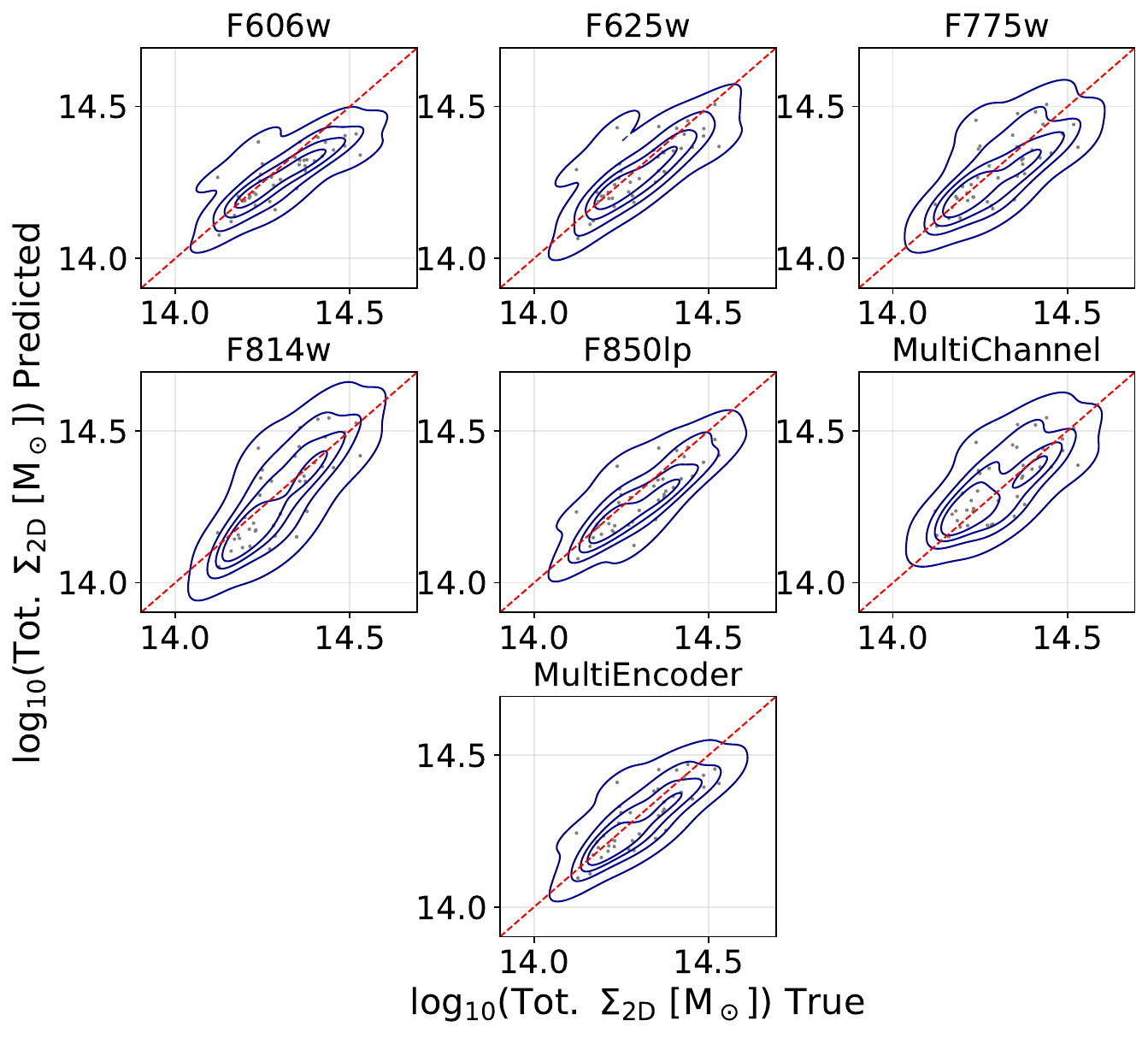}
    
    \includegraphics[width=0.7\linewidth]{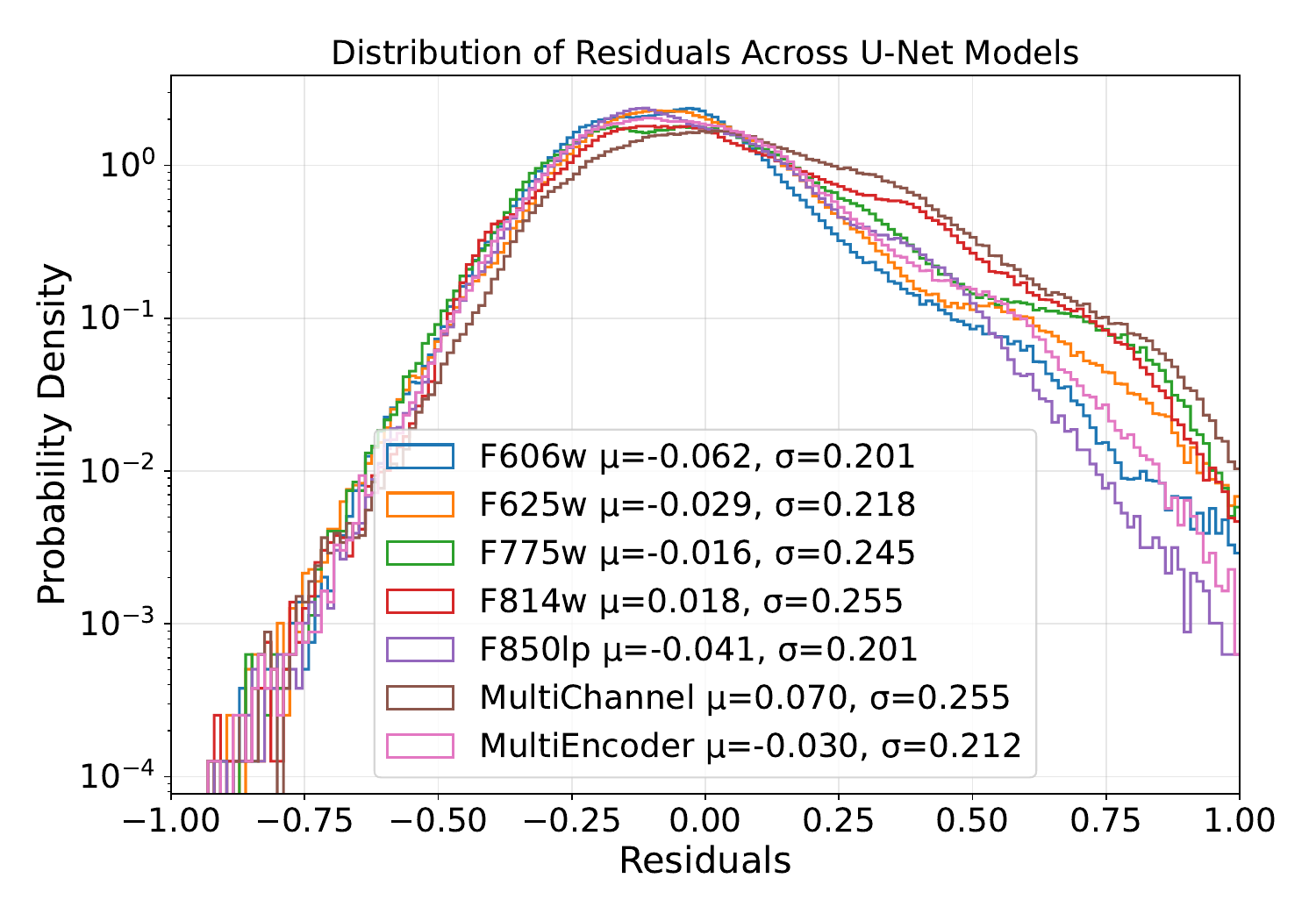}

\caption{Overall mass map reconstruction accuracy for TNG clusters. The top panel compares the total projected mass predicted by the U-Net models against the true values across the full TNG test dataset; the dashed line indicates the ideal 1:1 relation, and contours represent the kernel density estimation (KDE). The bottom panel displays the distribution of pixel-wise relative errors for the same sample, with the legend indicating the mean relative error and $\pm 1\sigma$ standard deviation for each model.}
\label{fig:Unet-Overall-Data-TNG}
\end{figure}

In Fig.~\ref{fig:Unet-Overall-Data-TNG}, we assess the global reconstruction accuracy of the U-Net models when applied to the TNG cluster sample. Fig. \ref{fig:Unet-Overall-Data-TNG} upper panel compares the total projected mass within the field of view predicted by each model with the corresponding true values from the TNG simulations. Overall, the predictions follow the ideal one-to-one relation with slightly larger scatter compared to The300 test set results, possibly due to domain shift. To further quantify the reconstruction quality, we perform a pixel-wise statistical analysis of the relative residuals, shown in Fig.~\ref{fig:Unet-Overall-Data-TNG} lower panel. The resulting distributions indicate a $\pm1\sigma$ scatter in the residuals at the $\sim 19$–$22\%$ level, comparable to that obtained in the The300 simulation test set analysis. These results demonstrate that the U-Net models retain robust predictive performance when applied to an independent simulation suite with different underlying physics and galaxy populations.

\section{Impact of Preprocessing on Model Performance }\label{sec:Preprocessing}
For the main text analysis, we apply a signed logarithmic transform (Eq.~\ref{eq:signed_log}) to the 
luminosity maps and a standard logarithmic transform (Eq.~\ref{eq:log_mass}) to the mass 
maps, followed by standardization using training set statistics. To assess the impact of 
these preprocessing schemes on model performance, we also explored two other alternative methods and 
compare all three using the multi-encoder model.

In the second strategy, negative luminosity values are discarded rather than sign-transformed. The luminosity maps are log-transformed only where the signal is positive,
\begin{equation}
    \tilde{x} =
    \begin{cases}
        \log_{10}(x) & x > 0, \\
        0            & x \leq 0,
    \end{cases}
    \label{eq:masked_log}
\end{equation}
and a binary mask describing the locations of valid pixels,
\begin{equation}
    \mathcal{M} =
    \begin{cases}
        1 & x > 0, \\
        0 & x \leq 0,
    \end{cases}
    \label{eq:mask}
\end{equation}
is concatenated as a second input channel with the luminosity map. Since invalid pixels are already set to zero in the log-transformed map, it provides a direct binary signal of pixel validity rather than requiring the network to infer invalid regions. The mass maps are transformed identically to the main text. No standardization is applied for this method.

In the third method, rather than zeroing invalid pixels, non-positive values are assigned a fixed very low negative value in log space,
\begin{equation}
    \tilde{x} =
    \begin{cases}
        \log_{10}(x) & x > 0,   \\
        -10          & x \leq 0,
    \end{cases}
    \label{eq:floor_log}
\end{equation}
with the same transform applied to both luminosity and mass maps. The floor value of $-10$ 
lies well below the typical range of the log-transformed observed luminosity signal. Again, no standardization is applied for this method.

The three schemes therefore differ in two key respects: how much information they preserve about negative pixels, and how they signal pixel validity to the network. The signed logarithm is the only method that retains the magnitude of negative luminosity values as physically meaningful signal, while the masked and floor-clipped variants treat them as missing data. The additional standardization applied in the first method further distinguishes it from the two methods explored here.

\begin{figure}[tbp]
    \centering
    \includegraphics[width=0.5\linewidth]{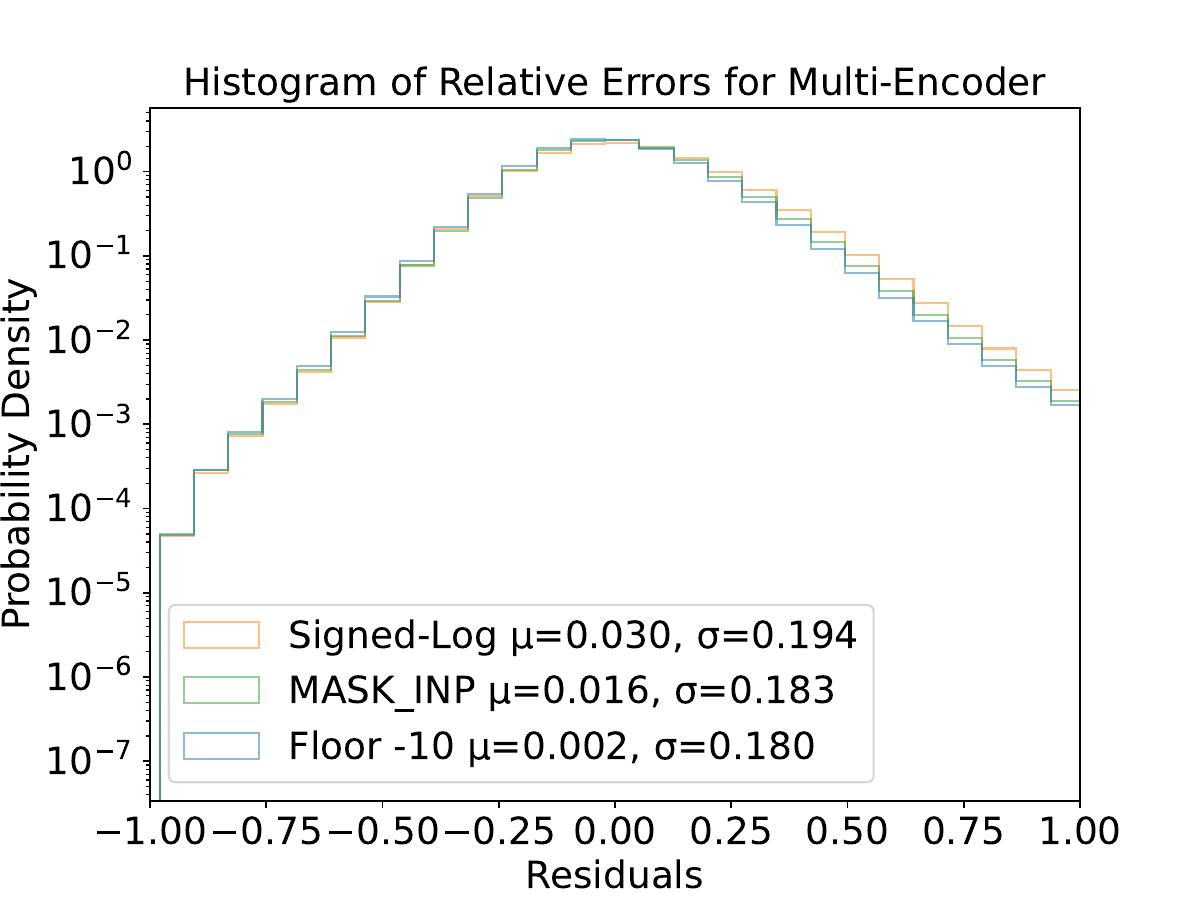}
    \includegraphics[width=0.5\linewidth]{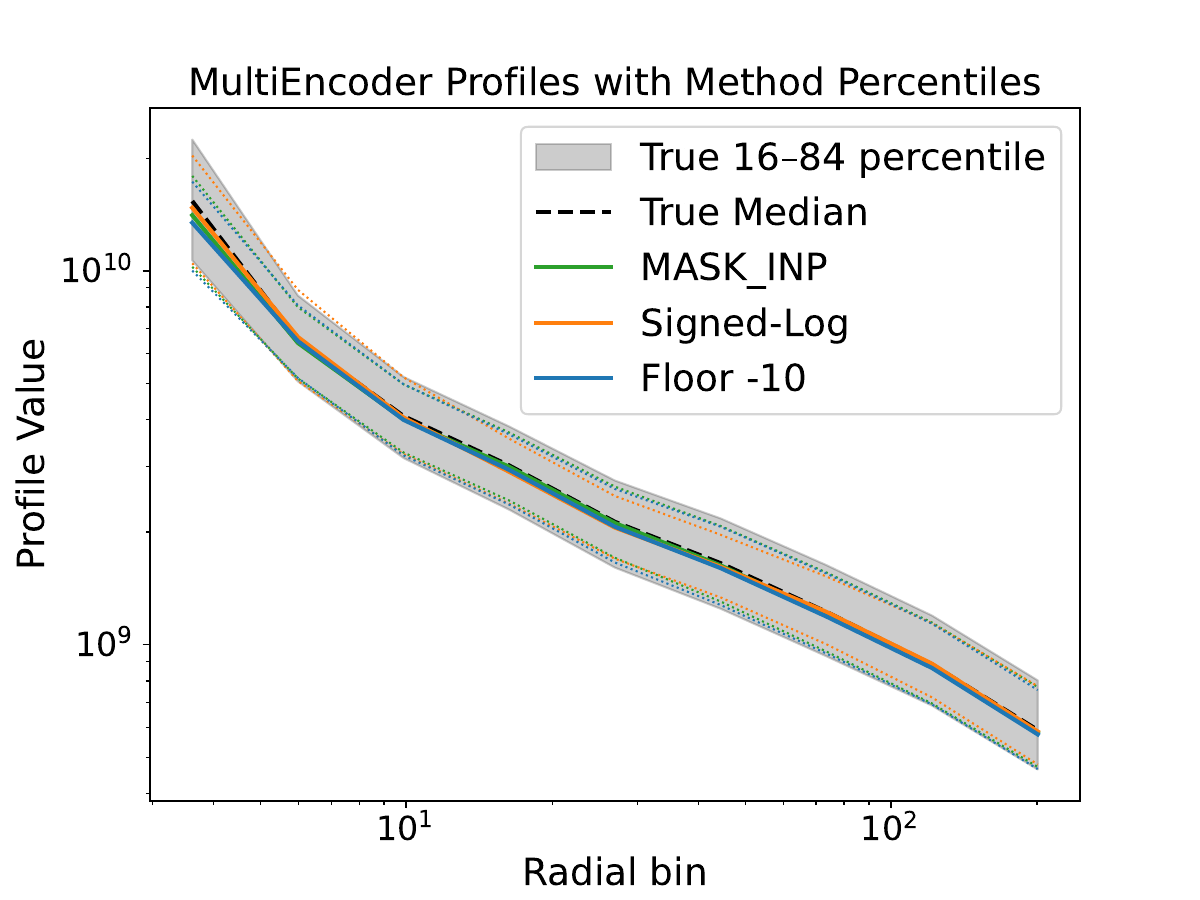}
    \includegraphics[width=0.5\linewidth]{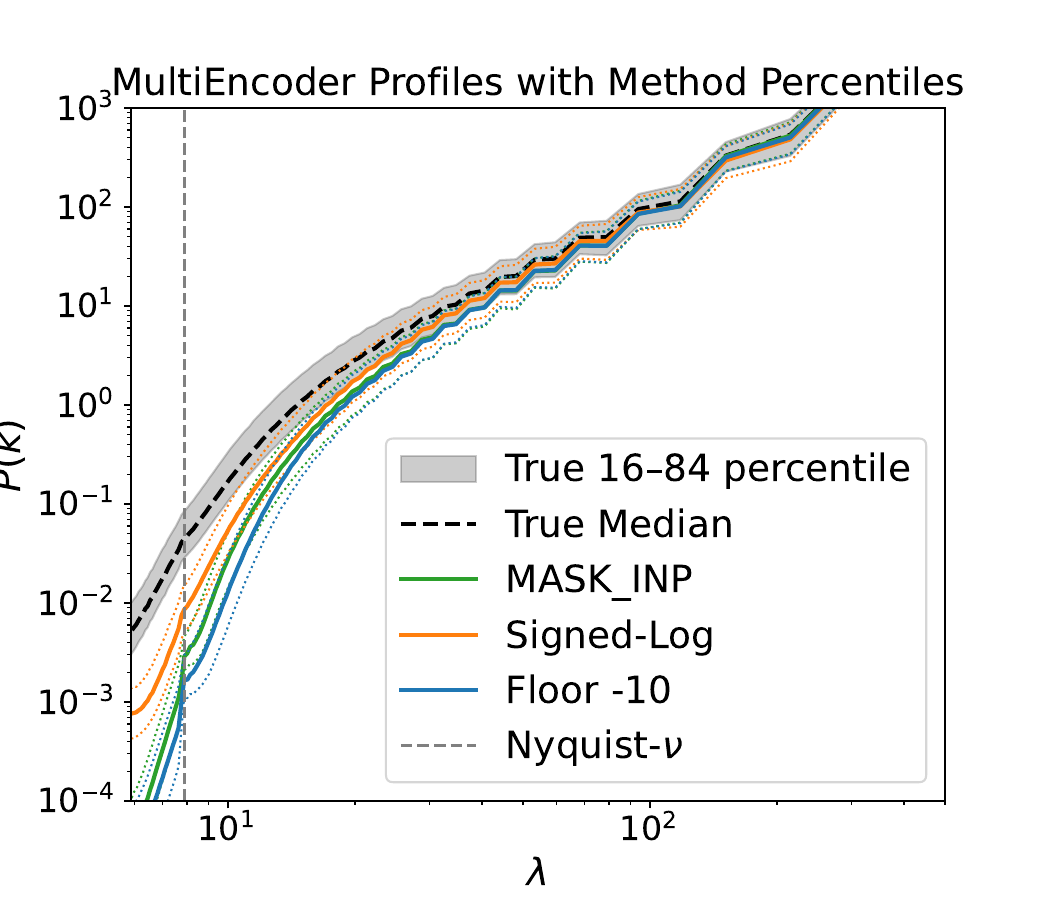}
    \caption{Comparison of the three luminosity map preprocessing schemes-signed logarithmic transform (main text), masked logarithm, and floor-clipped logarithm — evaluated on the full test set using the multi-encoder U-Net. First Panel: relative error between the true and predicted surface mass density maps Second Panel: azimuthally averaged surface mass density profile. Third Panel: power spectrum of the predicted maps compared to the ground truth.}
    \label{fig:Methods_Preprocessing}
\end{figure}

Fig.~\ref{fig:Methods_Preprocessing} compares the performance of the multi-encoder U-Net trained with the three luminosity map preprocessing schemes, evaluated on the full test set. The comparison is based on three metrics: the relative error between the true and predicted surface mass density, the azimuthally averaged surface mass density profile, and the power spectrum. While all three preprocessing methods yield broadly consistent results in terms of the relative error and the surface mass density 
profile, the signed logarithmic transform produces power spectra that are most consistent with the test set median and its percentile distribution, recovering small-scale structure more faithfully than the masked or floor-clipped variants. This motivates our choice of the signed logarithmic transform as the default preprocessing scheme in the main analysis.

\section{Substructures identification and mass estimation }\label{sec:peak_finder}
For the subhalo statistics study using mass maps derived from lensing and U-Net models for the CLASH clusters, we developed a subhalo identification and mass estimation algorithm. The aim is to identify subhalos directly from the mass maps and to define a criterion for assigning an adaptive aperture for their two-dimensional mass estimation.

We begin by first addressing the problem of aperture estimation for subhalos when their locations are known \textit{apriori}. For this, we used the available AHF halo catalog from the The300 simulation. We considered the training-set mass maps constructed with the line of sight along the $Z$-axis, and applied the same selection to the AHF catalog used in generating the projected mass maps along this direction, in order to identify the corresponding subhalos in the maps. We selected all subhalos gravitationally bound to the main host halo within the field of view, along with any additional objects falling within the same projected field of view and line-of-sight depth, in order to account for projected structures along the line of sight.

\subsection{Adaptive Aperture algorithm for Substructures}
For an identified subhalo with peak position $p_i = (x_i, y_i)$ in the two-dimensional mass map, we define an adaptive aperture radius $R_{i,\mathrm{ap}}$ centered on the peak. The aperture is calculated to scale with the relative prominence of the peak while also account for both local background variations and spatial distribution of neighboring subhalos.

We define a reference surface density at each peak as a weighted combination of a local background estimate and a global mean over all peaks,
\begin{equation}
    M_{i,\mathrm{ref}} = w\, M_{i,\mathrm{local}} + (1 - w)\, \bar{M}_\mathrm{sub},
\end{equation}
where $M_{i,\mathrm{local}}$ is the local background surface density around the peak, $\bar{M}_\mathrm{sub}$ is the mean surface density evaluated at all peak locations, and we adopt a fixed weight $w = 0.5$.
The aperture scale is defined as
\begin{equation}
    R_{i,\mathrm{phys}} = \alpha \left(\frac{M_{i,\mathrm{peak}}}{M_{i,\mathrm{ref}}}\right)^{\gamma} \frac{d_{\mathrm{nn}}}{2},
\end{equation}
where $M_{i,\mathrm{peak}}$ is the projected surface density at the peak and $d_{\mathrm{nn}}$ is the distance to the nearest neighboring peak.
To avoid nonphysical small apertures, we introduce a smooth minimum scale $R_\beta$, such that the final aperture radius is given by
\begin{equation}
    R_{i,\mathrm{ap}} =
    \begin{cases}
    \sqrt{R_{i,\mathrm{phys}}^2 + R_\beta^2}, & \text{if } R_{i,\mathrm{phys}} < R_\beta, \\
    R_{i,\mathrm{phys}}, & \text{otherwise}. \label{eq:aperture}
    \end{cases}
\end{equation}
The above procedure that subhalos in dense environments are assigned apertures relative to their local background, while also controlling excessive overlap through the dependence on the nearest-neighbor distance. The parameter $\gamma$ controls the sensitivity of the aperture size to the peak contrast, and $R_\beta$ acts as a regularizer. The enclosed mass is computed by integrating the surface density within a circular aperture of radius $R_{i,\mathrm{ap}}$. To achieve subpixel accuracy, the mass map is interpolated onto a high-resolution grid within the aperture using interpolation, and the integral is evaluated numerically over the selected pixels. The free parameters $\{\alpha, \gamma, R_\beta\}$ are calibrated on a training set of 500 simulated cluster mass maps drawn from the full sample of 3273 clusters. We employ a Bayesian optimization scheme to determine the parameter combination that best reproduces the known three-dimensional subhalo masses $M_{500c}$ from the AHF catalog, by jointly matching the differential and cumulative subhalo mass functions. This procedure yields the parameters values $\alpha = 0.21, \gamma = 1.2,~\mathrm{and}~R_\beta = 0.95 $ .

\begin{figure}[tbp] 
    \centering
    \includegraphics[width=\linewidth]{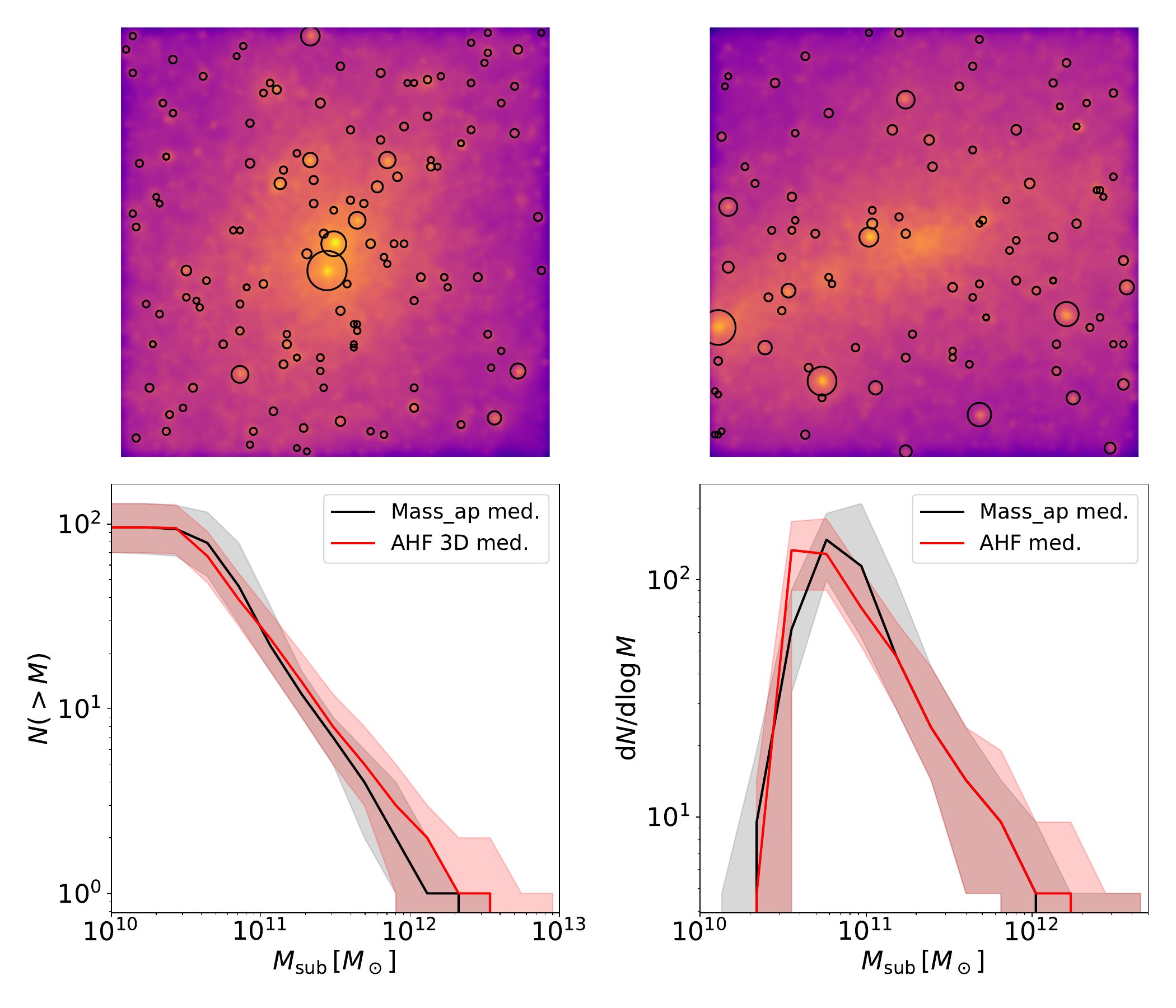}
   \caption{Validation of the adaptive aperture algorithm.  Top: Projected mass maps with adaptive apertures (black circles) overlaid using a priori subhalo positions. Bottom-left: Cumulative subhalo mass function $N(>M)$ from 2D aperture masses (grey) and 3D AHF catalog masses (red). Bottom-right: Corresponding differential subhalo mass function. Envelopes and lines indicate the $16^{\rm th}\text{--}84^{\rm th}$ percentile range and medians across 3,273 simulated clusters.}
    \label{fig:aperture constraint}
\end{figure}

The performance of the adaptive aperture algorithm is illustrated in Fig.~\ref{fig:aperture constraint}, where we validate the method using a subset of 1200 cluster mass maps drawn from the The300 simulation, under the assumption that the true subhalo positions are known \textit{apriori} from the AHF catalog. In the top panels, we present a visual representation of the method, showing two representative projected mass maps with the corresponding adaptive apertures overlaid. In the lower panels, we compare the subhalo mass functions derived from the two-dimensional aperture masses with those obtained from the three-dimensional AHF catalog. The cumulative and differential subhalo mass functions are shown, with shaded regions indicating the 16th--84th percentile range across the full sample, and solid lines representing the median relations. We find good agreement between the two mass functions, particularly at the massive end, in both shape and normalization. This demonstrates that the calibrated aperture prescription successfully recovers the underlying subhalo mass function, indicating that the algorithm is well tuned.

\subsection{Substructures Identification from Mass Maps}
To identify substructures directly from projected mass maps, we developed a multistage substructure detection algorithm designed to robustly extract density peaks across range of spatial scales. The algorithm mainly uses the following three steps for making detection:
\begin{enumerate}
    \item Background estimation and standardization: We construct a radially varying background for each mass map in order to isolate local density fluctuations associated with potential substructures by removing the overall smooth cluster component. This is achieved using a combination of median filtering and Gaussian smoothing at different spatial scales. 
    Specifically, we first apply two median filters: a fine-scale filter with a kernel size proportional to the base smoothing scale $\sigma_0$, and a coarse-scale filter with twice this kernel size. Each of these maps is then smoothed with Gaussian kernels of widths $\sigma_0$ and $1.5\,\sigma_0$, respectively. 
    To account for radial variations in the cluster profile, the two background estimates are smoothly blended using a sigmoid weighting function, defined as $w_{\mathrm{outer}} = 1 / \left(1 + \exp\left[-\frac{r - r_{\mathrm{mid}}}{r_{\mathrm{width}}}\right]\right)$.
    The spatial radial scale $r_{\mathrm{mid}}$ and ${r_{\mathrm{width}}}$ controls the transition from fine background near the center to a much coarser background at the outskirts for the mass map of the cluster. The final background is then given by $M_{\mathrm{bkg}} = (1 - w_{\mathrm{outer}})\, M_{\mathrm{inner}} + w_{\mathrm{outer}}\, M_{\mathrm{outer}}$. The resulting background map is subtracted from the original mass map to produce a residual map that highlights local over-densities. This residual map is subsequently standardized by subtracting its mean and dividing by its 1$\sigma$ standard deviation, and is used as input for the peak detection stage.
    \item Multi-scale peak detection : For the next step, we employ a multiscale detection scheme on the background-subtracted and standardized map. The standardized map is smoothed at different spatial scales using a sequence of logarithmically spaced Gaussian kernels, defined as $\sigma_{s} = \sigma_{0} \mathcal{f}^{s}$, where $\mathcal{f} = 1.25$ is the scale factor and $s$ denotes the smoothing step, ranging from 0 to 5. This results in a total of six smoothing scales used for substructure detection.  
    Peaks are initially detected at the finest smoothing scale, i.e. at $\sigma_{0}$, using a local maximum finder from \texttt{skimage}, with a minimum separation proportional to $\sigma_{0}$ and an absolute detection threshold $\alpha_{\mathrm{det}}$. The identified candidate peaks are then validated using the multiscale scheme. At each smoothing scale, the maximum value within a small neighborhood around each candidate peak is evaluated, allowing for slight positional drift. A peak is considered persistent at a given scale if its value remains above a fraction $\eta$ of its base-scale amplitude. We assign a score of $+1$ whenever this condition is satisfied. A candidate is accepted as a genuine detection only if it persists in at least 50\% of the smoothing scales.
    The motivation for this multi-scale scheme is to retain genuine detections while rejecting spurious noise fluctuations, as true subhalo peaks persist across multiple smoothing scales, whereas noisy peaks are washed out. The free parameters smoothing scale $\sigma_{0}$, the detection threshold $\alpha_{\mathrm{det}}$, and the persistence threshold $\eta$, are constrained through calibration with the AHF catalog.
    \item Signal-to-noise (SNR) filtering : To further suppress spurious detections, we compute a local SNR ratio for each candidate peak. The noise is estimated from an annular region centered on the peak, with inner and outer radii adapted to the distance to the nearest neighboring peak in order to avoid contamination from nearby structures. Specifically, the noise is estimated as 1.4826 $\times$ the median absolute deviation within the annulus, and the SNR was computed using this noise, mass at the peak value and mass within the annular region. A potential substructure is retained if it satisfies the afore-mentioned persistence criterion and minimum SNR threshold of $\alpha_{snr}$. The final output of the algorithm is the set of detected substructures position in the two-dimensional mass map.
\end{enumerate}
The hyperparameters $\{\sigma_0, \alpha_{\mathrm{det}}, \eta, \alpha_{\mathrm{snr}}\}$ are optimized using simulated cluster mass maps, where the true substructures (halos) are taken from the AHF catalog. We use \texttt{Hyperopt} with Bayesian optimization (TPE algorithm) to explore the parameter space.  
For each set of parameters, the detection algorithm is run on the small sample of pre-selected mass maps, and the predicted subhalos are compared with the AHF halos. The performance is evaluated using a combined loss that includes agreement with the true mass function $(dN/d\log M)$, along with precision and recall. The optimizer then tries new parameter values, focusing more on those that give lower loss. In this way, the algorithm is tuned to recover both the correct number of subhalos and their positions as given by AHF. The Bayesian optimization constrained these hyper-parameters values to $\sigma_0 = 0.75, \alpha_{\mathrm{det}} = 1.01, \eta = 0.32,~\mathrm{and}~\alpha_{\mathrm{snr}} = 1.90.$.These parameters balance completeness and precision in the detected subhalo population.

When this hyper-parameter peak detection algorithm is evaluated on 2000 randomly selected maps from the The300 sample, it results in a median recall of $0.750 \pm 0.149$ and a median precision of $0.881 \pm 0.102$. The recall measures the fraction of true AHF subhalo peaks that are successfully recovered, while precision quantifies the fraction of detected peaks that correspond to genuine subhalos. This implies that approximately 25 per cent of the true subhalo peaks are missed by the algorithm, while about 12 per cent of the detections are false positives. Restricting to subhalos above the resolution limit of $10^{11},M_\odot$, the algorithm achieves an average completeness of $0.952 \pm 0.060$, requiring detections to lie within 2 pixels ($\approx 8$ kpc) of the true AHF positions. Overall, this demonstrates that the calibrated peak detection algorithm reliably recovers the underlying subhalo population, particularly at the high-mass end, and supports its application to both simulated data and observed CLASH cluster mass maps
\begin{figure}[tbp] 
    \centering
    \includegraphics[width=.8\linewidth]{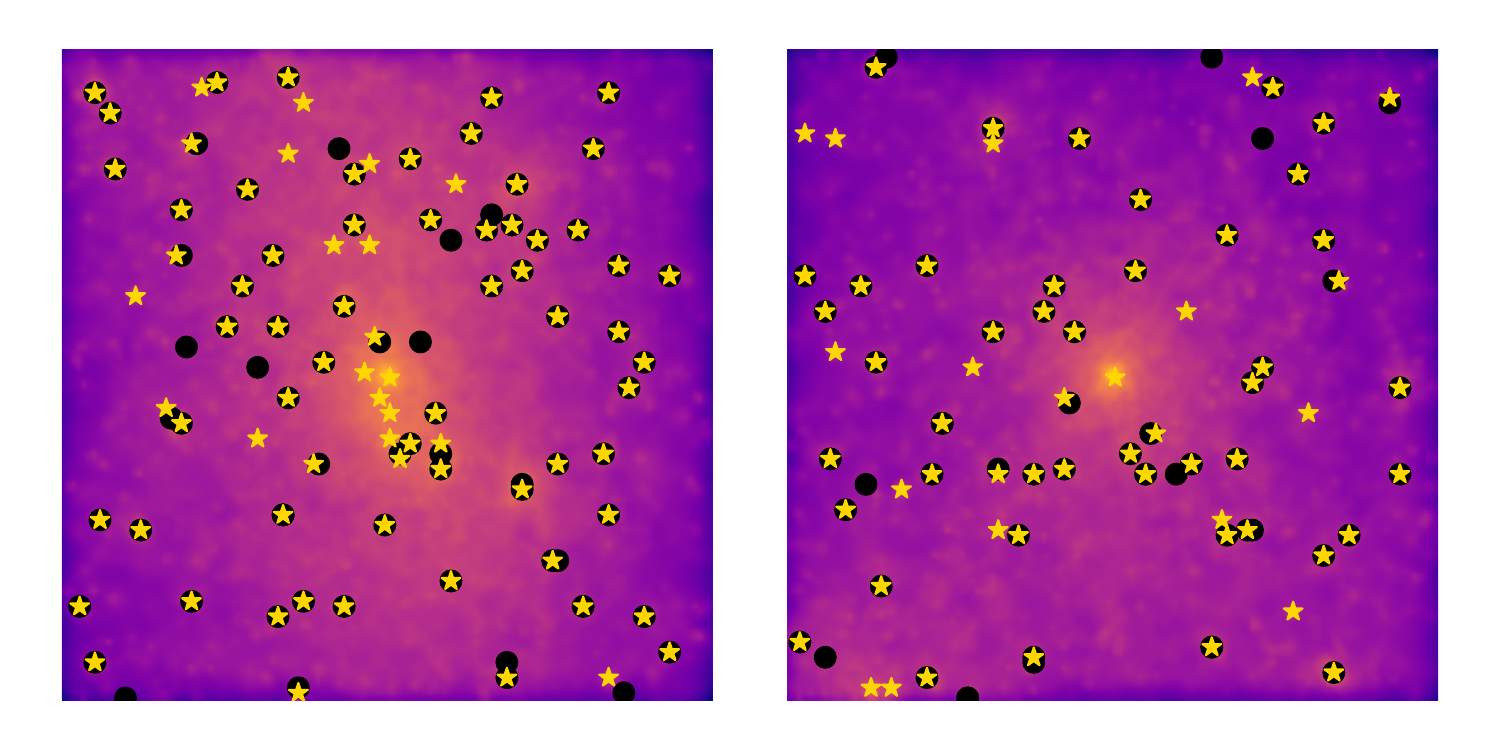}
    \includegraphics[width=.49\linewidth]{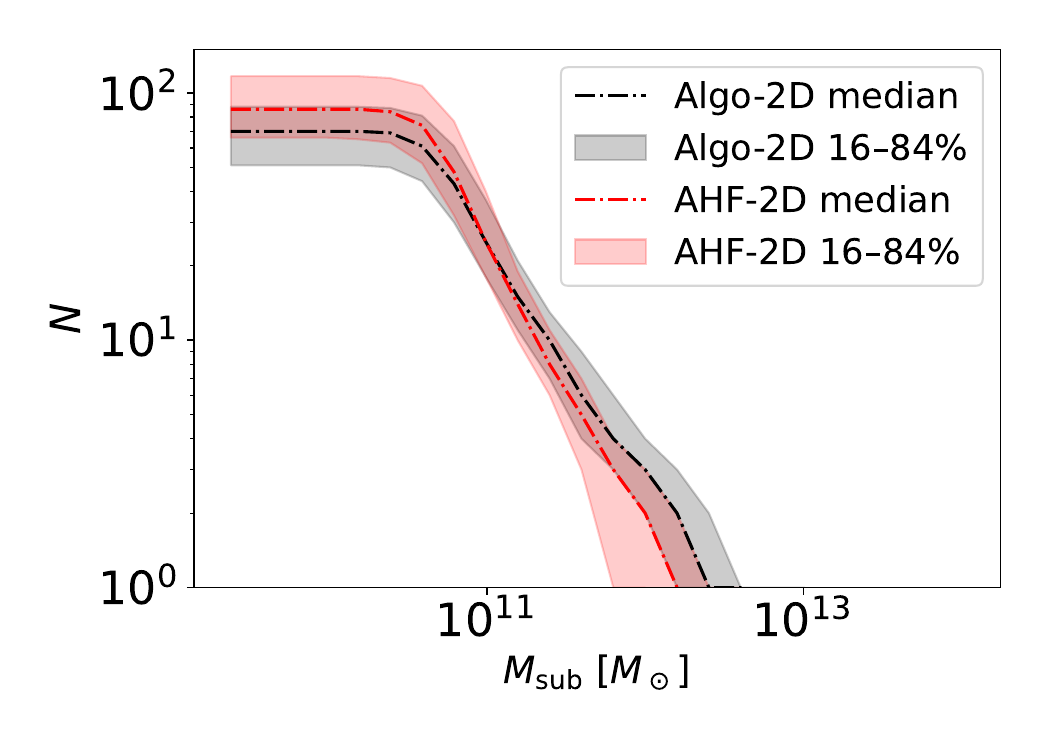} 
    \includegraphics[width=.49\linewidth]{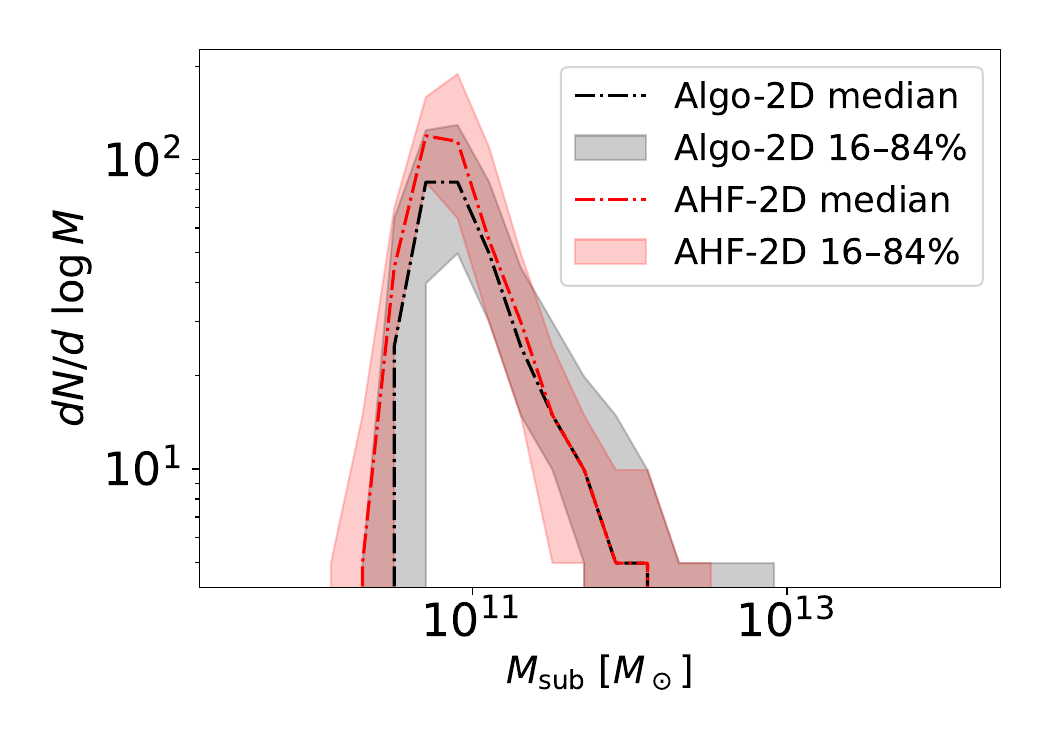}
\caption{Subhalo peak detection algorithm performance. 
Top: Projected mass maps with true AHF subhalo positions (black circles) and algorithm-detected peaks (gold stars). The algorithm recovers most true subhalos, including outskirt substructures, with minimal false positives or missed detections. Bottom-left: Cumulative subhalo mass function from aperture masses of algorithm-detected peaks (black, Algo-2D) and true AHF peaks (red, AHF-2D). Bottom-right: Corresponding differential subhalo mass function. 
Envelopes and lines indicate the $16^{\rm th}\text{--}84^{\rm th}$ percentile range and medians across the full sample.}    \label{fig:peak_finder_algorithm_pics}
\end{figure}

\begin{figure}[tbp] 
    \centering
    \includegraphics[angle=90,width=\linewidth]{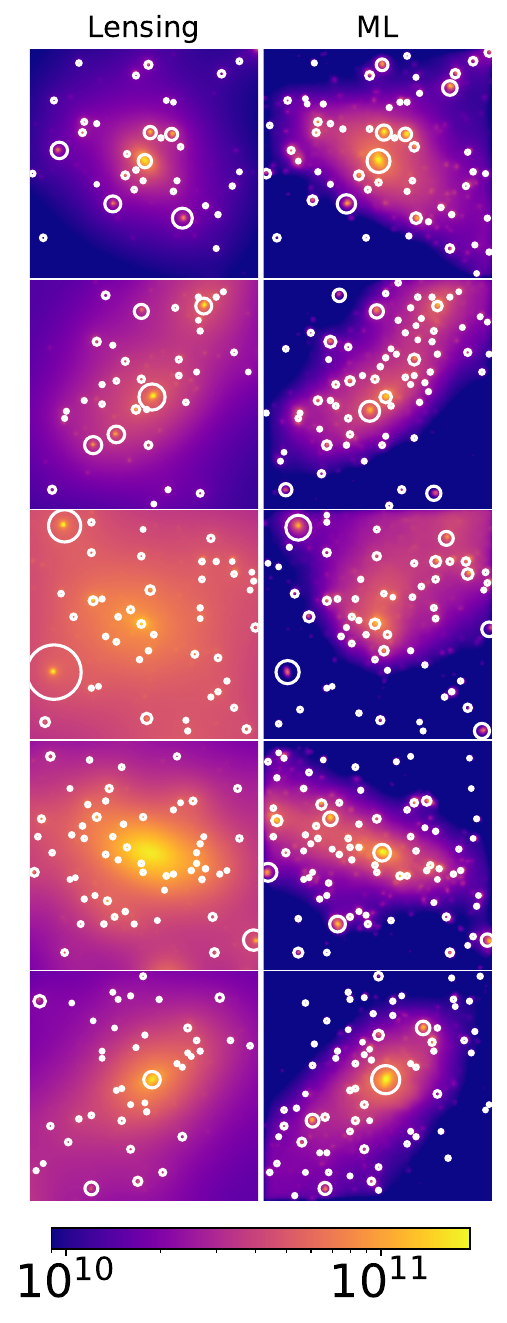}
    \caption{Substructures in CLASH sample. Projected mass maps of the five CLASH clusters with spectroscopic redshift catalogs, with identified subhalo peaks and their adaptive apertures overlaid as white circles. The upper panels show the lensing-derived mass maps and the lower panels show the corresponding U-Net-predicted mass maps. The radius of each identified substructure is given by (Eq.~\ref{eq:aperture}).}
    \label{fig:CLASH-Apertures}
\end{figure}
Fig.~\ref{fig:peak_finder_algorithm_pics} presents the performance of the calibrated peak detection algorithm evaluated on the full The300 simulation sample. The top panels show two representative projected mass maps, with the true AHF subhalo positions and the algorithm-detected peaks overlaid. The algorithm successfully recovers the majority of true subhalo locations, with only a small number of missed detections and false positives, mainly in the dense central regions. The bottom panels show the cumulative and differential subhalo mass functions derived from the aperture masses of the detected peaks (Algo-2D) and the true AHF peaks (AHF-2D), shown as the median and 16th–84th percentile range across the full sample. The two mass functions are in good agreement in both shape and normalization, with the median line lying within scatter over the full subhalos mass range. 
In Fig.~\ref{fig:CLASH-Apertures}, we show the identified substructures and their apertures in the lensing maps (top panel) and the multi-encoder U-Net predicted maps (bottom panel) for the five main CLASH clusters discussed in Section~\ref{sec:Results} of the main manuscript.

\end{appendix}
\end{document}